\documentclass{aa_old}
\usepackage{psfig}

\begin{document}

\title{Unveiling the nature of {\it INTEGRAL} objects through optical 
spectroscopy. VI. A multi-observatory identification campaign\thanks{Based on 
observations collected at the following observatories: ESO (La Silla, Chile),
partly under program 079.A-0171(A); Astronomical Observatory of Bologna in 
Loiano (Italy); Astronomical Observatory of Asiago (Italy); Cerro Tololo 
Interamerican Observatory (Chile); Complejo Astron\'omico El Leoncito (San
Juan, Argentina); South African Astronomical Observatory (Sutherland, 
South Africa); Observatorio del Roque de los Muchachos 
of the Instituto de Astrofísica de Canarias (Canary Islands, Spain); 
Anglo-Australian Observatory (Siding Spring, Australia); Apache Point 
Observatory (New Mexico, USA).}}

\author{N. Masetti\inst{1},
E. Mason\inst{2},
L. Morelli\inst{3},
S.A. Cellone\inst{4,5},
V.A. McBride\inst{6},
E. Palazzi\inst{1},
L. Bassani\inst{1}, \\
A. Bazzano\inst{7}, 
A.J. Bird\inst{6},
P.A. Charles\inst{6,8},
A.J. Dean\inst{6},
G. Galaz\inst{9},
N. Gehrels\inst{10},
R. Landi\inst{1},
A. Malizia\inst{1}, \\
D. Minniti\inst{9},
F. Panessa\inst{7},
G.E. Romero\inst{4,11},
J.B. Stephen\inst{1},
P. Ubertini\inst{7} and
R. Walter\inst{12}
}

\institute{
INAF -- Istituto di Astrofisica Spaziale e Fisica Cosmica di 
Bologna, Via Gobetti 101, I-40129 Bologna, Italy
\and
European Southern Observatory, Alonso de Cordova 3107, Vitacura,
Santiago, Chile
\and
Dipartimento di Astronomia, Universit\`a di Padova,
Vicolo dell'Osservatorio 3, I-35122 Padua, Italy
\and
Facultad de Ciencias Astron\'omicas y Geof\'{\i}sicas, Universidad Nacional
de La Plata, Paseo del Bosque, B1900FWA La Plata, Argentina
\and
IALP (CONICET--UNLP), Argentina
\and
School of Physics \& Astronomy, University of Southampton, Southampton, 
Hampshire, SO17 1BJ, United Kingdom  
\and
INAF -- Istituto di Astrofisica Spaziale e Fisica Cosmica di
Roma, Via Fosso del Cavaliere 100, I-00133 Rome, Italy
\and
South African Astronomical Observatory, P.O. Box 9, Observatory 7935,
South Africa
\and
Departamento de Astronom\'{i}a y Astrof\'{i}sica, Pontificia Universidad 
Cat\'olica de Chile, Casilla 306, Santiago 22, Chile
\and
NASA/Goddard Space Flight Center, Greenbelt, MD 20771, USA
\and
Instituto Argentino de Radioastronom\'{\i}a (CONICET), C.C. 5, 1894 Villa
Elisa, Buenos Aires, Argentina
\and
INTEGRAL Science Data Centre, Chemin d'Ecogia 16, CH-1290 Versoix,
Switzerland
}

\offprints{N. Masetti (\texttt{masetti@iasfbo.inaf.it)}}
\date{Received 27 December 2007; accepted 5 February 2008}

\abstract{
Using 8 telescopes in the northern and southern hemispheres, 
plus archival data from two on-line sky surveys, we performed a systematic
optical spectroscopic study of 39 putative counterparts of unidentified or 
poorly studied {\it INTEGRAL} sources in order to determine or at least
better assess their nature. This was implemented within the framework of 
our campaign to reveal the nature of newly-discovered and/or 
unidentified sources detected by {\it INTEGRAL}.
Our results show that 29 of these objects are active galactic nuclei 
(13 of which are of Seyfert 1 type, 15 are Seyfert 2 galaxies and one
is possibly a BL Lac object) with redshifts between 0.011 and 0.316, 7 
are X--ray binaries (5 with high-mass companions and 2 with low-mass 
secondaries), one is a magnetic cataclysmic variable, one is a symbiotic 
star and one is possibly an active star. Thus, the large majority (74\%) 
of the identifications in this sample belongs to the AGN class.
When possible, the main physical parameters for these hard X--ray sources 
were also computed using the multiwavelength information available in the 
literature. These identifications further underscore the importance of 
{\it INTEGRAL} in studying the hard X--ray spectra of all classes of
X--ray emitting objects, and the effectiveness of a strategy of 
multi-catalogue cross-correlation plus optical spectroscopy to securely 
pinpoint the actual nature of still unidentified hard X--ray sources.
}

\titlerunning{A multi-observatory identification campaign of IGR sources}
\authorrunning{N. Masetti et al.}

\maketitle

\section{Introduction}

One of the  main objectives of the {\it  INTEGRAL} mission (Winkler et
al.  2003) is to survey the whole sky in the hard X--ray band above
20 keV. This makes use of the unique imaging capability of the 
IBIS instrument (Ubertini et al. 2003) which allows the detection of 
sources at the mCrab level with a typical localization accuracy of 
2-3 arcmin (Gros et al. 2003). 

The IBIS survey allowed pinpointing, for the first time, extragalactic 
sources in the so-called `Zone of Avoidance', which hampers observations 
in soft X--rays along the Galactic plane due to the presence of gas and 
dust. Moreover, this survey is expanding our knowledge about Galactic 
X--ray binaries, by showing the existence of a new class of heavily 
absorbed supergiant massive X--ray binaries (e.g., Walter et al. 2004), 
by allowing the discovery of supergiant fast X--ray 
transients (e.g., Sguera et al. 2006; Leyder et al. 2007), by doubling the 
number of known high-mass X--ray binaries (HMXBs; see Walter 2007), and by 
detecting a number of new magnetic cataclysmic variables (CVs; e.g., 
Barlow et al. 2006; Bonnet-Bidaud et al. 2007).

Up to now, IBIS detected more than 500 sources in the hard
X--rays between 20  and 100 keV (see e.g., Bird  et al. 2007; Krivonos
et al. 2007; see also Bodaghee et al. 2007).  Most of  these sources are 
known Galactic X--ray binaries
($\sim$35\%  of the  total number  of detected  objects),  followed by
active     galactic    nuclei     (AGNs,    $\sim$28\%)     and    CVs
($\sim$5\%). However,  a large number of the  remaining objects (about
27\% of all the IBIS detections) has no obvious counterpart
at other  wavelengths and  therefore cannot immediately  be associated
with   any  known   class  of   high-energy  emitting   objects.   The
multiwavelength study  of these unidentified sources  is thus critical
to determine their nature.  Therefore,  in 2004 we started a multisite
observational  campaign   devoted  to  the   identification  of  these
unidentified  objects  through optical  spectroscopy (see Masetti  et
al. 2004, 2006a,b,c,d; hereafter Papers I-V). Our results showed that
about  half of these  objects are  nearby ($z  \la$ 0.1)  AGNs (Papers
I-V), while a non-negligible fraction ($\sim$15\%) of objects belongs
to the class of magnetic CVs (Paper V and references therein).

In parallel, as a service to the scientific community, we maintain a web 
page\footnote{{\tt http://www.iasfbo.inaf.it/IGR/main.html}} reporting 
information on {\it INTEGRAL} sources identified via optical or 
near-infrared observations.

Continuing our  effort to identify unknown {\it  INTEGRAL} sources, we
here present optical  spectroscopy of  39 more  objects  which were
detected by  IBIS but are  still unidentified, unclassified  or poorly
studied.  Their   spectra  have  been  obtained   at  eight  different
telescopes around the world or retrieved from two public spectroscopic 
archives.

The paper is structured as follows: in Sect. 2 we explain the criteria
used to  select the sample of  {\it INTEGRAL} and  optical objects for
the present observational  campaign.  In Sect. 3 a  description of the
observations and of  the employed telescopes is given.  Sect. 4
reports and discusses the results,  divided into four broad classes of
sources  (CVs, X--ray binaries, AGNs  and peculiar  sources), together
with a  statistical outline of  the identifications of  {\it INTEGRAL}
sources obtained until now. Conclusions are drawn in Sect. 5.

\begin{figure*}
\caption{From left to  right and top to bottom:  optical images of the
fields  of IGR  J00040+7020, IGR  J00256+6821, IGR  J01528$-$0326, IGR
J02343+3229, IGR J02504+5443, IGR J03334+3718, IGR J06117$-$6625, IGR
J06292+4858 and  IGR J07437$-$5147.  The optical counterparts  of the
{\it INTEGRAL} sources are indicated with tick marks.  Field sizes are
5$'$$\times$5$'$ and are extracted  from the DSS-II-Red survey. In all
cases, North is up and East to the left.}
\end{figure*}

\begin{figure*}
\caption{As  Fig. 1,  but for  the  fields of  IGR J08023$-$6954,  IGR
J09446$-$2636, IGR J09523$-$6231,  IGR J11366$-$6002, IGR J12131+0700,
IGR   J13038+5348,  IGR   J13109$-$5552,  IGR   J13149+4422   and  IGR
J14298$-$6715. }
\end{figure*}

\begin{figure*}
\caption{As  Fig. 1,  but for  the  fields of  IGR J14331$-$6112,  IGR
J14471$-$6414,  IGR J14561$-$3738,  IC 4518a, IGR  J15161$-$3827, IGR
J15539$-$6142,   IGR  J16024$-$6107,   IGR   J16056$-$6110  and   IGR
J16385$-$2057.  The   field  of  IGR   J14471$-$6414  is 3$'$$\times$3$'$  
in  size and  has  been  obtained  at the  ESO  3.6m telescope using 
EFOSC2 and the $R$ filter.}
\end{figure*}

\begin{figure*}
\caption{As  Fig. 1,  but  for  the fields  of  IGR J16500$-$3307,  2E
1739.1$-$1210,   1RXS  J174607.8$-$213333,  SAX   J1802.7$-$2017,  IGR
J18048$-$1455,    SAX   J1818.6$-$1703,    IGR    J18483$-$0311,   IGR
J19405$-$3016  and Swift J2000.6+3210.  The fields   of    SAX   
J1802.7$-$2017   and    IGR   J18048$-$1455   are 3$'$$\times$3$'$ in size  
and were obtained at the  ESO 3.6m telescope with EFOSC2 plus  the $R$ 
filter.}
\end{figure*}

\begin{figure*}
\caption{As Fig. 1, but for the fields of IGR J21272+4241
(left panel), IGR J23308+7120 (middle panel) and IGR
J23542+5842 (right panel).}
\end{figure*}

\section{Sample selection}

In order to continue our program (Papers I-V) of identification of the
{\it INTEGRAL} sources  with unknown or poorly known  nature, we first
collected all  unidentified or  unclassified objects belonging  to the
3$^{\rm rd}$ IBIS  Survey (Bird et al. 2007),  the IBIS All-Sky Survey
(Krivonos et al. 2007), and  the Galactic Center Survey (Revnivtsev et
al. 2004a). Further information on some of these sources can be found
in Kuiper et al. (2006) and Keek et  al. (2006). 

We then cross-correlated the IBIS  positions with those in the catalogues
of soft ($<$10 keV) X--ray  sources. This allowed to reduce the X--ray
error  box  size  to   typically  less  than  $\la$10  arcsecs.   More
specifically,  for  the present  sample,  we  selected {\it  INTEGRAL}
objects which have, within their  IBIS error box, single source detections 
by either  {\it ROSAT}  (Voges et  al.  1999),  or {\it  Swift}/XRT ({\tt
http://www.asdc.asi.it};  see also  Landi et  al. 2007b,c,d,f and 
Rodriguez et al. 2007),  or {\it
Chandra}  ({\tt http://cxc.harvard.edu}),  or  {\it XMM-Newton}  ({\tt
http://heasarc.gsfc.nasa.gov}; see also Saxton  et al. 2008 and Walter
et al. 2006).  This approach was chosen as Stephen et al. (2005, 2006)
showed,  from statistical considerations,  that these  are very  likely 
the soft X--ray  counterparts of the {\it INTEGRAL}  sources.  Indeed, our
previous  observations,  which  were   based  on  the  same  selection
criteria, have sucessfully identified 40 selected targets (see Papers I-V).

For  the  cross-correlation searches,  we  adopted 90\%  confidence
level  {\it INTEGRAL}/IBIS  error circles.  This corresponds to  error box
radii of 3$\farcm$5 and 5$'$ for the IBIS All-Sky Survey (Krivonos et
al. 2007) and the Galactic Center Survey (Revnivtsev  et al. 2004a),
respectively.  For the 3$^{\rm rd}$ IBIS Survey sources we adopted the
error box radii reported by Bird et al. (2007).

After this first selection of objects with arcsec-sized X--ray 
positions, we next chose out of them a subsample of cases for which, when 
these refined error boxes were overlaid onto the corresponding
DSS-II-Red          survey\footnote{available          at         {\tt
http://archive.eso.org/dss/dss}}  images, they were found to contain
a single or few (3 at most) possible optical  counterparts with magnitudes 
$R \la$ 20,  that is, objects for which optical spectroscopy could be 
obtained with reasonable S/N ratio at telescopes of small and medium size
(up to 4 metres).

In this way we selected 35 sources. However,  we note that, for object 
IGR J08023$-$6954, we found only a marginal ($<$3$\sigma$) {\it
Swift}/XRT detection within the IBIS error circle, positionally consistent
with one putative optical counterpart. For IGR J12131+0700, instead, two
{\it Swift}/XRT soft X--ray sources were found within  the IBIS error
circle (Landi et al 2007e). Here, we focus only on the XRT source \#1
(according to the notation of Landi et al. 2007e) given  that it is
the brightest one and it has an optical counterpart.

Three  additional  sources  (IGR  J06292+4858, IGR J07437$-$5137
and IGR J14561$-$3738) were added to our sample though they do not
have a soft X--ray counterpart. The latter two cases were
chosen because the IBIS error circle encompasses a single radio
source reported in the NVSS (Condon  et al. 1998)
or SUMSS (Mauch et al. 2003) catalogues, as well as a
far-infrared IRAS source (IRAS 1988) consistent with the radio
position, and in addition a single optical object could be found
at the far-infrared and radio locations. For IGR J06292+4858,
instead, we chose a conspicuous optical galaxy within the {\it INTEGRAL}
error box. We in any case refer the reader to Paper III for the
caveats and the shortcomings of choosing, within the IBIS error box,
``peculiar" sources which are not readily associated with an
arcsec-sized soft X--ray position.

A further caveat that is to be mentioned is that some sources of our 
sample and extracted from the catalogue of Bird et al. (2007), namely IGR 
J06292+4858, IGR J12131+0700, IGR J15161$-$3827, IGR J16056$-$6110, IGR 
J19405$-$3016 and IGR J21272+4241, have statistical significance of their 
IBIS detection between 4.5$\sigma$ and 4.9$\sigma$. While {\it INTEGRAL} 
sources detected with significance above 5$\sigma$ have a probability of 
less than 1\% of being spurious (see Bird et al. 2007 and Krivonos et al. 
2007), detections with lower significance might actually have a larger 
probability of being not real. Therefore, we caution the reader that, 
because of this, the hard X--ray detection of the above sources may be 
spurious and that the proposed optical identification may not correspond 
to an actual object emitting at high energies above 20 keV.

With the approach illustrated above, we could eventually 
select 38 unidentified, unclassified or poorly studied {\it INTEGRAL} 
sources that are associated to arcsec-sized soft X--ray or radio error 
boxes, plus one object possibly associated with a peculiar optical source 
in the IBIS error circle.

Figures 1-5 report the optical fields of the 39 sources of the selected 
sample. The corresponding optical counterparts are indicated with tick 
marks. The list of identified {\it INTEGRAL} sources is reported in Table 
1 (which we thoroughly describe in the next Section).

We here put forward that, for the high-energy sources with more than one 
optical candidate, we spectroscopically observed all objects with 
magnitude $R\la$ 20 within the longer-wavelength arcsec-sized error 
circle. However, in the following we will report only on their firm or 
likely optical counterparts, recognized via their peculiar spectral 
features (basically, the presence of emission lines). All other candidates 
are discarded because their spectra do not show any peculiarity (in 
general they are recognized as Galactic stars) and will not be considered 
further.

In our final sample one can also find the {\it INTEGRAL} objects mentioned 
by Torres et al. (2004), Negueruela \& Smith (2006) and Burenin et al. 
(2006a,b; see also Halpern 2006). These objects, although already 
identified by means of preliminary reports, still have fragmentary 
longer-wavelength information.  Our observations are meant to confirm the 
identification and to improve classification and knowledge of these hard 
X--ray sources.

\section{Optical spectroscopy}

The data collected and presented in this work are the result of a 
multisite campaign that involved the following observatories and 
telescopes in the past 2 years: 

\begin{itemize}
\item the 1.5m at the Cerro Tololo Interamerican Observatory (CTIO), Chile;
\item the 1.52m ``Cassini'' telescope of the Astronomical Observatory of 
Bologna, in Loiano, Italy; 
\item the 1.8m ``Copernicus'' telescope at the Astrophysical Observatory 
of Asiago, in Asiago, Italy, 
\item the 1.9m ``Radcliffe'' telescope at the South African Astronomical 
Observatory (SAAO), in Sutherland, South Africa; 
\item the 2.15m ``Jorge Sahade'' telescope at the Complejo Astronomico el 
Leoncito (CASLEO) in Argentina;
\item the 3.5m ``New Technology Telescope'' (NTT) and the 3.6m telescope 
at the ESO-La Silla Observatory, Chile; and
\item the 4.2m ``William Herschel Telescope'' (WHT) at the Roque de Los 
Muchachos Observatory in La Palma, Spain. 
\end{itemize}

The data taken by at these telescopes were reduced following standard 
procedures using IRAF\footnote{IRAF is the Image Reduction and Analysis 
Facility made available to the astronomical community by the National 
Optical Astronomy Observatories, which are operated by AURA, Inc., under 
contract with the U.S. National Science Foundation. It is available at 
{\tt http://iraf.noao.edu/}}.  Calibration frames (flat fields and lamps 
for wavelength calibration) were taken on the day preceeding or following 
the observing night.  The wavelength calibration uncertainty was 
$\sim$0.5~\AA~for all cases; this was checked using the positions of 
background night sky lines.  Flux calibration was performed using 
catalogued spectrophotometric standards.

Moreover, further spectroscopic data were retrieved from two different 
astronomical archives. In particular, we retrieved 2 pipeline reduced 
spectra from the Sloan Digitized Sky Survey\footnote{{\tt 
http://www.sdss.org}} (SDSS, Adelman-McCarthy et al. 2005) archive, and 4 
from the Six-degree Field Galaxy Survey\footnote{{\tt 
http://www.aao.gov.au/local/www/6df/}} (6dFGS) archive (Jones et al. 
2004). As the 6dFGS archive provides spectra which are not calibrated in 
flux, we used the optical photometric information in Jones et al. (2005) 
and Doyle et al. (2005) to calibrate the 6dFGS data presented here.

We report in Table~1 the  detalied log of observations.  In particular
we list  in column  1 the  name of the  observed {\it INTEGRAL}  sources. 
In columns 2 and 3 we report the object coordinates, extracted 
from the 2MASS catalogue (with an accuracy of $\leq$0$\farcs$1, according 
to Skrutskie et al. 2006) or from the DSS-II-Red astrometry (which has an 
accuracy of $\sim$1$''$). In column 4 we list the telescope and  the  
instrument used for the observations. The characteristic of each 
spectrograph are given in columns 5 and 6. Column 7 provides the 
observation date and the UT time at mid-exposure, while column 8 reports 
the exposure times and the number of spectral pointings.

As a complement to the information on the putative counterparts
of IGR J14331$-$6112, SAX J1802.7$-$2017 and IGR J18048$-$1455,
we analyzed optical $R$-band frames acquired in parallel with the
corresponding spectroscopic pointings on these sources.
The field of SAX J1802.7$-$2017 was observed with NTT plus EMMI on
28 July 2007 (start time: 02:01 UT; duration: 10 s) under a seeing
of 1$\farcs$5; the 2$\times$2-rebinned CCDs of EMMI secured a plate scale
of 0$\farcs$33/pix, and a useful field of 9$\farcm$9$\times$9$\farcm$1.
A 20-second image of the field of IGR J14331$-$6112 was
obtained on 2007 June 22 (start time: 04:21 UT) with the 3.6m ESO
telescope plus EFOSC2 under a seeing of 1$\farcs$2; for the field of
IGR J18048$-$1455, a 10-second image was acquired on 2007 June 23
(start time: 06:13 UT) with the same setup as above under a seeing
of 1$\farcs$7. These two images, 2$\times$2 binned, had a scale of
0$\farcs$31/pix and covered a field of 5$\farcm$2$\times$5$\farcm$2.

The corresponding imaging frames were corrected for bias and flat-field in 
the usual fashion and calibrated using nearby USNO-A2.0\footnote{available 
at \\ {\tt http://archive.eso.org/skycat/servers/usnoa/}} stars. 
Simple aperture photometry,
within the MIDAS\footnote{\texttt{http://www.eso.org/projects/esomidas}}
package, was then used to measure the $R$-band magnitude of the putative
optical counterparts of IGR J14331$-$6112, SAX J1802.7$-$2017 and IGR
J18048$-$1455.

\begin{table*}[th!]
\caption[]{Log of the spectroscopic observations presented in this paper
(see text for details). If not otherwise indicated, source coordinates
are extracted from the 2MASS catalogue and have an accuracy better than 0$\farcs$1.}
\scriptsize
\begin{center}
\begin{tabular}{llllcccr}
\noalign{\smallskip}
\hline
\hline
\noalign{\smallskip}
\multicolumn{1}{c}{Object} & \multicolumn{1}{c}{RA} & \multicolumn{1}{c}{Dec} & 
\multicolumn{1}{c}{Telescope+instrument} & $\lambda$ range & Disp. & \multicolumn{1}{c}{UT Date \& Time}  & Exposure \\
 & \multicolumn{1}{c}{(J2000)} & \multicolumn{1}{c}{(J2000)} & & (\AA) & (\AA/pix) & 
\multicolumn{1}{c}{at mid-exposure} & time (s)  \\

\noalign{\smallskip}
\hline
\noalign{\smallskip}

IGR J00040+7020          & 00:04:01.92 & +70:19:18.5   & Copernicus+AFOSC & 4000-8000 & 4.2 & 27 Nov 2006, 20:23 & 3$\times$1800  \\
IGR J00256+6821          & 00:25:32.5$^\dagger$ & +68:21:44$^\dagger$        & Cassini+BFOSC & 3500-8000 & 4.0 & 18 Nov 2006, 21:42 & 3$\times$1800  \\
IGR J01528$-$0326        & 01:52:49.00 & $-$03:26:48.5 & AAT+6dF    & 3900-7600 & 1.6 & 02 Dec 2002, 11:07 & 1200+600        \\
IGR J02343+3229          & 02:34:20.10 & +32:30:20.0   & Cassini+BFOSC & 3500-8000 & 4.0 & 03 Oct 2006, 23:19 & 3$\times$1800  \\
IGR J02504+5443          & 02:50:42.59 & +54:42:17.7   & Copernicus+AFOSC & 4000-8000 & 4.2 & 14 Nov 2006, 01:25 & 2$\times$2400 \\
IGR J03334+3718          & 03:33:18.79 & +37:18:11.1   & Cassini+BFOSC & 3500-8000 & 4.0 & 02 Oct 2006,  02:33 & 2$\times$1800  \\
IGR J06117$-$6625        & 06:11:48.34 & $-$66:24:33.7 & CTIO 1.5m+RC Spc.   & 3300-10500 & 5.7 & 19 Feb 2007, 04:45 & 3$\times$1800  \\
IGR J06292+4858$^*$      & 06:29:13.57 & +49:01:24.9   & Cassini+BFOSC & 3500-8000 & 4.0 & 13 Dec 2006, 23:52 & 2$\times$1800  \\
IGR J07437$-$5137$^*$    & 07:43:31.71 & $-$51:40:56.7 & CTIO 1.5m+RC Spc.   & 3300-10500 & 5.7 & 20 Feb 2007, 02:55 & 2$\times$1800  \\
IGR J08023$-$6954$^*$    & 08:02:41.64 & $-$69:53:37.7 & CTIO 1.5m+RC Spc.   & 3300-10500 & 5.7 & 20 Feb 2007, 01:57 & 2$\times$900   \\
IGR J09446$-$2636        & 09:44:37.02 & $-$26:33:55.4 & AAT+6dF    & 3900-7600 & 1.6 & 17 Mar 2004, 12:00 & 1200+600       \\
IGR J09523$-$6231        & 09:52:20.7$^\dagger$ & $-$62:32:37$^\dagger$      & 3.6m+EFOSC   & 3685-9315 & 2.8 & 24 Jun 2007, 00:27 & 2$\times$1800   \\
IGR J11366$-$6002        & 11:36:42.04 & $-$60:03:06.6 & NTT+EMMI    & 3300-9050 & 2.8 & 29 Jul 2007, 23:15 & 2$\times$300  \\
IGR J12131+0700          & 12:12:49.81 & +06:59:45.1   & SDSS+CCD Spc. & 3800-9200 & 1.0 & 16 Jan 2005, 12:32 & 4000           \\
IGR J13038+5348          & 13:03:59.43 & +53:47:30.1   & Cassini+BFOSC & 3500-8000 & 4.0 & 13 Jan 2007, 04:47 & 2$\times$1200  \\
IGR J13109$-$5552        & 13:10:43.35 & $-$55:52:11.4 & 3.6m+EFOSC   & 3685-9315 & 2.8 & 21 Jun 2007, 03:44 & 2$\times$600  \\
IGR J13149+4422          & 13:15:17.25 & +44:24:25.9   & SDSS+CCD Spc. & 3800-9200 & 1.0 & 25 Mar 2004, 09:11 & 1920           \\
IGR J14298$-$6715        & 14:29:59.81 & $-$67:14:44.8 & 3.6m+EFOSC   & 3685-9315 & 2.8 & 23 Jun 2007, 02:26 & 2$\times$360   \\
IGR J14331$-$6112        & 14:33:08.33 & $-$61:15:39.7 & 3.6m+EFOSC   & 3685-9315 & 2.8 & 22 Jun 2007, 04:41 & 2$\times$900   \\
IGR J14471$-$6414        & 14:46:28.26 & $-$64:16:24.3 & 3.6m+EFOSC   & 3685-9315 & 2.8 & 22 Jun 2007, 05:34 & 2$\times$1200  \\
IGR J14561$-$3738$^*$    & 14:56:08.43 & $-$37:38:52.4 & Jorge Sahade+REOSC & 3890-7360 & 3.4 & 11 Mar 2007, 05:37 & 2$\times$1200  \\
IC 4518a                 & 14:57:41.16 & $-$43:07:55.2 & CTIO 1.5m+RC Spc.   & 3300-10500 & 5.7 & 19 Feb 2007, 08:44 & 2$\times$1200  \\
IGR J15161$-$3827        & 15:15:59.70 & $-$38:25:46.8 & AAT+6dF    & 3900-7600 & 1.6 & 11 Mar 2003, 16:30 & 1200+600       \\
IGR J15539$-$6142        & 15:53:35.28 & $-$61:40:58.4 & Jorge Sahade+REOSC & 3890-7360 & 3.4 & 11 Mar 2007, 07:46 & 2$\times$1200  \\
IGR J16024$-$6107        & 16:01:48.23 & $-$61:08:54.7 & Radcliffe+Grating Spc.   & 3850-7200 & 2.3 & 26 Apr 2007, 01:15 & 2$\times$1200  \\
IGR J16056$-$6110        & 16:05:51.17 & $-$61:11:44.0 & Jorge Sahade+REOSC & 3890-7360 & 3.4 & 12 Mar 2007, 05:16 & 2$\times$1800  \\
IGR J16385$-$2057        & 16:38:30.91 & $-$20:55:24.6 & AAT+6dF    & 3900-7600 & 1.6 & 08 Jun 2003, 14:05 & 1200+600       \\
IGR J16500$-$3307        & 16:49:55.64 & $-$33:07:02.0 & Jorge Sahade+REOSC & 3890-7360 & 3.4 & 12 Mar 2007, 08:02 & 2$\times$1800  \\
2E 1739.1$-$1210         & 17:41:55.25 & $-$12:11:56.6 & 3.6m+EFOSC   & 3685-9315 & 2.8 & 25 Jun 2007, 01:13 & 300            \\
1RXS J174607.8$-$213333  & 17:46:03.16 & $-$21:33:27.1 & CTIO 1.5m+RC Spc.   & 3300-10500 & 5.7 & 16 Jun 2007, 05:32 & 2$\times$1200  \\
SAX J1802.7$-$2017       & 18:02:41.94 & $-$20:17:17.2 & NTT+EMMI    & 3300-9050 & 2.8 & 28 Jul 2007, 01:27 & 2$\times$600   \\
IGR J18048$-$1455        & 18:04:38.92 & $-$14:56:47.4 & 3.6m+EFOSC   & 3685-9315 & 2.8 & 23 Jun 2007, 06:35 & 2$\times$1200  \\
SAX J1818.6$-$1703       & 18:18:37.90 & $-$17:02:47.9 & 3.6m+EFOSC   & 3685-9315 & 2.8 & 23 Jun 2007, 08:37 & 2$\times$1200  \\
IGR J18483$-$0311        & 18:48:17.20 & $-$03:10:16.8 & 3.6m+EFOSC   & 3685-9315 & 2.8 & 22 Jun 2007, 06:21 & 2$\times$1200  \\
IGR J19405$-$3016        & 19:40:15.07 & $-$30:15:52.2 & Radcliffe+Grating Spc.   & 3850-7200 & 2.3 & 30 Apr 2007, 04:10 & 2$\times$900   \\
Swift J2000.6+3210       & 20:00:21.85 & +32:11:23.2   & Cassini+BFOSC & 3500-8000 & 4.0 & 13 Jul 2006, 23:19 & 2$\times$1800  \\
IGR J21272+4241          & 21:27:18.51 & +42:39:11.2   & WHT+ISIS    & 5050-10300 & 1.8 & 01 Sep 2007, 02:40 & 2$\times$400   \\
IGR J23308+7120          & 23:30:37.68 & +71:22:46.6   & Cassini+BFOSC & 3500-8000 & 4.0 & 25 Jul 2007, 01:45 & 2$\times$1800  \\
IGR J23524+5842          & 23:52:22.11 & +58:45:30.7   & WHT+ISIS    & 5050-10300 & 1.8 & 01 Sep 2007, 03:10 & 2$\times$900   \\
\noalign{\smallskip}
\hline
\noalign{\smallskip}
\multicolumn{8}{l}{$^*$: source with tentative optical identification} \\
\multicolumn{8}{l}{$^\dagger$: coordinates extracted from the DSS-II-Red frames, having an 
accuracy of $\sim$1$''$.}\\
\noalign{\smallskip}
\hline
\hline
\noalign{\smallskip}
\end{tabular}
\end{center}
\end{table*}

\section{Results}

In this section we present the results of our spectroscopic campaign
described in Sect. 3. The optical magnitudes quoted below, if not 
otherwise stated, are extracted from the USNO-A2.0 catalogue.

As already done in Papers I-V, we use the following identification and
classification criteria for the optical spectra of the sources considered
in this work.

For the determination of the distance of compact Galactic X--ray sources,
in the case of CVs we assumed an absolute magnitude M$_V \sim$ 9 and 
an intrinsic color index $(V-R)_0 \sim$ 0 mag (Warner 1995), whereas for 
HMXBs, when applicable, we used the intrinsic stellar color indices 
and absolute magnitudes as reported in Lang (1992) and Wegner (1994).
For low-mass X--ray binaries (LMXBs), we considered $(V-R)_0$ $\sim$ 0 
$\sim$ M$_R$ (e.g., van Paradijs \& McClintock 1995).

For the emission-line AGN classification, we used the criteria 
of Veilleux \& Osterbrock (1987) and the line ratio diagnostics 
of Ho et al. (1993, 1997) and of Kauffmann et al. (2003); moreover, for 
the subclass assignation of Seyfert 1 nuclei, we used the 
H$_\beta$/[O {\sc iii}]$\lambda$5007 line flux ratio criterion as per 
Winkler (1992).

When possible in the cases of extragalactic objects, for the calculation 
of the intrinsic absorption in the host galaxy of a line-emitting AGN 
we first dereddened the H$_\alpha$ and H$_\beta$ line fluxes by applying 
a correction for the 
Galactic absorption along the source line of sight. This was done 
following the prescription for the computation of the Galactic color 
excess $E(B-V)_{\rm Gal}$ given by Schlegel et al. (1998), and 
considering the Galactic extinction law by Cardelli et al. (1989).
Then, we assumed an intrinsic H$_\alpha$/H$_\beta$ line ratio of 
2.86 (Osterbrock 1989) and we computed the color excess 
$E(B-V)_{\rm AGN}$ local to the AGN host, using again Cardelli et 
al.'s (1989) extinction law, from the comparison between the intrinsic 
line ratio and the one corrected for the Galactic reddening.

The spectra of the galaxies shown here were not corrected for
starlight contamination (see, e.g., Ho et al. 1993, 1997) given the
limited S/N and the spectral resolution. We do not consider this to 
affect any of our conclusions. 

In the following we assume a cosmology with $H_{\rm 0}$ = 65
km s$^{-1}$ Mpc$^{-1}$, $\Omega_{\Lambda}$ = 0.7 and $\Omega_{\rm m}$ =
0.3; the luminosity distances of the extragalactic objects reported in 
this paper are computed with these parameters using the Cosmology
Calculator of Wright (2006). Moreover, when not explicitly stated 
otherwise, for our X--ray flux estimates we will assume a Crab-like 
spectrum except for the {\it XMM-Newton} Slew Survey sources, for which 
we considered the 0.2--12 keV flux reported by Saxton et al. (2008). 
We also remark that the results presented here supersede the 
preliminary ones reported in Masetti et al. (2006e,f, 2007a,b).

The next Subsections report the object identifications divided 
into three broad classes (CVs, X--ray binaries and AGNs) ordered according 
to their increasing distance from Earth; two more sources (IGR 06292+4858 
and IGR J08023$-$6954) are discussed in a separate Subsection due to the 
peculiarities about their identification, as stressed in Sect. 2.

\subsection{CVs}

\begin{figure*}
\mbox{\psfig{file=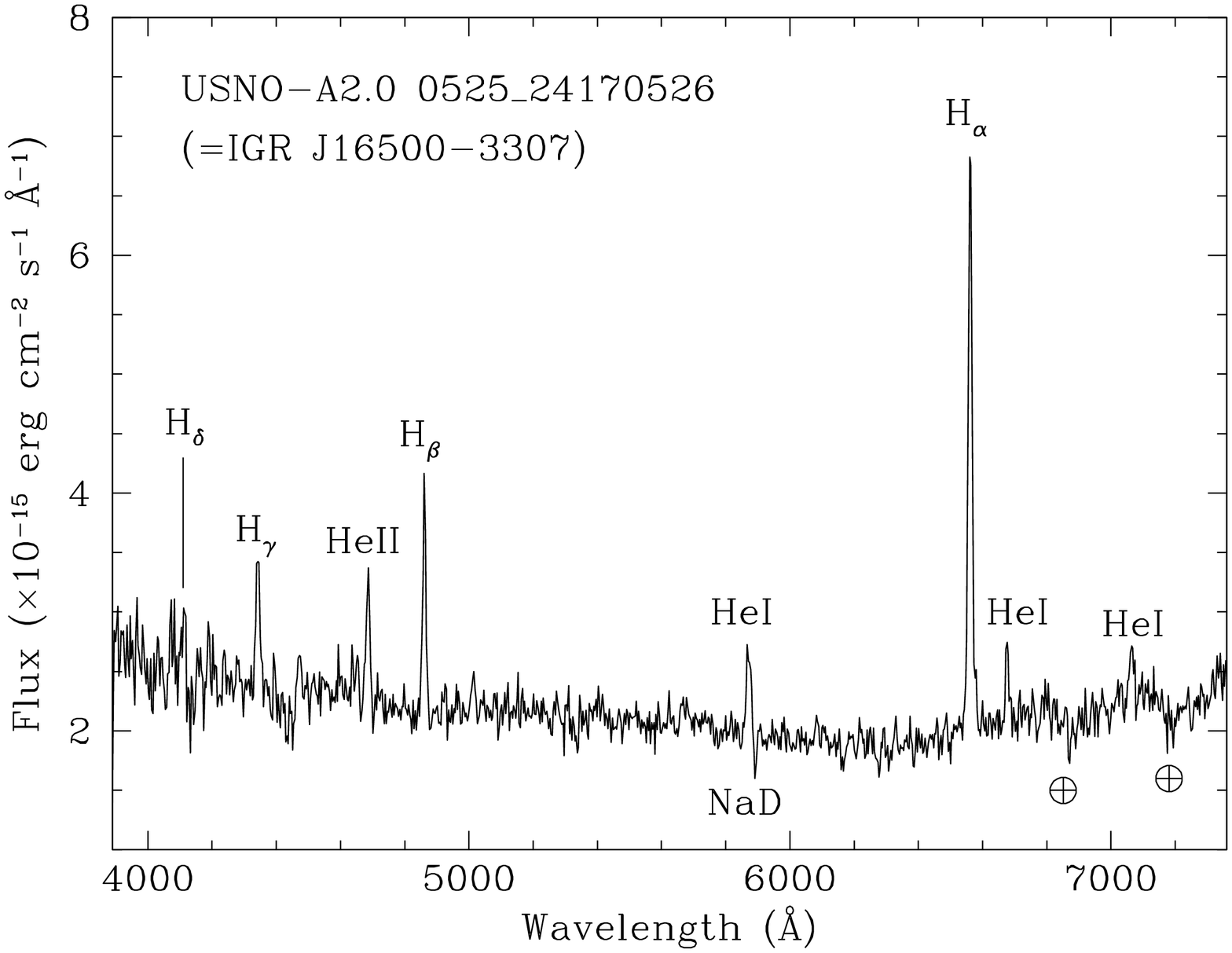,width=9cm,angle=270}}
\mbox{\psfig{file=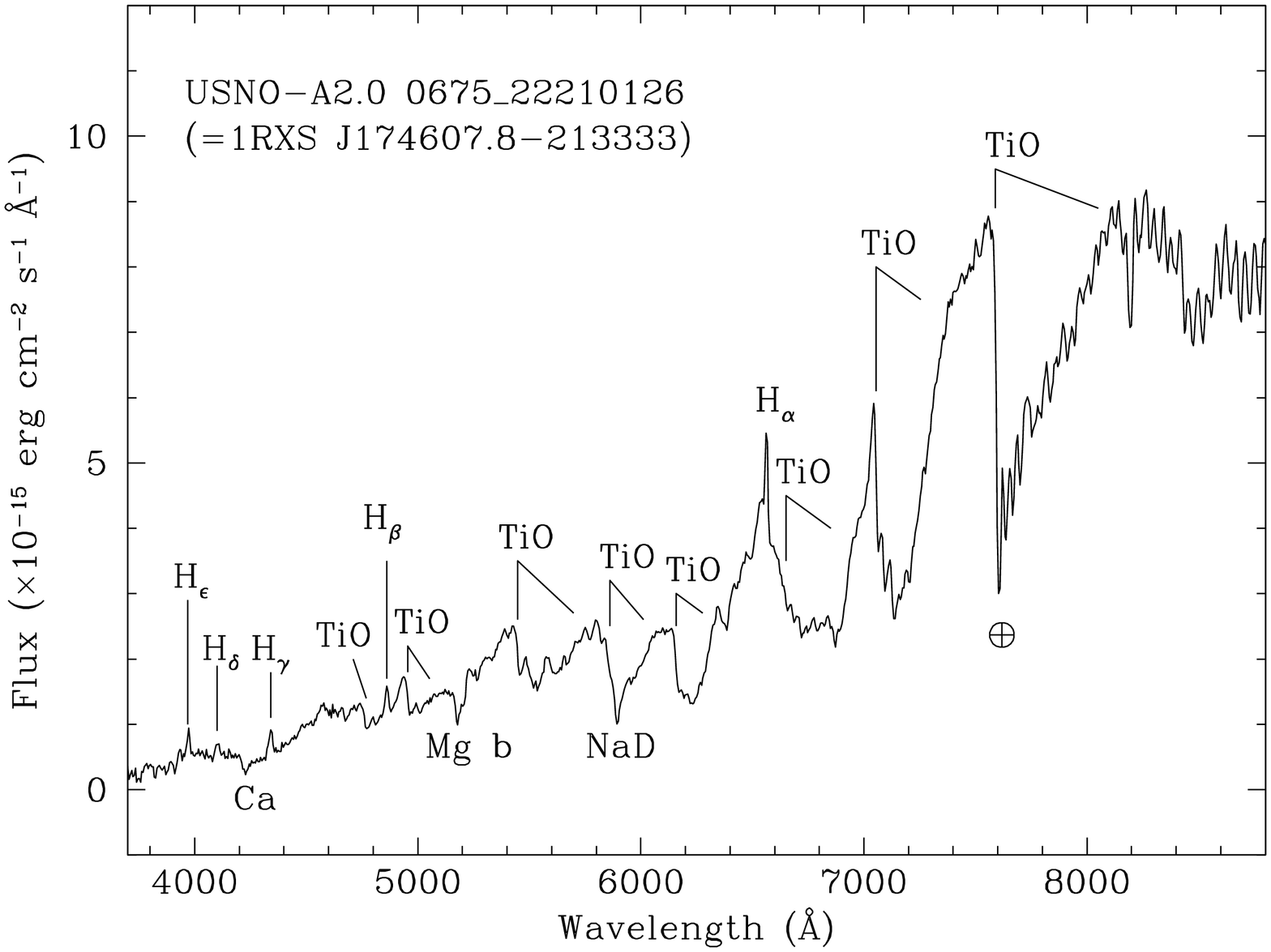,width=9cm,angle=270}}
\vspace{-.5cm}
\caption{Spectra (not corrected for the intervening Galactic absorption) 
of the optical counterparts of the CVs belonging to
the {\it INTEGRAL} sources sample presented in this paper.
For each spectrum the main spectral features are labeled. The 
symbol $\oplus$ indicates atmospheric telluric absorption bands.}
\end{figure*}

One object of our sample, IGR J16500$-$3307, was identified as a dwarf 
nova CV through the appearance of its optical spectrum (Fig. 6, left 
panel). It shows Balmer emissions up to at least H$_\delta$, as well as 
several He {\sc i} and He {\sc ii} lines in emission. All of the detected 
lines are consistent with being at $z$ = 0, indicating that this object 
belongs to our Galaxy. In addition, source 1RXS J174607.8$-$213333 was 
identified as a symbiotic star, given that its optical spectral continuum 
shows the typical characteristics of a red giant star with superimposed 
Balmer emissions at $z$ = 0 (Fig. 6, right panel).

The main spectral characteristics for the two objects, and the parameters 
which can be inferred from their optical and X--ray data, are 
listed in Table 2. The fact that, in the spectrum of IGR J16500$-$3307,
the He {\sc ii}$\lambda$4686/H$_\beta$ equivalent width (EW) ratio is 
$\ga$0.5 and the EWs of He {\sc ii} and H$_\beta$ are around
10 \AA~indicates that this source is a magnetic CV belonging to the 
intermediate polar (IP) subclass (see Warner 1995 and references therein). 
Its optical spectrum, moreover, closely resembles those of other CVs
detected with {\it INTEGRAL} (Papers IV and V), and which were classified 
as IPs. For this source, we determine the reddening along its line of 
sight as $A_V$ = 0.53 mag, inferred by comparing the observed 
H$_\alpha$/H$_\beta$ flux ratio with an assumed intrinsic one of 2.86 
(Osterbrock 1989) and then applying the Cardelli et al.'s (1989) 
Galactic extinction law.

Using the Bruzual-Persson-Gunn-Stryker\footnote{available at:\\ 
{\tt ftp://ftp.stsci.edu/cdbs/cdbs1/grid/bpgs/}} (Gunn \& Stryker 1983) 
and Jacoby-Hunter-Christian\footnote{available at:\\ 
{\tt ftp://ftp.stsci.edu/cdbs/cdbs1/grid/jacobi/}} (Jacoby et al. 1984) 
spectroscopy atlases, we constrained the spectral type of the optical 
counterpart of 1RXS J174607.8$-$213333 to be between M2\,III and
M4\,III. From this information, assuming the colors and absolute $V$ 
magnitude of a M3\,III star (Ducati et al. 2001; Lang 1992), we obtain 
a distance of $\sim$22 kpc for the source, which would place it 
on the other side of the Galaxy, beyond the Bulge. This suggests that 
some amount of absorption should be present along the line of sight and 
that this number should rather be used as an (admittedly loose) upper 
limit for the distance to this object.

The X--ray luminosities for the two objects were then computed using 
the fluxes reported in Bird et al. (2007) and Revnivtsev et al. (2004a).

We conclude this section by mentioning that, although 1RXS 
J174607.8$-$213333 is present in the survey of Revnivtsev et al. (2004a), 
it does not appear in the deeper IBIS surveys of Bird et al. (2007) and of 
Krivonos et al. (2007). This may suggest either some variability of this 
X--ray source, or that the {\it INTEGRAL} detection reported by Revnivtsev 
et al. (2004a) is spurious. Nevertheless, the {\it Swift}/XRT observation 
indicates that X--ray activity up to 10 keV is in any case produced by 
the symbiotic system identified here.

\begin{table*}
\caption[]{Synoptic table containing the main results concerning the 2 CVs
(see Fig. 6) identified in the present sample of {\it INTEGRAL} sources.}
\scriptsize
\begin{center}
\begin{tabular}{lccccccccr}
\noalign{\smallskip}
\hline
\hline
\noalign{\smallskip}
\multicolumn{1}{c}{Object} & \multicolumn{2}{c}{H$_\alpha$} & 
\multicolumn{2}{c}{H$_\beta$} & \multicolumn{2}{c}{He {\sc ii} $\lambda$4686} & 
$R$ & $d$ & \multicolumn{1}{c}{$L_{\rm X}$} \\
\cline{2-7}
\noalign{\smallskip} 
 & EW & Flux & EW & Flux & EW & Flux & mag & (pc) & \\

\noalign{\smallskip}
\hline
\noalign{\smallskip}

IGR J16500$-$3307 & 35.2$\pm$0.2 & 7.1$\pm$0.4 & 9.9$\pm$1.0 & 2.1$\pm$0.2 & 7.1$\pm$0.7 & 
1.59$\pm$0.16 & 16.0 & $\sim$210 & 0.022 (0.1--2.4) \\
 & & & & & & & & & 1.0 (20--100) \\ 

& & & & & & & & & \\ 

1RXS J174607.8$-$213333 & 5.2$\pm$0.5 & 2.1$\pm$0.2 & 7.1$\pm$0.7 & 0.84$\pm$0.08 & $<$1 & $<$0.1 & 
14.7 & $<$21800 & $<$10500 (18--60) \\

\noalign{\smallskip} 
\hline
\noalign{\smallskip} 
\multicolumn{10}{l}{Note: EWs are expressed in \AA, line fluxes are
in units of 10$^{-14}$ erg cm$^{-2}$ s$^{-1}$, whereas X--ray luminosities
are in units of 10$^{32}$ erg s$^{-1}$}\\
\multicolumn{10}{l}{and the reference band (between brackets) is expressed in keV.}\\
\noalign{\smallskip} 
\hline
\hline
\noalign{\smallskip} 
\end{tabular} 
\end{center}
\end{table*}

\subsection{X--ray binaries}

\subsubsection{HMXBs}

\begin{figure*}
\mbox{\psfig{file=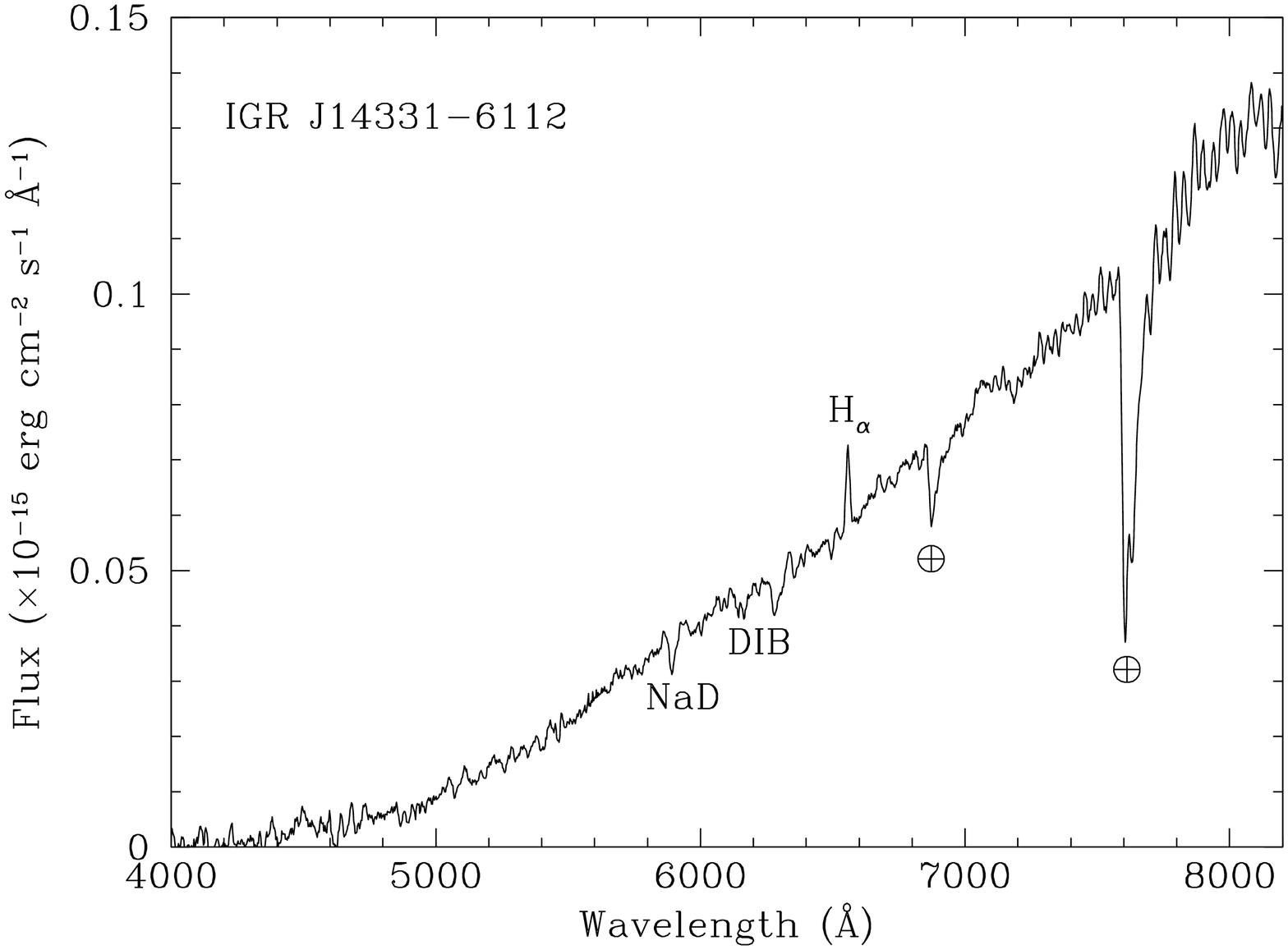,width=9cm,angle=270}}
\mbox{\psfig{file=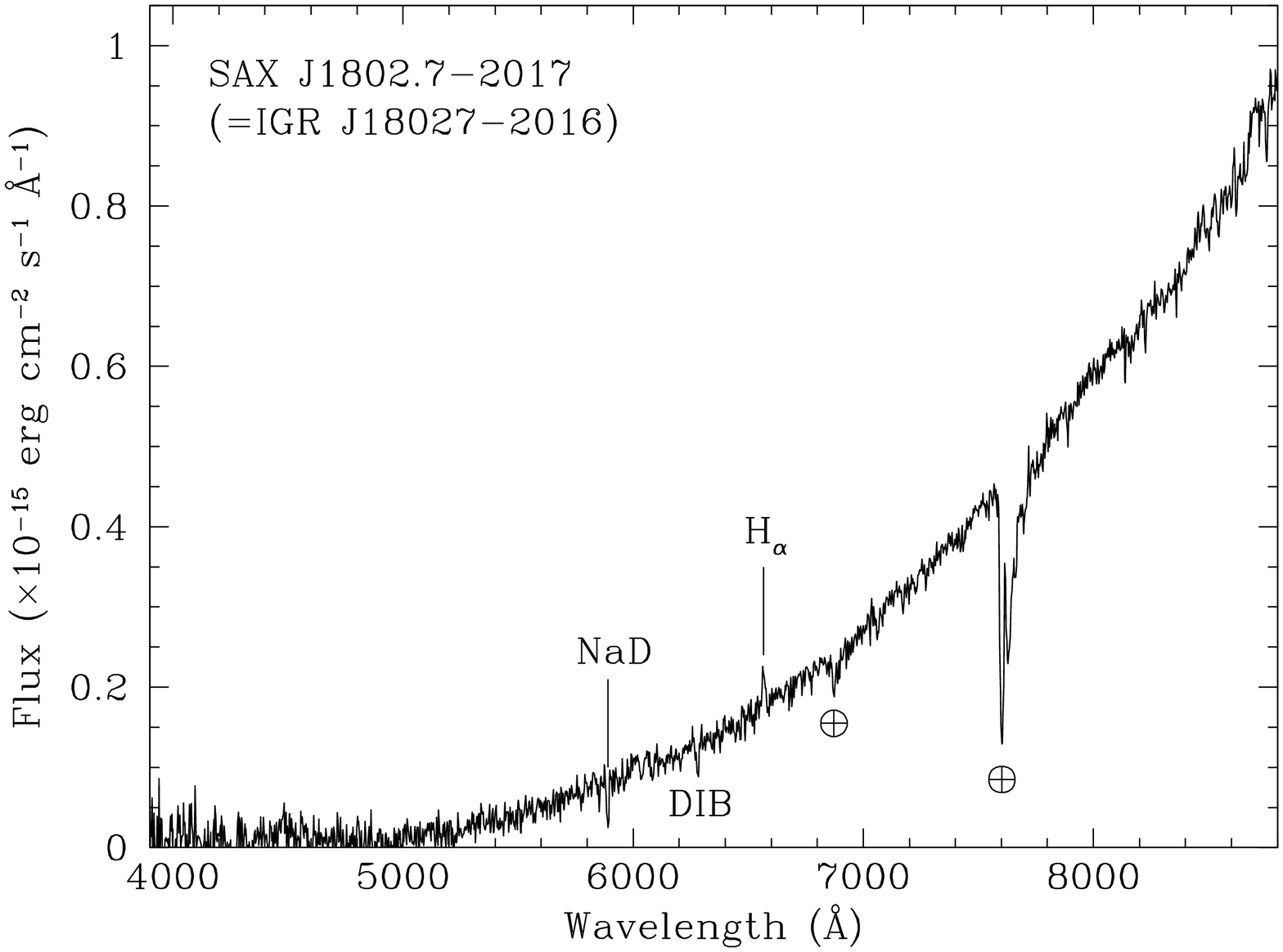,width=9cm,angle=270}}

\vspace{-.9cm}
\mbox{\psfig{file=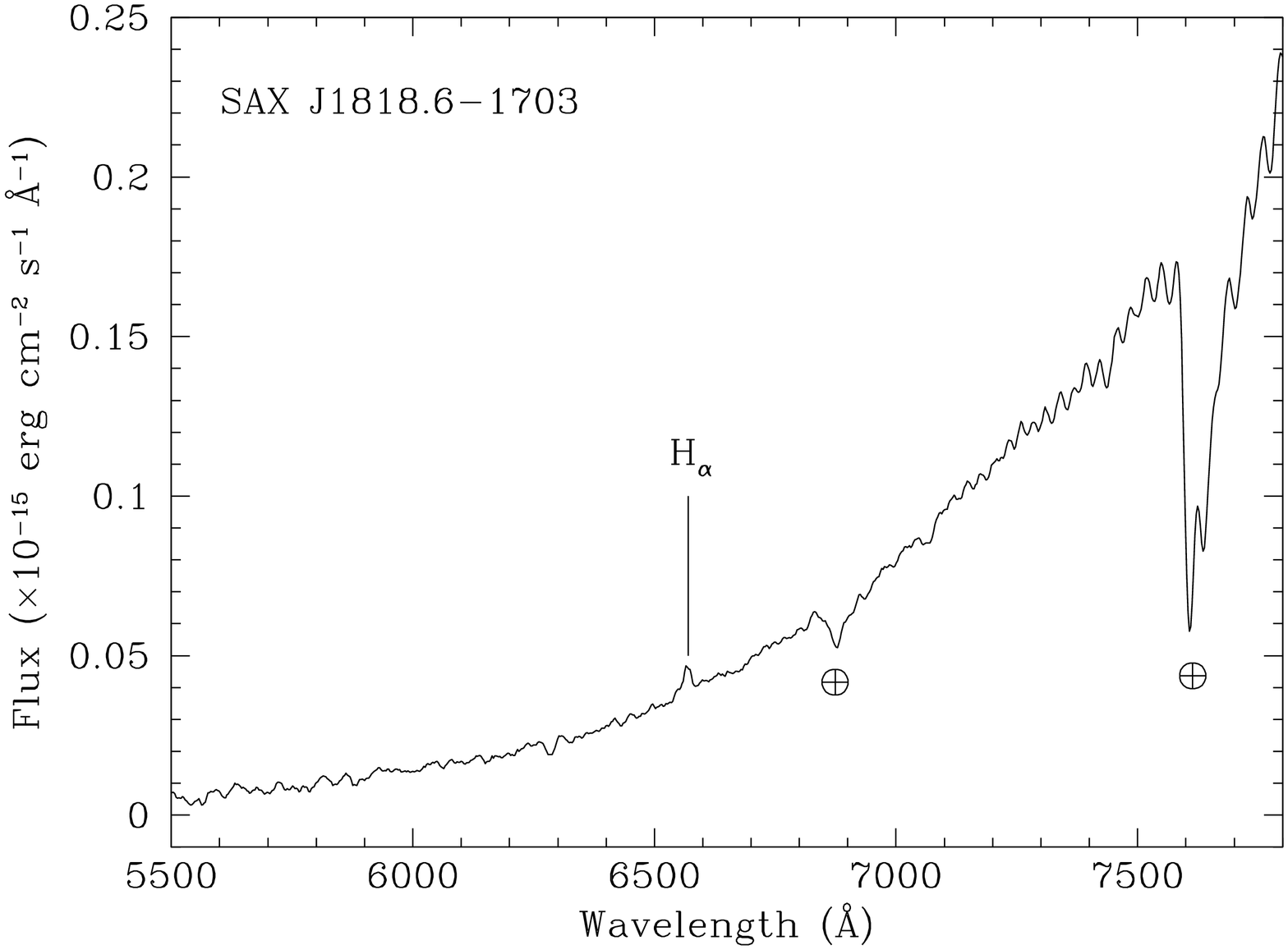,width=9cm,angle=270}}
\mbox{\psfig{file=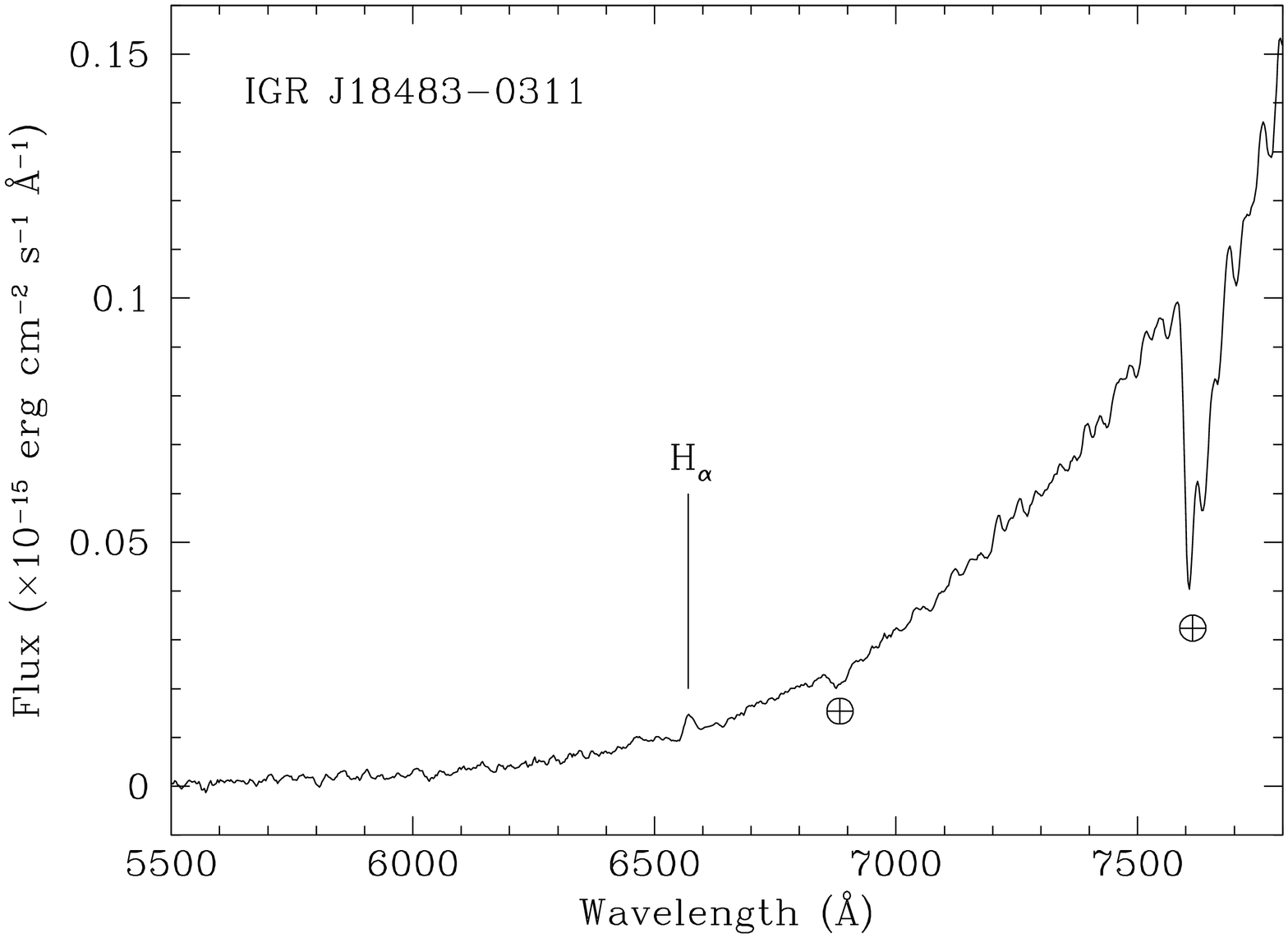,width=9cm,angle=270}}

\vspace{-.9cm}
\parbox{9cm}{
\psfig{file=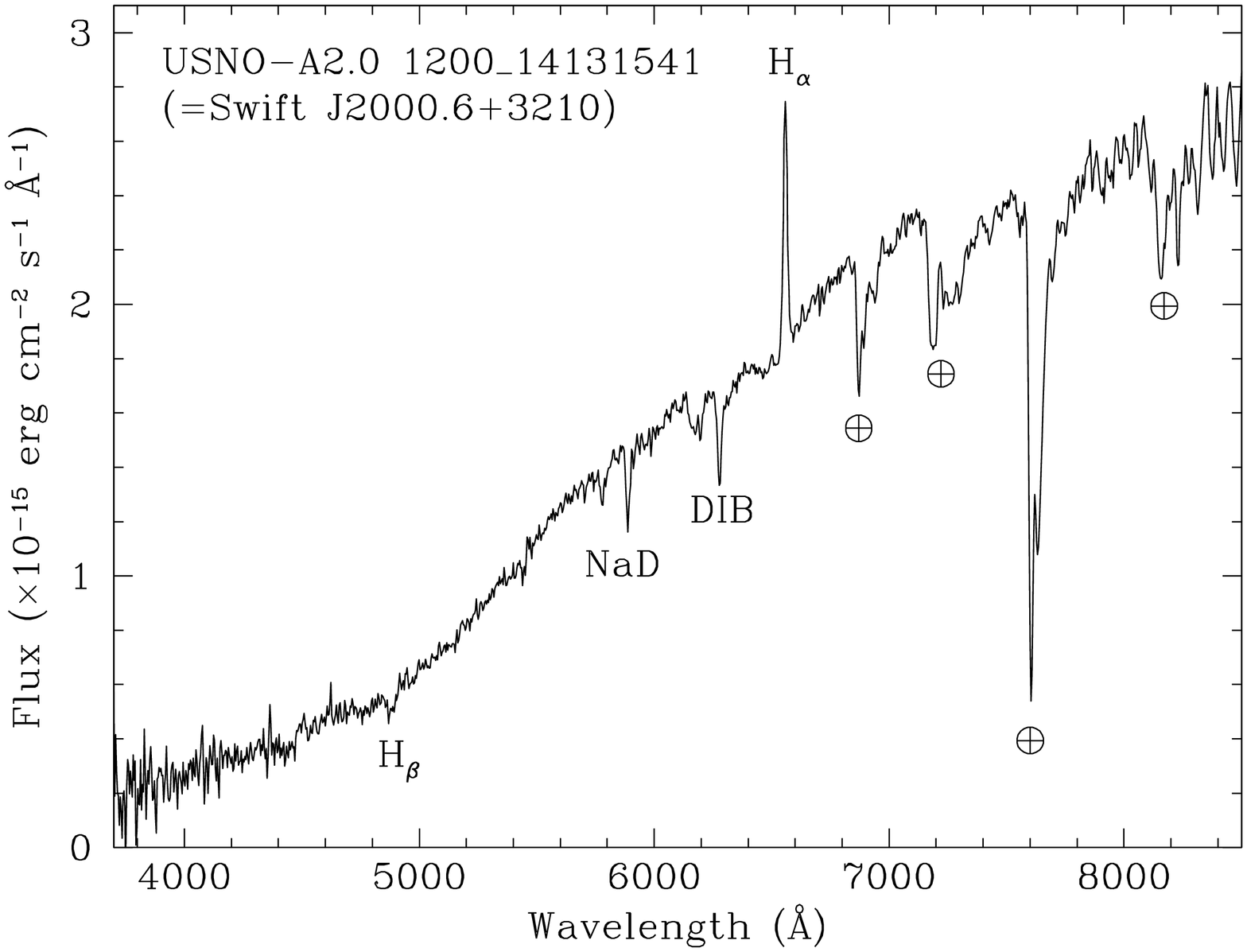,width=9cm,angle=270}
}
\hspace{0.5cm}
\parbox{7.5cm}{
\vspace{-.5cm}
\caption{Spectra (not corrected for the intervening Galactic absorption) 
of the optical counterparts of the HMXBs belonging to
the {\it INTEGRAL} sources sample presented in this paper.
For each spectrum the main spectral features are labeled. The 
symbol $\oplus$ indicates atmospheric telluric absorption bands.
The ESO 3.6m spectra have been smoothed using a Gaussian filter with 
$\sigma$ = 5 \AA.}
}
\end{figure*}

\begin{table*}
\caption[]{Synoptic table containing the main results concerning the 5 HMXBs
(see Fig. 7) identified or observed in the present sample of {\it INTEGRAL} 
sources.}
\scriptsize
\begin{center}
\begin{tabular}{lccccccr}
\noalign{\smallskip}
\hline
\hline
\noalign{\smallskip}
\multicolumn{1}{c}{Object} & \multicolumn{2}{c}{H$_\alpha$} & 
Optical & $A_V$ & $d$ & Spectral & \multicolumn{1}{c}{$L_{\rm X}$} \\
\cline{2-3}
\noalign{\smallskip} 
 & EW & Flux & mag. & (mag) & (kpc) & type & \\

\noalign{\smallskip}
\hline
\noalign{\smallskip}

IGR J14331$-$6112 & 4.6$\pm$0.5 & 0.027$\pm$0.003 & 18.1 ($R$) & $\approx$6.5 & $\approx$10 & 
early B\,III or mid B\,V & 0.054 (2--10) \\
 & & & & & & & 0.15 (20--100) \\

 & & & & & & & \\ 

SAX J1802.7$-$2017 & 4.2$\pm$0.6 & 0.074$\pm$0.011 & 16.9 ($R$) & $\sim$8.3 & $\approx$10 & 
early B\,III & 1.1$^{\rm a}$ (2--10) \\
 & & & & & & & 0.75 (20--100) \\

 & & & & & & & \\ 

SAX J1818.6$-$1703 & 3.2$\pm$0.6$^{\rm b}$ & 0.0124$\pm$0.0013$^{\rm b}$ & 17.4 ($R$) & $\sim$14 & $\sim$2.5 & 
OB supergiant & 0.0056$^{\rm c}$ (0.5--10) \\
 & & & & & & & 0.018$^{\rm c}$ (20--100) \\

 & & & & & & & \\ 

IGR J18483$-$0311 & 8.5$\pm$2.6 & 0.0089$\pm$0.0027 & 19.3$^{\rm d}$ ($R$) & $\approx$13$^{\rm d}$ & 
$\sim$3.5$^{\rm d}$ & OB giant & 0.0063$^{\rm c}$ (1--7) \\
 & & & & & & & 0.088$^{\rm c}$ (20--100) \\

 & & & & & & & \\ 

Swift J2000.6+3210 & 10.2$\pm$0.5 & 1.88$\pm$0.09 & 16.1 ($R$) & $\sim$4.0 & $\approx$8 & early B\,V or 
mid B\,III & $\approx$0.038 (0.2--12) \\
 & & & & & & & $\approx$0.25 (20--100) \\ 

\noalign{\smallskip} 
\hline
\noalign{\smallskip}
\multicolumn{8}{l}{Note: EWs are expressed in \AA, line fluxes are
in units of 10$^{-14}$ erg cm$^{-2}$ s$^{-1}$, whereas X--ray luminosities
are in units of 10$^{36}$ erg s$^{-1}$} \\
\multicolumn{8}{l}{and the reference band (between brackets) is expressed in keV.}\\
\multicolumn{8}{l}{$^{\rm a}$: luminosity estimate from Hill et al. (2005); $^{\rm b}$: sum of two 
components;} \\
\multicolumn{8}{l}{$^{\rm c}$: low-state luminosity values; $^{\rm d}$: from Sguera et al. (2007)} \\
\noalign{\smallskip} 
\hline
\hline
\end{tabular} 
\end{center} 
\end{table*}

We classify 5 of the {\it INTEGRAL} sources of our sample as HMXBs by 
their overall spectral appearance (see Fig. 7), which is typical of this 
class of objects (see e.g. Papers III and V), with narrow H$_\alpha$ emission 
at a wavelength consistent with that of the laboratory restframe, 
superimposed on an intrinsically blue continuum with Balmer absorptions. 
Practically in all cases, however, the stellar 
continuum appears strongly reddened and sometimes almost undetected 
bluewards of 5000 \AA, implying the presence of substantial interstellar 
dust along the line of sight. This also is quite typical of HMXBs 
detected with {\it INTEGRAL} (e.g., Paper V) and indicates that these 
objects are relatively far from Earth.

Table 3 collects the relevant optical spectral information on these 5
sources, along with their main parameters inferred from the available 
optical and X--ray data. Luminosities for these objects were calculated 
using the X--ray fluxes of Bird et al. (2007), Saxton et al. (2008), 
in 't Zand et al. (2006), Sguera et al. (2007) and Landi et al. 
(2007b).

Below we report on results concerning the optical counterparts of these 5 
hard X--ray sources.

The optical spectrum of the counterpart of SAX J1818.6$-$1703 seems to
show a faint, double-peaked H$_\alpha$ emission, with the blue peak twice
as faint with respect to the red one. The two peaks appear to be separated 
by $\Delta \lambda \sim$ 8.5 \AA. 

The EW of the H$_\alpha$ emission line detected in the optical spectrum
of IGR J18483$-$0311 appears too large for a supergiant secondary star 
(see Leitherer 1988). Thus, together with the use of the results reported 
in Sguera et al. (2007), we infer that this HMXB hosts a secondary star of 
intermediate luminosity class. This allows us to determine distance and 
X--ray luminosities for this source as reported in Table 3.

Our $R$-band photometry of the counterparts of IGR J14331$-$6112 and SAX 
J1802.7$-$2017 yields magnitudes $R$ = 18.1$\pm$0.1 and $R$ = 
16.9$\pm$0.1, respectively. Unfortunately, due to a general lack of 
reliable multifilter optical photometry for the counterparts of the HMXBs 
considered here (in particular for IGR J14331$-$6112 and Swift 
J2000.6+3210), no precise information concerning distance, spectral type 
and X--ray luminosity can be determined.

For SAX J1802.7$-$2017 this problem can be circumvented by using the 
information in Hill et al. (2005). Indeed, using the mass and radius 
estimates of these authors and comparing them with the tabulated values 
for early-type stars in Lang (1992), we conclude that the optical 
companion in this wind-fed X--ray system is likely an early B giant. We 
note this is at odds with the companion star classification of Hill et al. 
(2005), who suggested an early-type supergiant star on the basis of the 
position of the X--ray source in the Corbet's (1986) $P_{\rm 
orb}$--$P_{\rm spin}$ diagram. However, were it a HMXB with an OB 
supergiant, it would be placed at a distance of $\sim$18 kpc, on the other 
side of the Galaxy. This would make it hardly detectable at optical 
wavelengths because of strong reddening due to the presence of large 
amounts of intervening Galactic dust.

Concerning SAX J1818.6$-$1703, we derived the distance using the 
magnitudes reported in Negueruela \& Smith (2006) and assuming that
the counterpart is an OB supergiant as suggested by Negueruela et al.
(2007).

In the cases of IGR J14331$-$6112 and Swift J2000.6+3210, however, 
by considering the absolute magnitudes of early-type stars and by 
applying the method described in Paper III for the classification of 
source 2RXP J130159.6$-$635806, we obtained the constraints for 
distance, reddening, spectral type and X--ray luminosity shown in 
Table 3.

Sguera et al. (2007) found for IGR J18483$-$0311, from the comparison 
between the line-of sight absorptions derived from the optical reddening 
$A_V$ and from the hydrogen column density $N_{\rm H}$ inferred from the 
source X--ray spectra, that the latter one appears to be larger.
We can make a similar comparison considering the two more HMXBs reported 
in this Section for which an $N_{\rm H}$ measurement is available, 
that is, SAX J1802.7$-$2017 (Walter et al. 2006) and SAX J1818.6$-$7103
(in 't Zand et al. 2006). Using the empirical foumula of Predehl \& 
Schmitt (1995), we find that in the two cases the X--ray absorption 
is two to five times larger than that determined in the optical.
This is often observed in absorbed HMXBs detected
with {\it INTEGRAL} (e.g., Chaty 2007) and suggests the presence 
of additional absorbing material in the vicinity of the X--ray source, 
likely due to the accretion stream flowing onto the compact object in 
these X--ray systems.

We conclude this Section by noting that none of these sources is 
associated with a radio source. This means that none of them is likely a 
jet-emitting HMXB (i.e., a microquasar).

\subsubsection{LMXBs}

\begin{figure*}
\mbox{\psfig{file=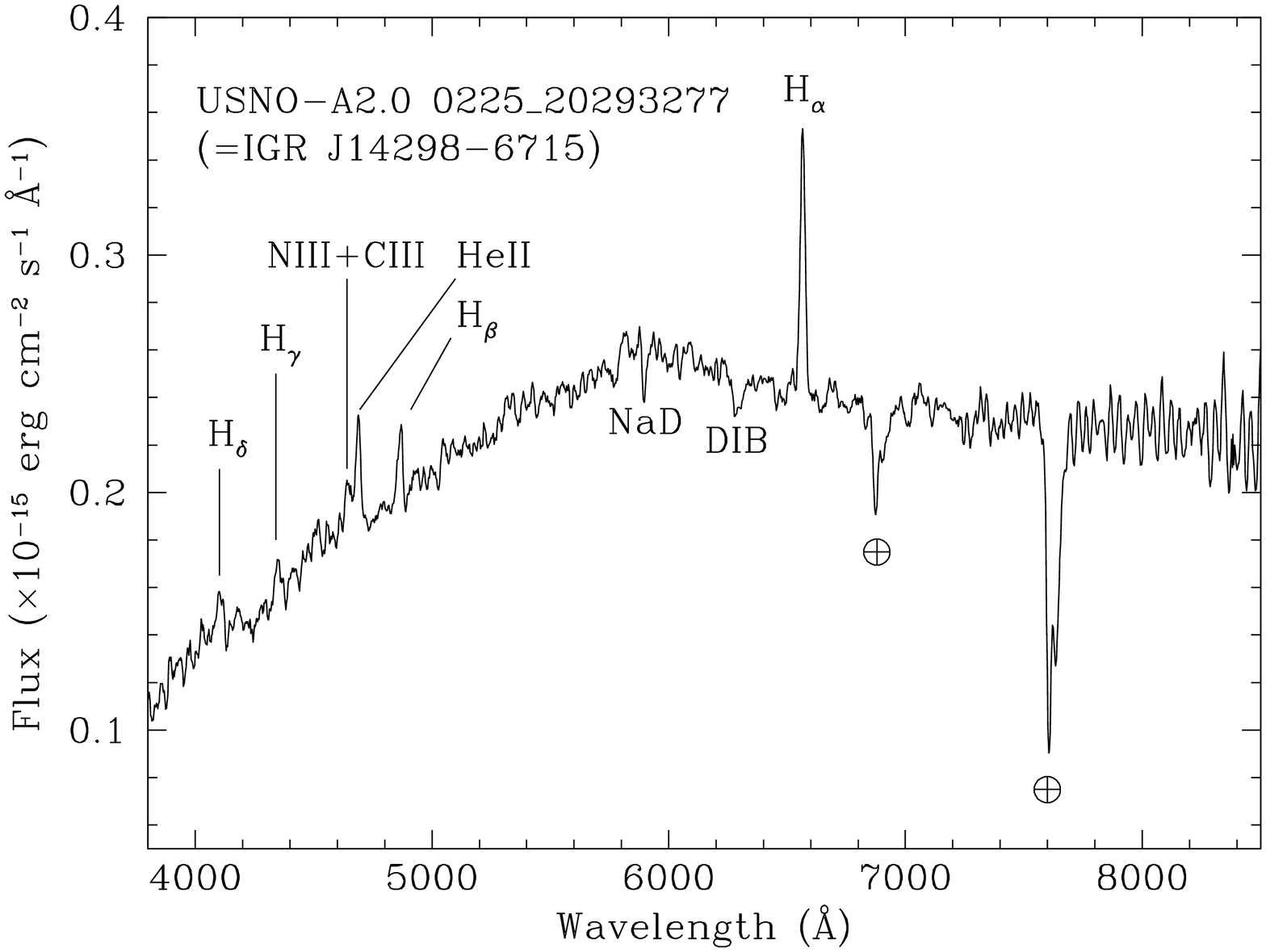,width=9cm,angle=270}}
\mbox{\psfig{file=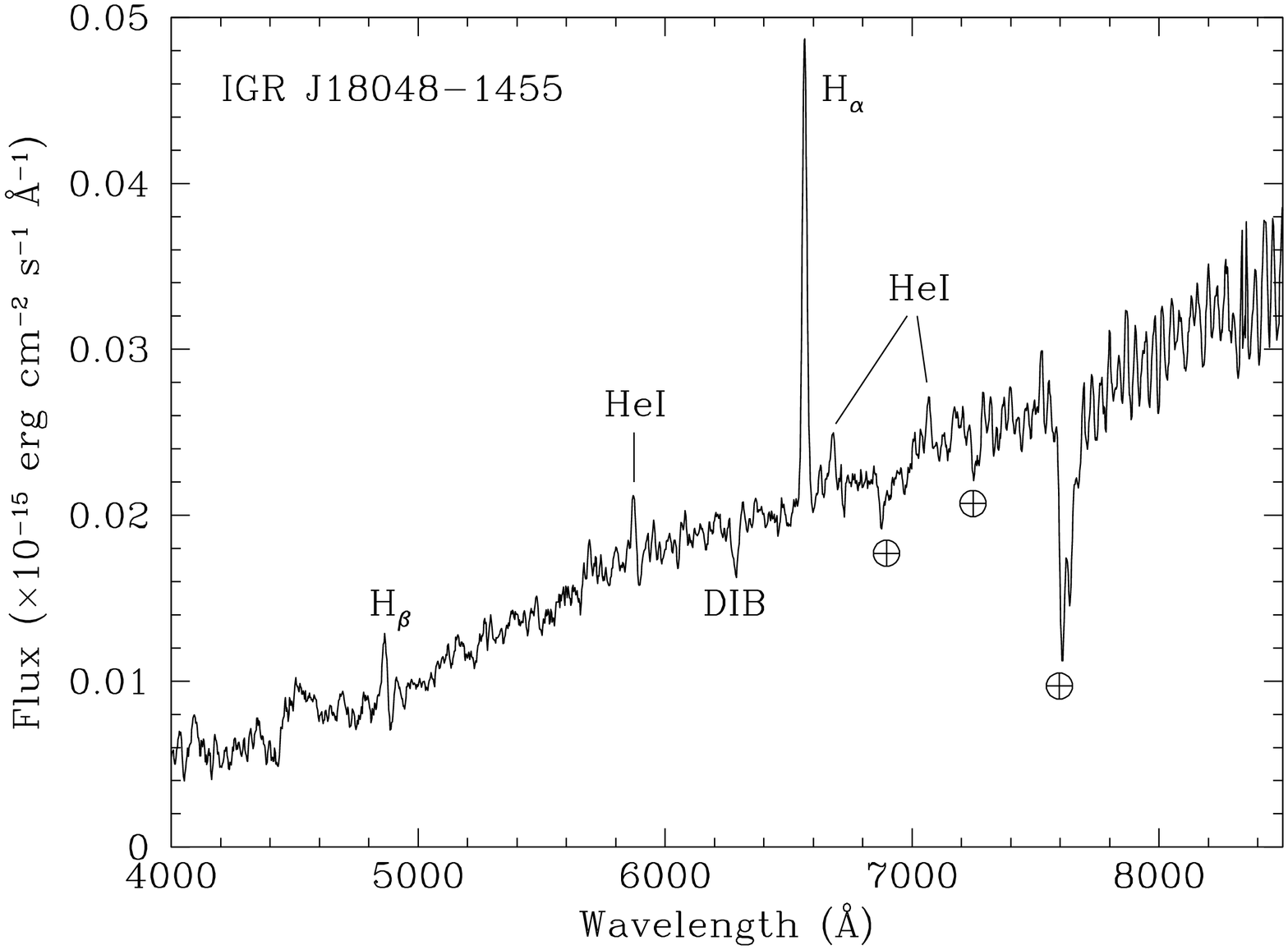,width=9cm,angle=270}}
\vspace{-.5cm}
\caption{Spectra (not corrected for the intervening Galactic absorption) 
of the optical counterparts of the LMXBs belonging to
the {\it INTEGRAL} sources sample presented in this paper.
For each spectrum the main spectral features are labeled. The 
symbol $\oplus$ indicates atmospheric telluric absorption bands.
Both spectra have been smoothed using a Gaussian filter with
$\sigma$ = 5 \AA.}
\end{figure*}

\begin{table*}
\caption[]{Synoptic table containing the main results concerning the 2 LMXBs
(see Fig. 8) identified in the present sample of {\it INTEGRAL} sources.}
\scriptsize
\begin{center}
\begin{tabular}{lccccccccr}
\noalign{\smallskip}
\hline
\hline
\noalign{\smallskip}
\multicolumn{1}{c}{Object} & \multicolumn{2}{c}{H$_\alpha$} & 
\multicolumn{2}{c}{H$_\beta$} & \multicolumn{2}{c}{He {\sc ii} $\lambda$4686} & 
$R$ & $d$ & \multicolumn{1}{c}{$L_{\rm X}$} \\
\cline{2-7}
\noalign{\smallskip} 
 & EW & Flux & EW & Flux & EW & Flux & mag & (kpc) & \\

\noalign{\smallskip}
\hline
\noalign{\smallskip}

IGR J14298$-$6715 & 12.0$\pm$0.6 & 28.9$\pm$1.4 & 4.1$\pm$0.4 & 8.1$\pm$0.8 & 8.0$\pm$0.8 & 
14.6$\pm$1.5 & 16.4 ($R$) & $\sim$10 & 0.31 (2--10) \\
 & & & & & & & & & 1.5 (20--100) \\ 

 & & & & & & & & & \\ 

IGR J18048$-$1455 & 29.4$\pm$1.5 & 6.2$\pm$0.3 & 9.6$\pm$2.9 & 0.85$\pm$0.26 & $<$14 & $<$1.0 
& 18.7 ($R$) & $\sim$7 & 0.72 (20--100) \\

\noalign{\smallskip} 
\hline
\noalign{\smallskip}
\multicolumn{10}{l}{Note: EWs are expressed in \AA, line fluxes are
in units of 10$^{-16}$ erg cm$^{-2}$ s$^{-1}$, whereas X--ray luminosities
are in units of 10$^{35}$ erg s$^{-1}$} \\ 
\multicolumn{10}{l}{and the reference band (between brackets) is expressed in keV.} \\
\noalign{\smallskip} 
\hline
\hline
\noalign{\smallskip} 
\end{tabular} 
\end{center}
\end{table*}

Two hard X--ray sources of our sample, IGR J14298$-$6715 and IGR 
J18048$-$1455, show Balmer and helium emission lines at $z$ = 0
superimposed on a reddened continuum (see Fig. 8). This again tells 
us that these two objects are Galactic X--ray systems. Furthermore, we 
measure a magnitude $R$ = 18.7$\pm$0.1 for the counterpart of 
IGR J18048$-$1455.

The shape of the optical continuum of these two sources does not show 
features of a red giant star (as in the case of 1RXS J174607.8$-$213333), 
and thus suggests that these objects are not intrinsically red but 
actually substantially absorbed; therefore, they are likely quite far from 
Earth. Because of this, we can exclude that they are CVs.

We can also rule out an HMXB nature for them because their optical and 
2MASS near-infrared (NIR) magnitudes do not fit any star of early spectral 
type, not even if absorption along the line of sight is taken into account 
(see below). We thus suggest that these two hard X--ray sources are LMXBs.

Confirmation of the presence of reddening along the line of sight of these
sources comes from the observed H$_\alpha$/H$_\beta$ line ratio for the two 
cases, which implies $A_V$ = 0.56 mag and 2.96 mag for IGR J14298$-$6715 and 
IGR J18048$-$1455, respectively. It should be noted, however, that if we 
correct the magnitudes of the counterparts with these figures, we obtain
distances which are exceedingly large ($>$15 kpc).
This suggests that the assumption we made on the intrinsic 
H$_\alpha$/H$_\beta$ line ratio in these sources is not correct; 
we thus consider it safer to use the Galactic color excess measure along 
the lines of sight, that is, $E(B-V)_{\rm Gal}$ = 0.516 mag for IGR 
J14298$-$6715 and $E(B-V)_{\rm Gal}$ = 1.787 mag for IGR J18048$-$1455. 
Assuming these estimates, optical and 2MASS magnitudes become compatible 
with those of persistent LMXBs, and we obtain the distances and X--ray 
luminosities reported in Table 4. In this table, the X--ray luminosities 
were computed using the fluxes reported in Bird et al. (2007) and in
Landi et al. (2007b).

We thus here revise the classification of IGR J18048$-$1455 as a HMXB given 
by Burenin et al. (2006a); we also note that the authors report a value for 
the H$_\alpha$ EW of this source which is half of the one we measure in our 
optical spectrum. This hints to long-term variability of H$_\alpha$ emission 
in this object, which is not unusual in LMXBs.

\subsection{AGNs}

\begin{figure*}[th!]
\mbox{\psfig{file=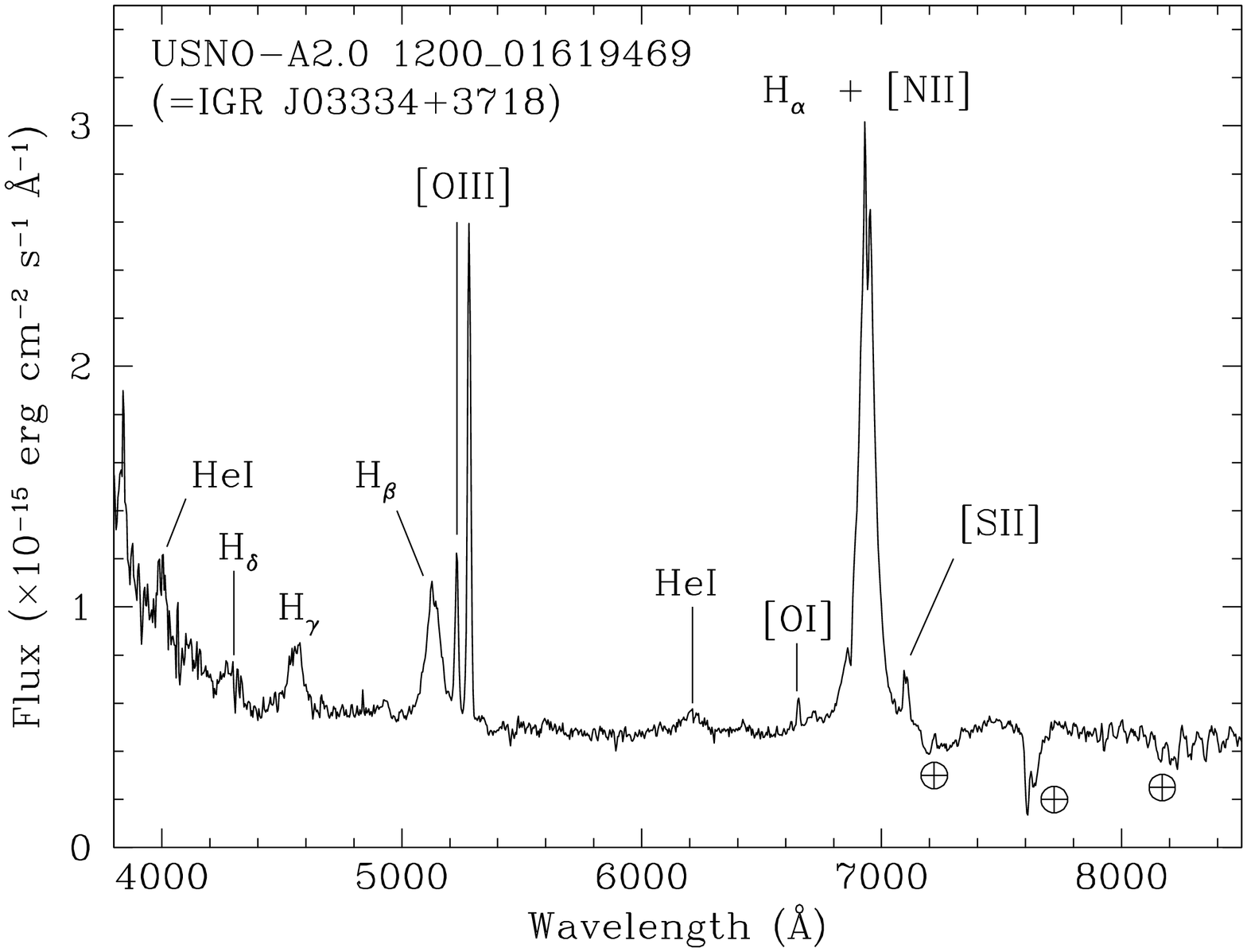,width=9cm,angle=270}}
\mbox{\psfig{file=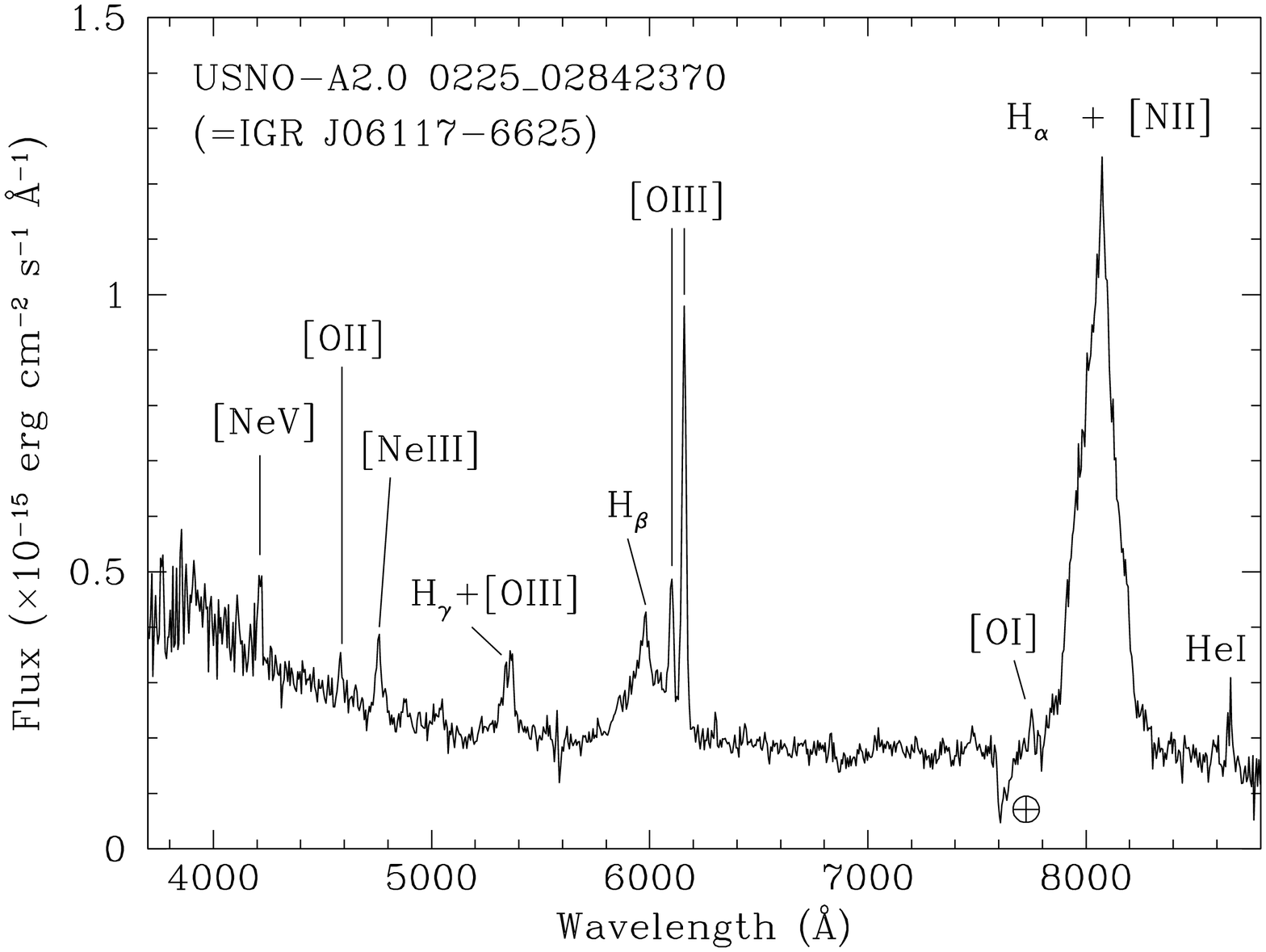,width=9cm,angle=270}}

\vspace{-.9cm}
\mbox{\psfig{file=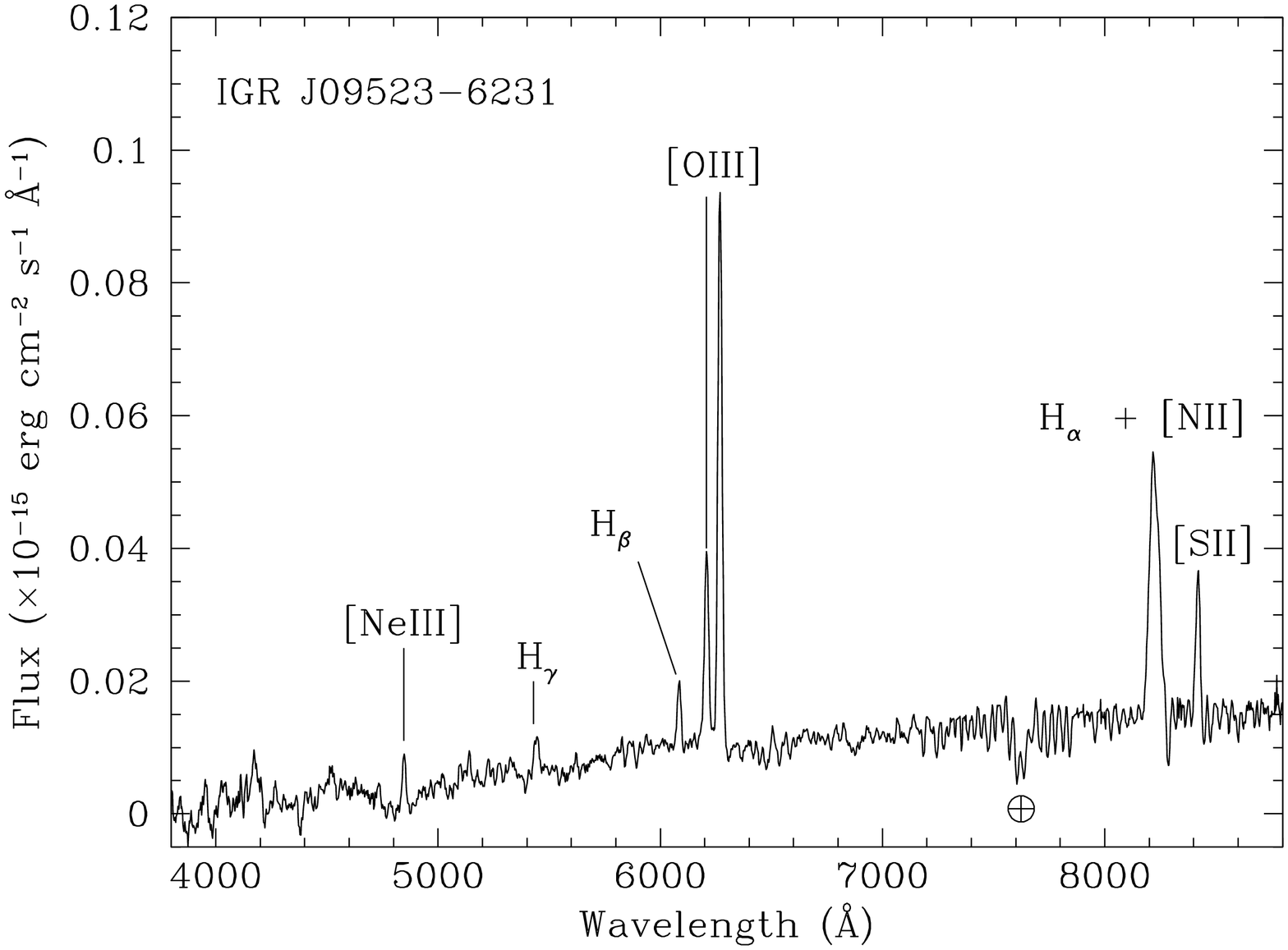,width=9cm,angle=270}}
\mbox{\psfig{file=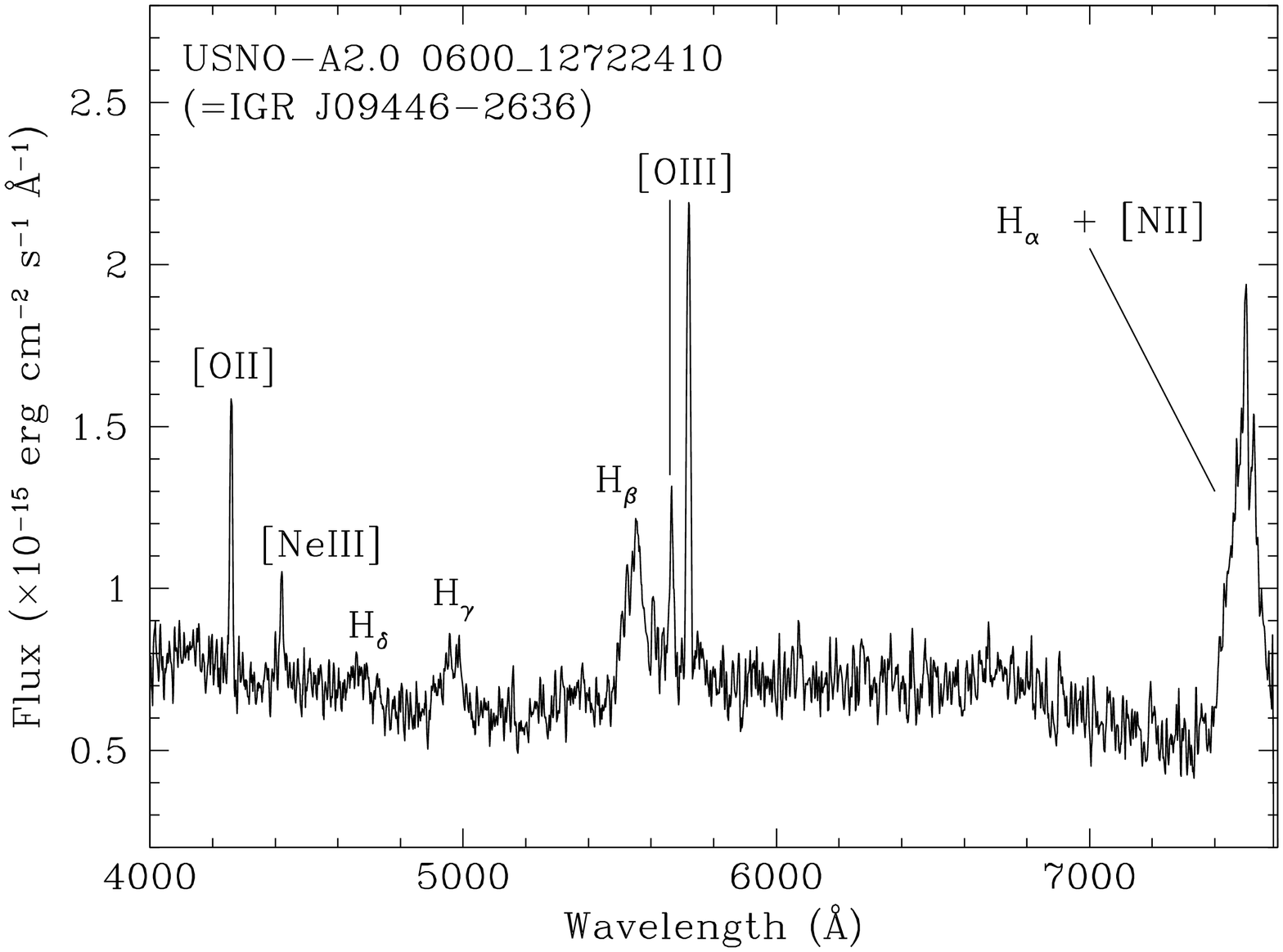,width=9cm,angle=270}}

\vspace{-.9cm}
\mbox{\psfig{file=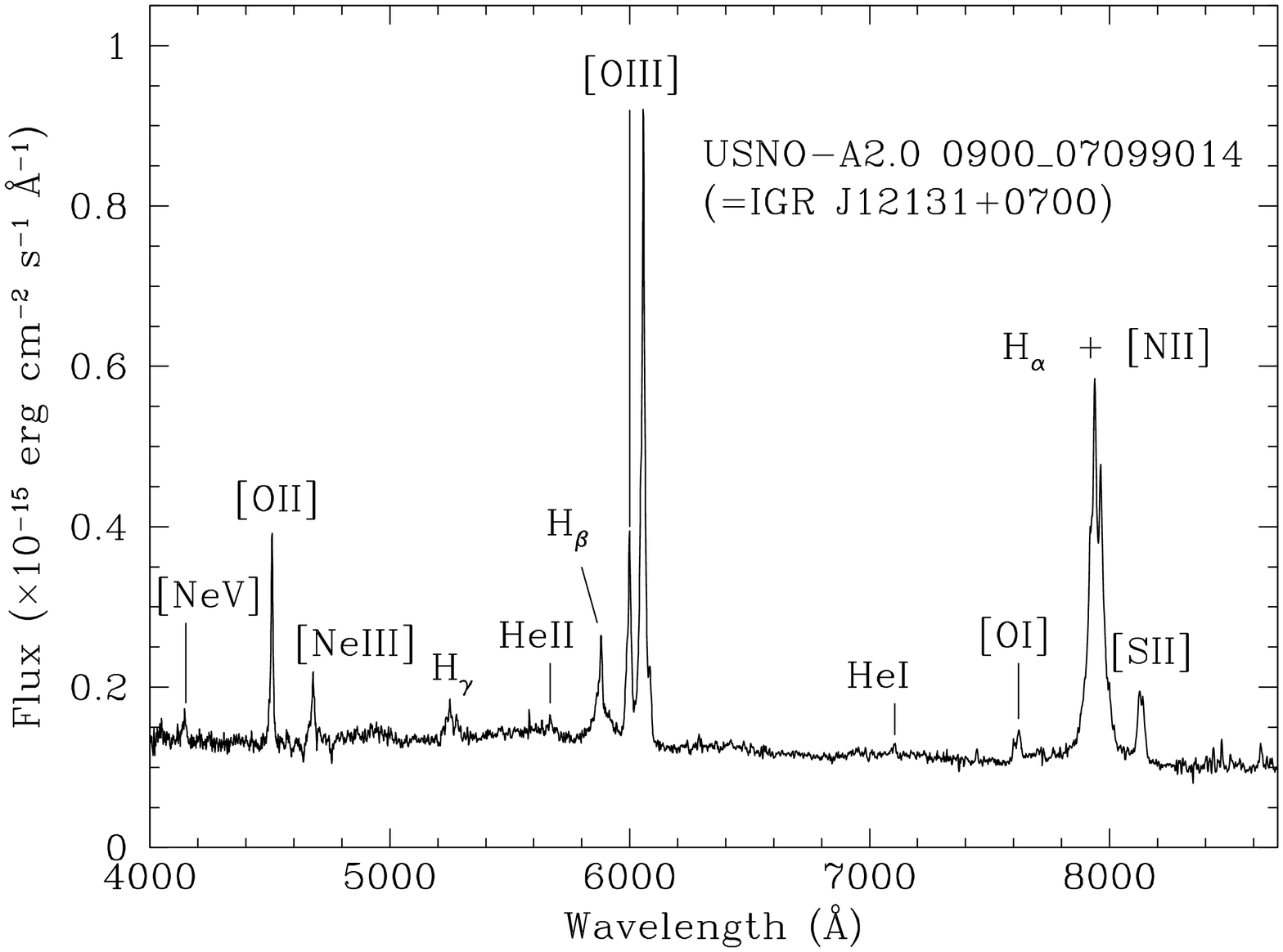,width=9cm,angle=270}}
\mbox{\psfig{file=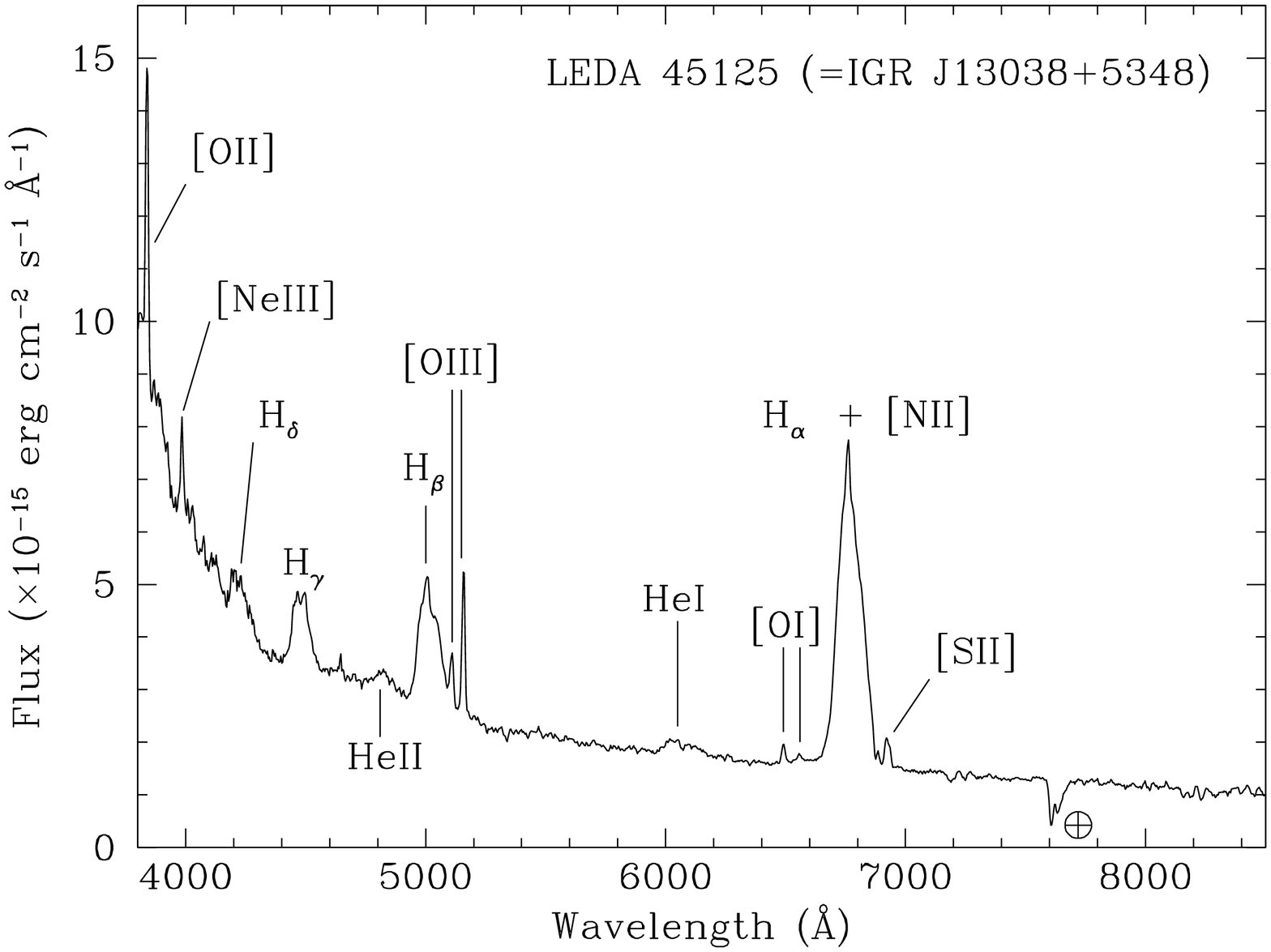,width=9cm,angle=270}}

\vspace{-.9cm}
\parbox{9cm}{
\psfig{file=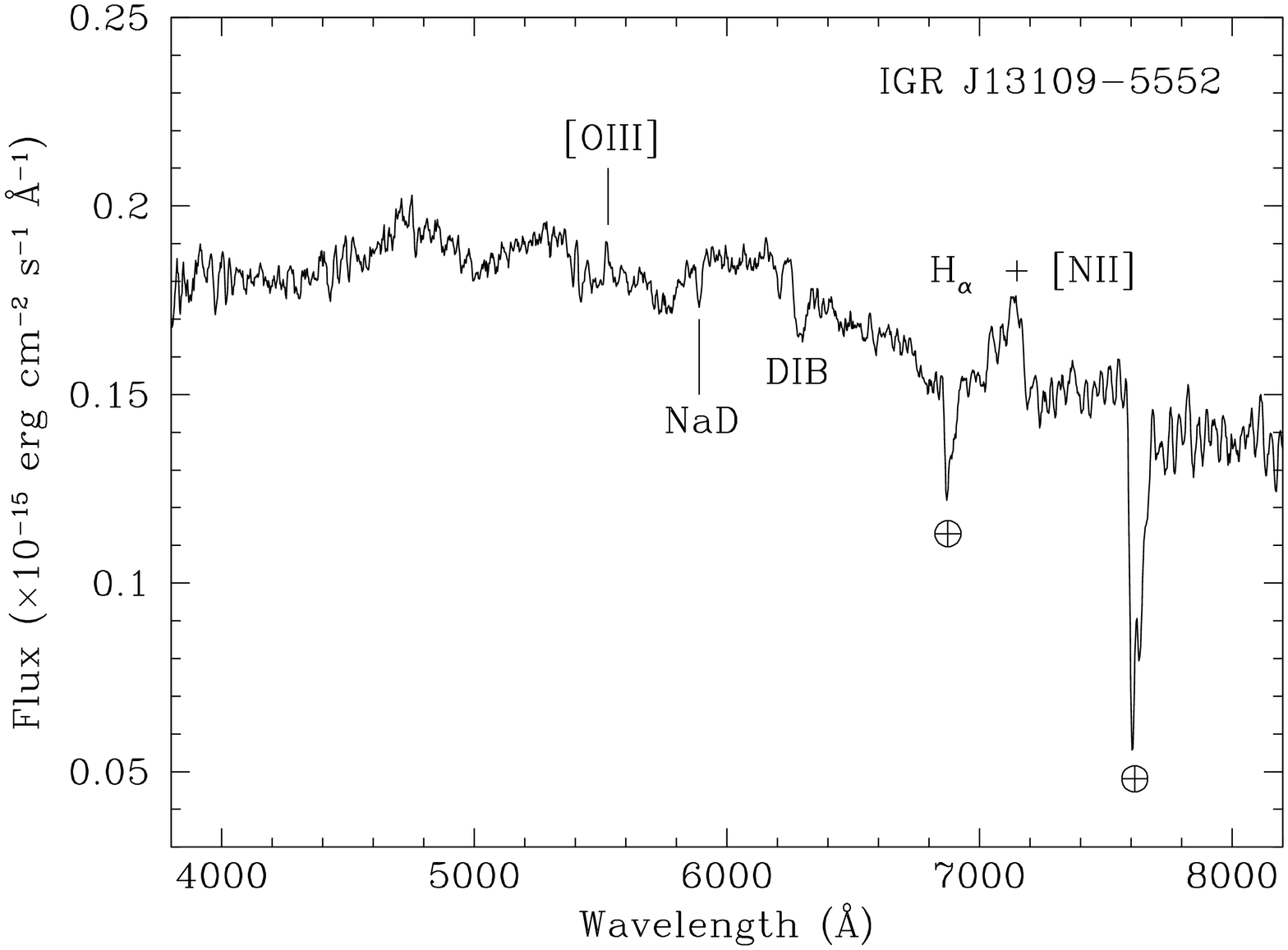,width=9cm,angle=270}
}
\hspace{.5cm}
\parbox{7.5cm}{
\vspace{1.5cm}
\caption{Spectra (not corrected for the intervening Galactic absorption) 
of the optical counterparts of 7 broad emission line AGNs belonging to 
the {\it INTEGRAL} sources sample presented in this paper.
For each spectrum the main spectral features are labeled. The 
symbol $\oplus$ indicates atmospheric telluric absorption bands.
The ESO 3.6m spectra have been smoothed using a Gaussian filter with
$\sigma$ = 5 \AA.
\vspace{1.5cm}
}
}
\end{figure*}

\begin{figure*}
\mbox{\psfig{file=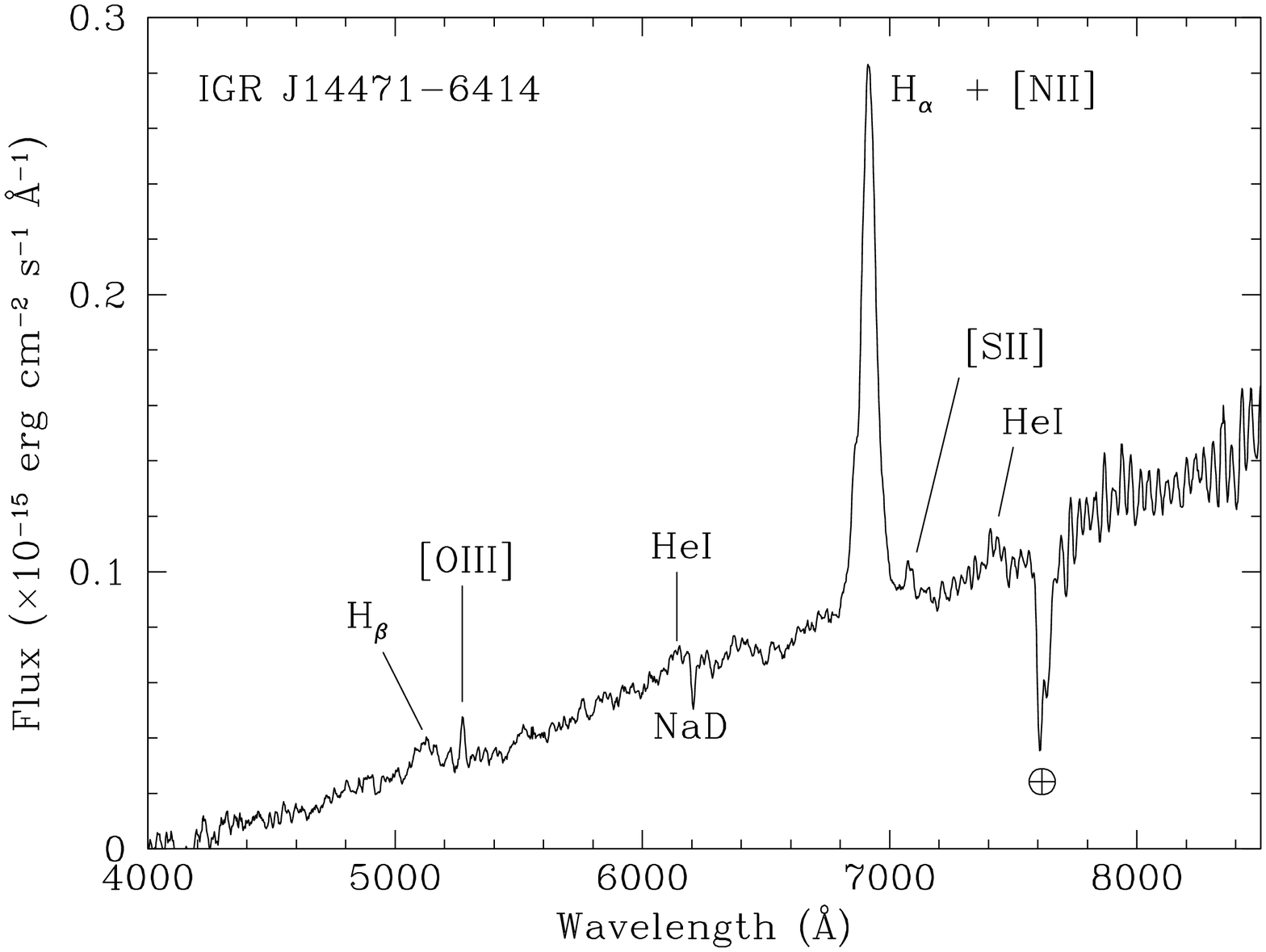,width=9cm,angle=270}}
\mbox{\psfig{file=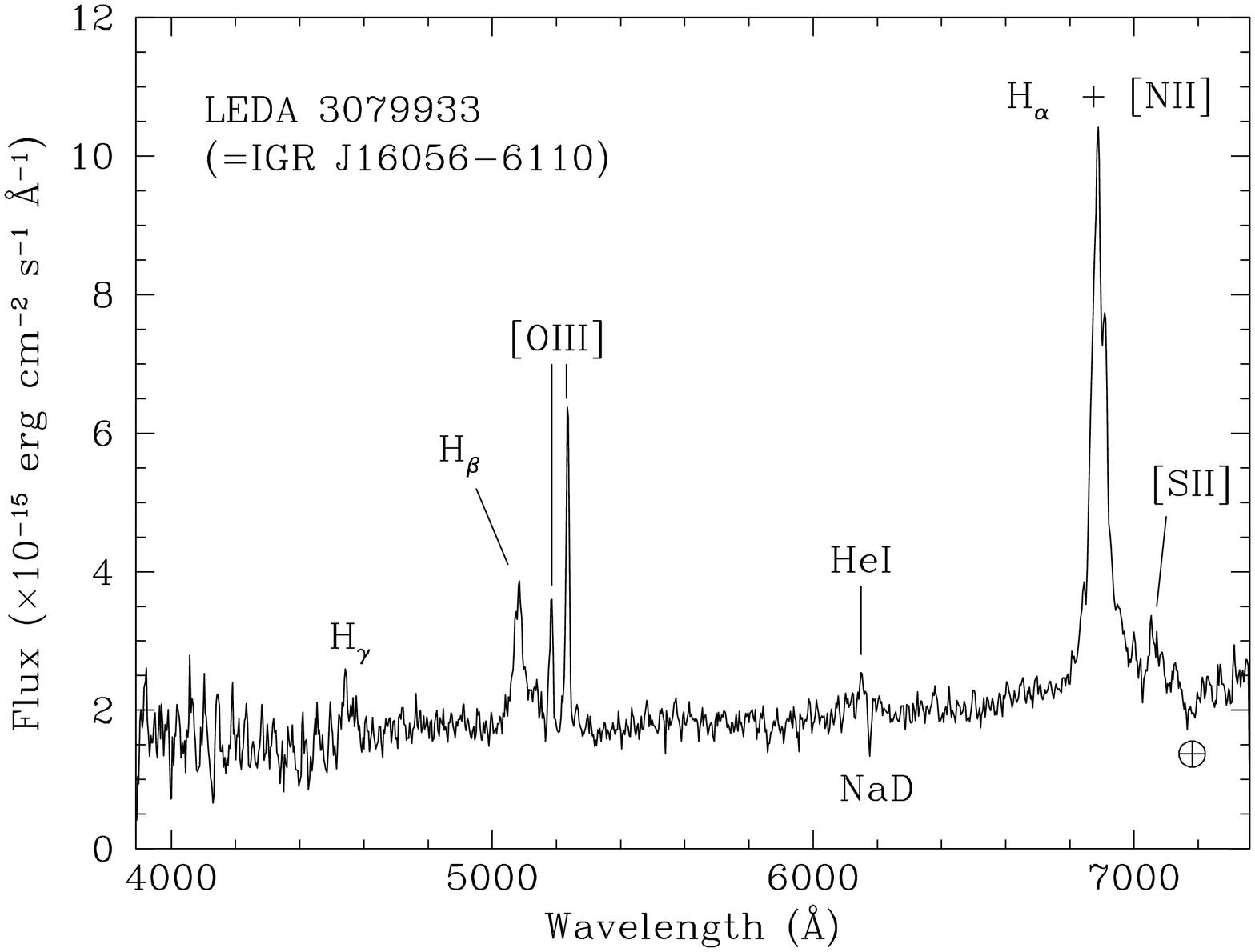,width=9cm,angle=270}}

\vspace{-.9cm}
\mbox{\psfig{file=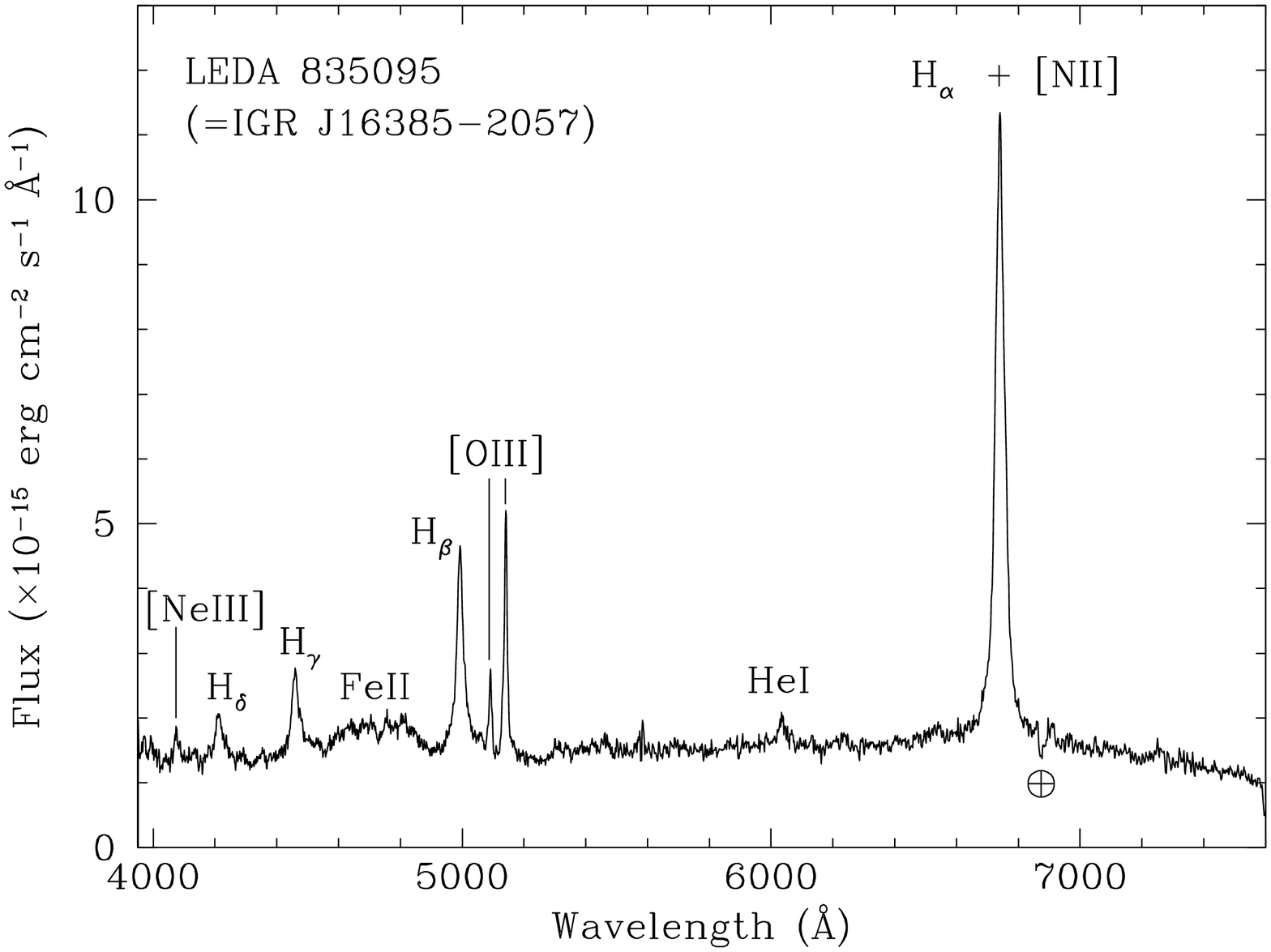,width=9cm,angle=270}}
\mbox{\psfig{file=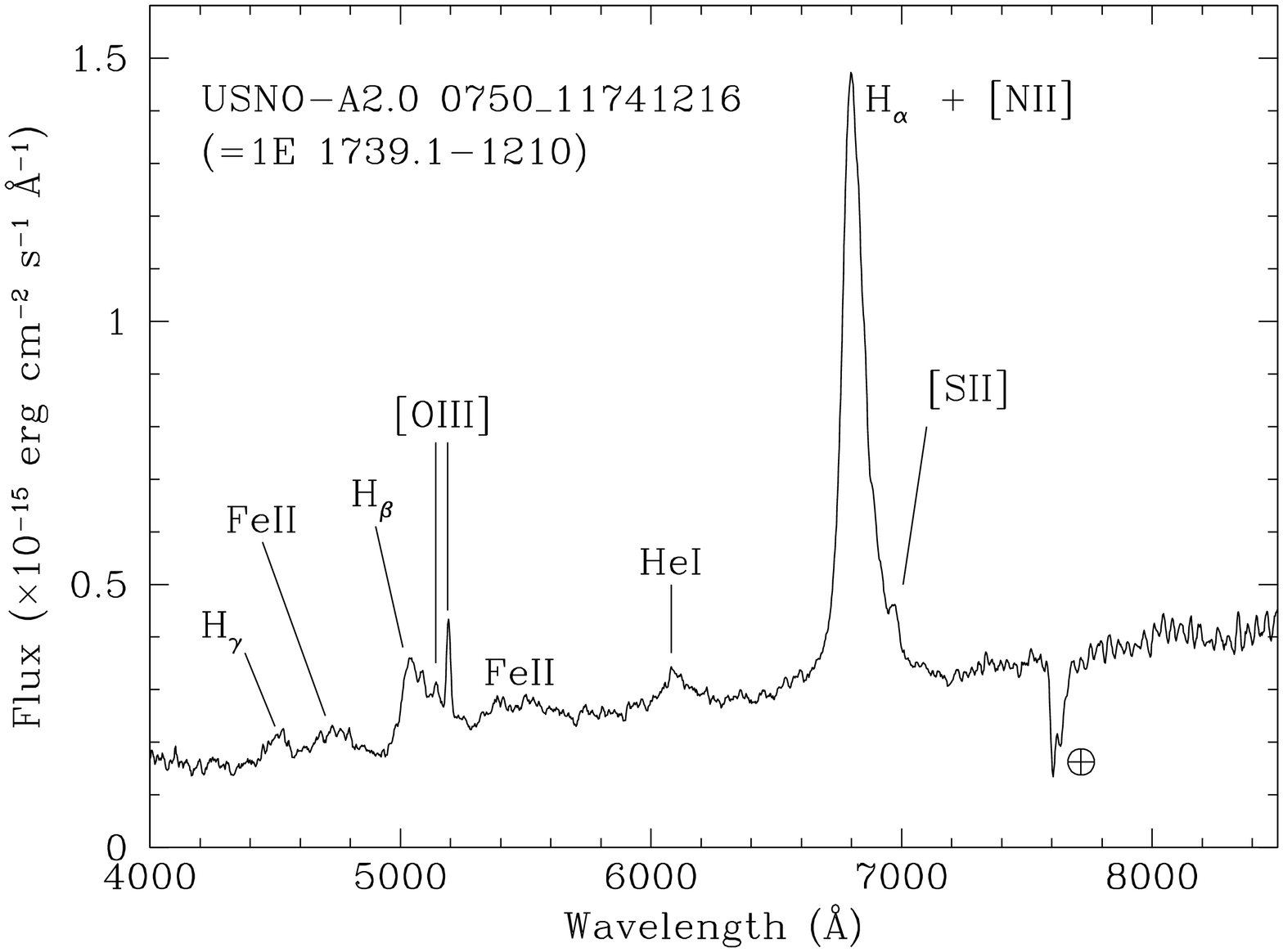,width=9cm,angle=270}}

\vspace{-.9cm}
\mbox{\psfig{file=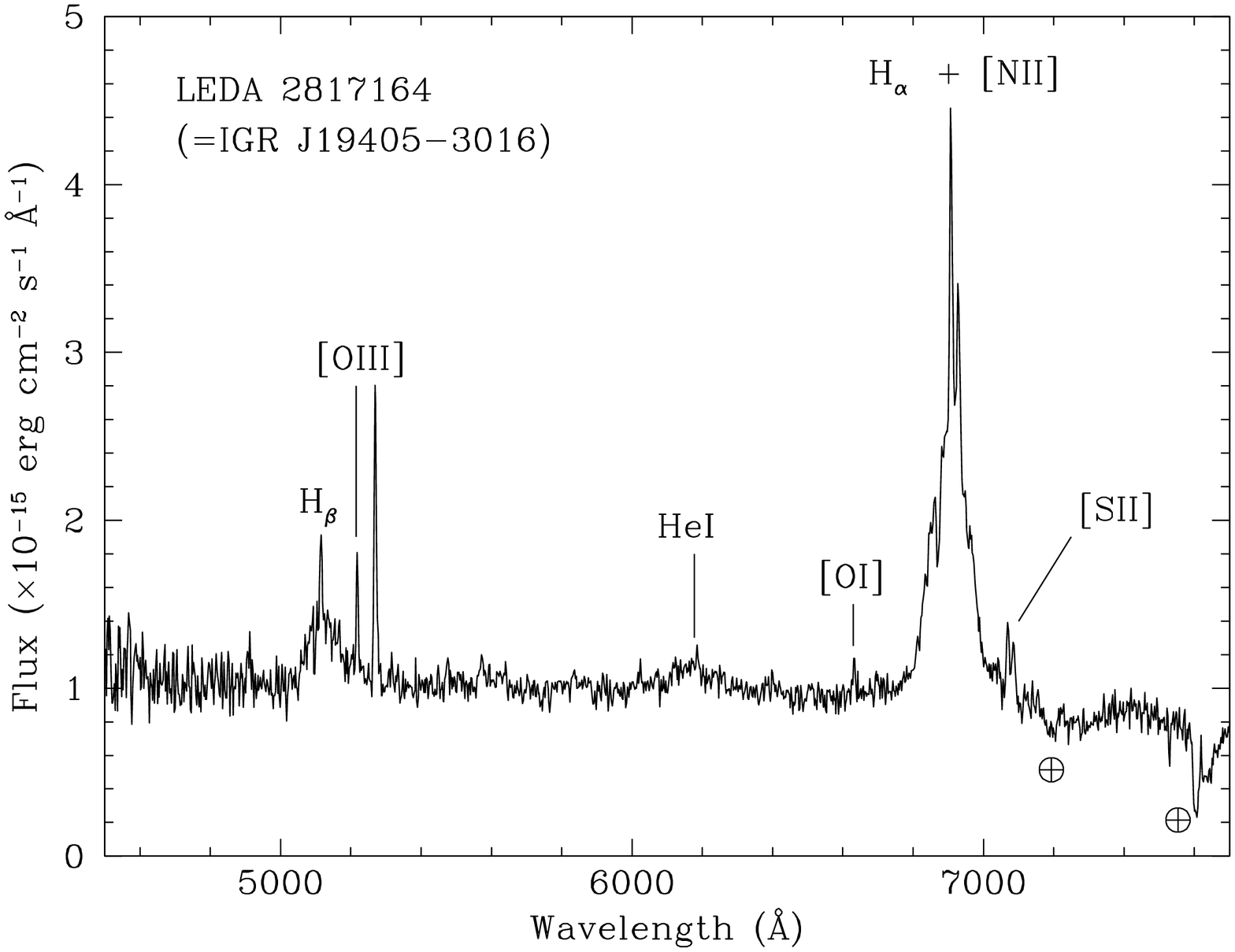,width=9cm,angle=270}}
\mbox{\psfig{file=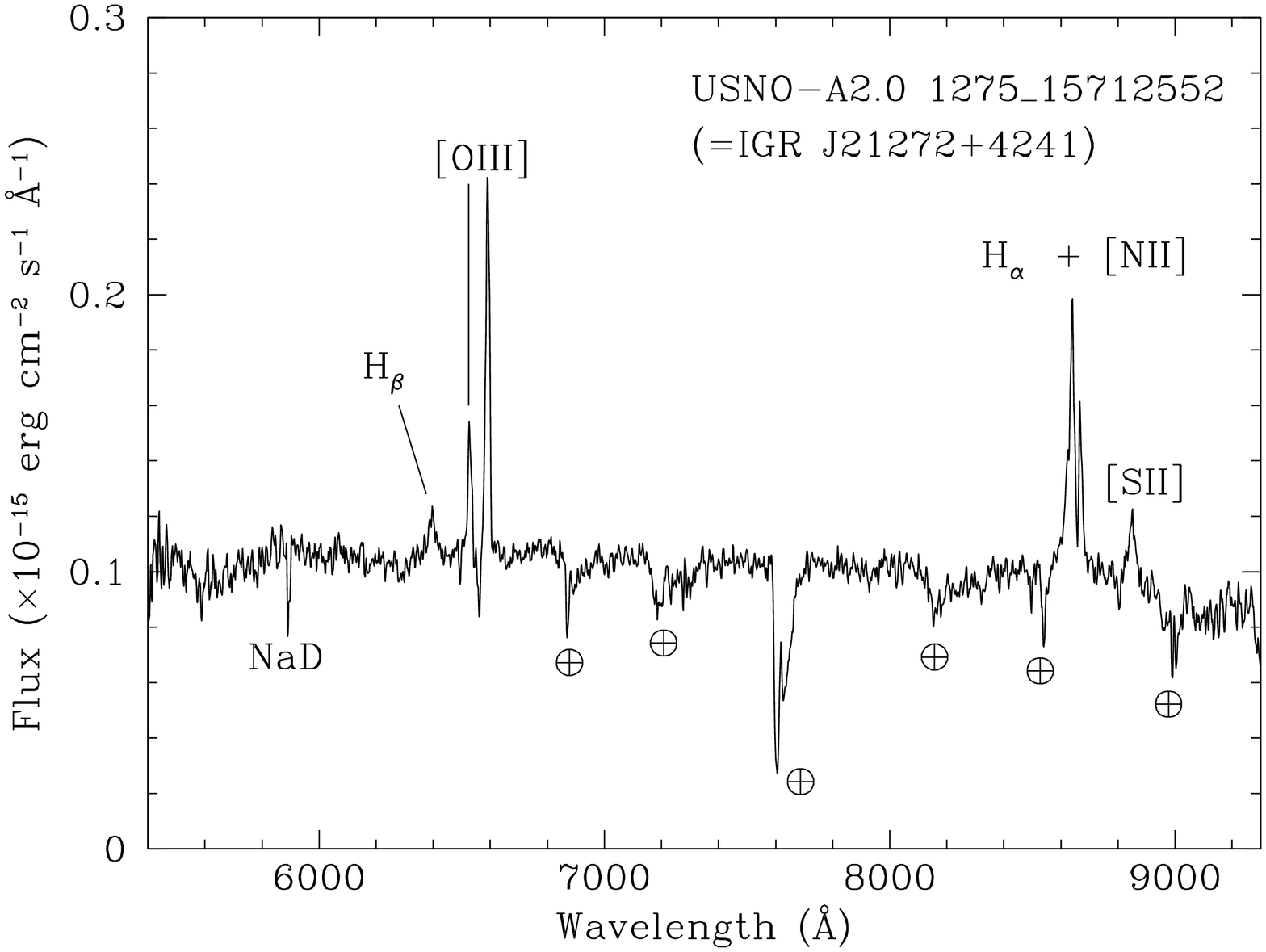,width=9cm,angle=270}}
\vspace{-0.5cm}
\caption{Spectra (not corrected for the intervening Galactic absorption)
of the optical counterparts of the remaining 6 broad emission line AGNs 
belonging to the {\it INTEGRAL} sources sample presented in this paper.
For each spectrum the main spectral features are labeled. The
symbol $\oplus$ indicates atmospheric telluric absorption bands.
The ESO 3.6m spectra have been smoothed using a Gaussian filter with
$\sigma$ = 5 \AA.}
\end{figure*}

\begin{figure*}
\mbox{\psfig{file=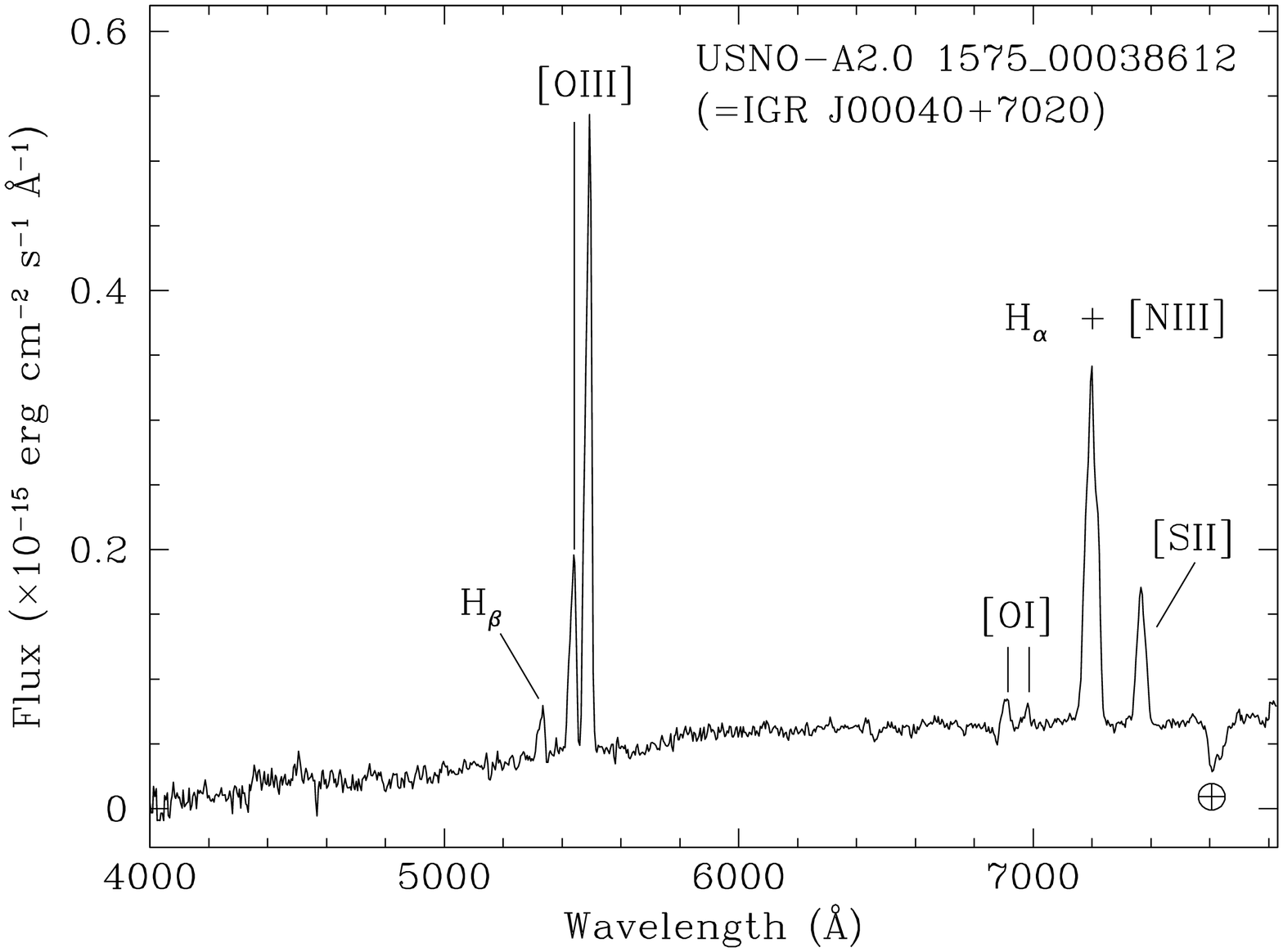,width=9cm,angle=270}}
\mbox{\psfig{file=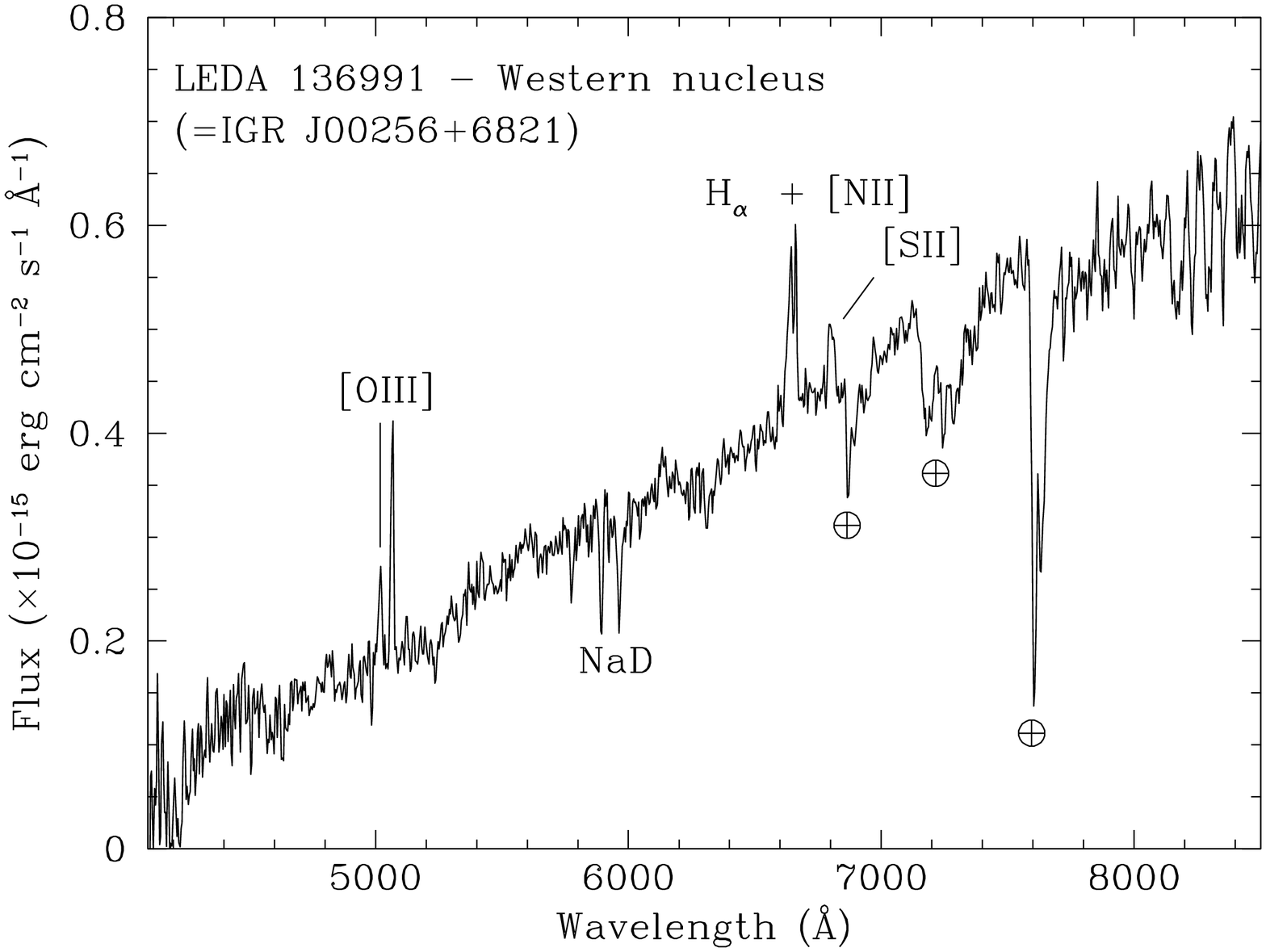,width=9cm,angle=270}}

\vspace{-.9cm}
\mbox{\psfig{file=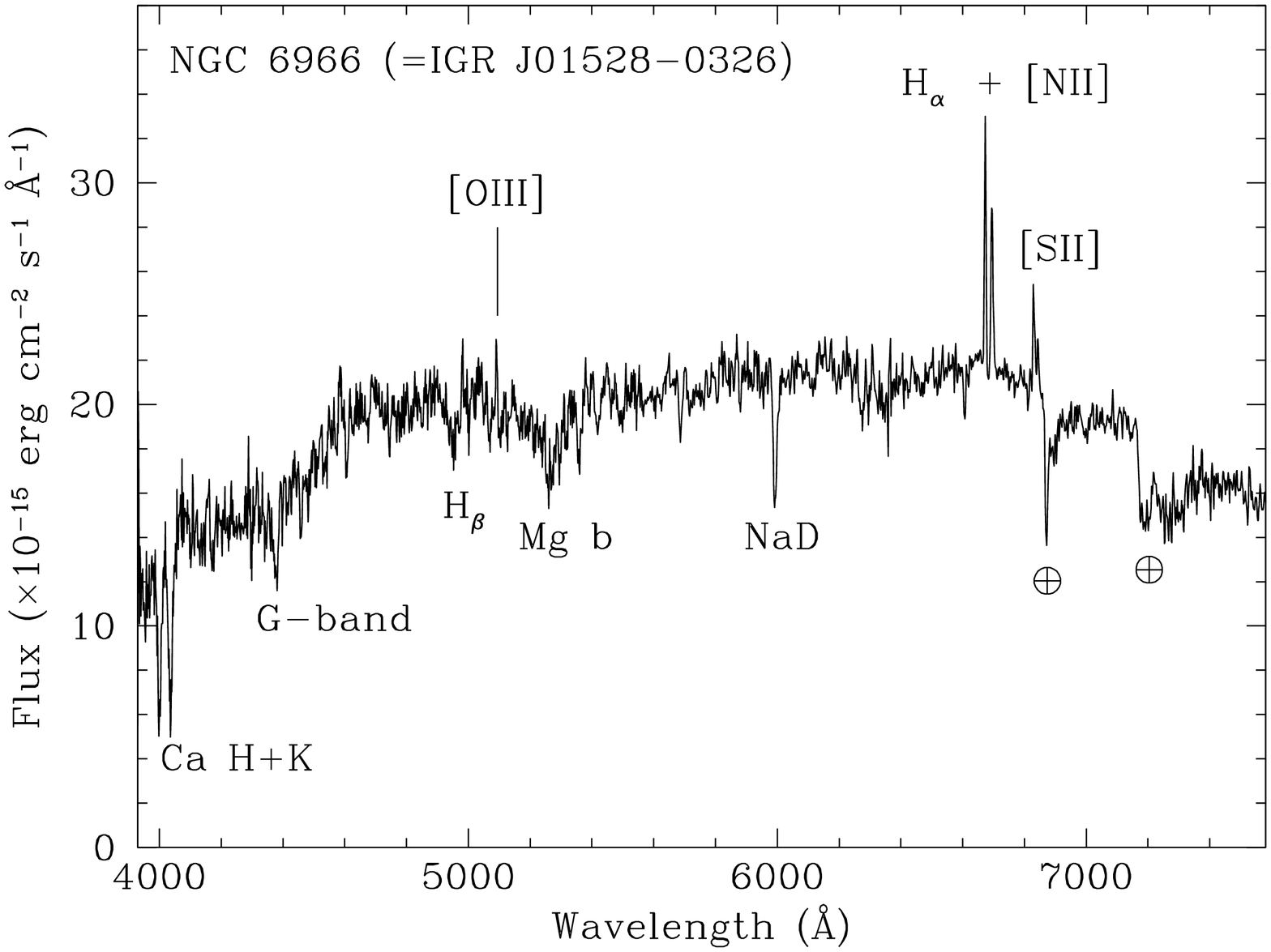,width=9cm,angle=270}}
\mbox{\psfig{file=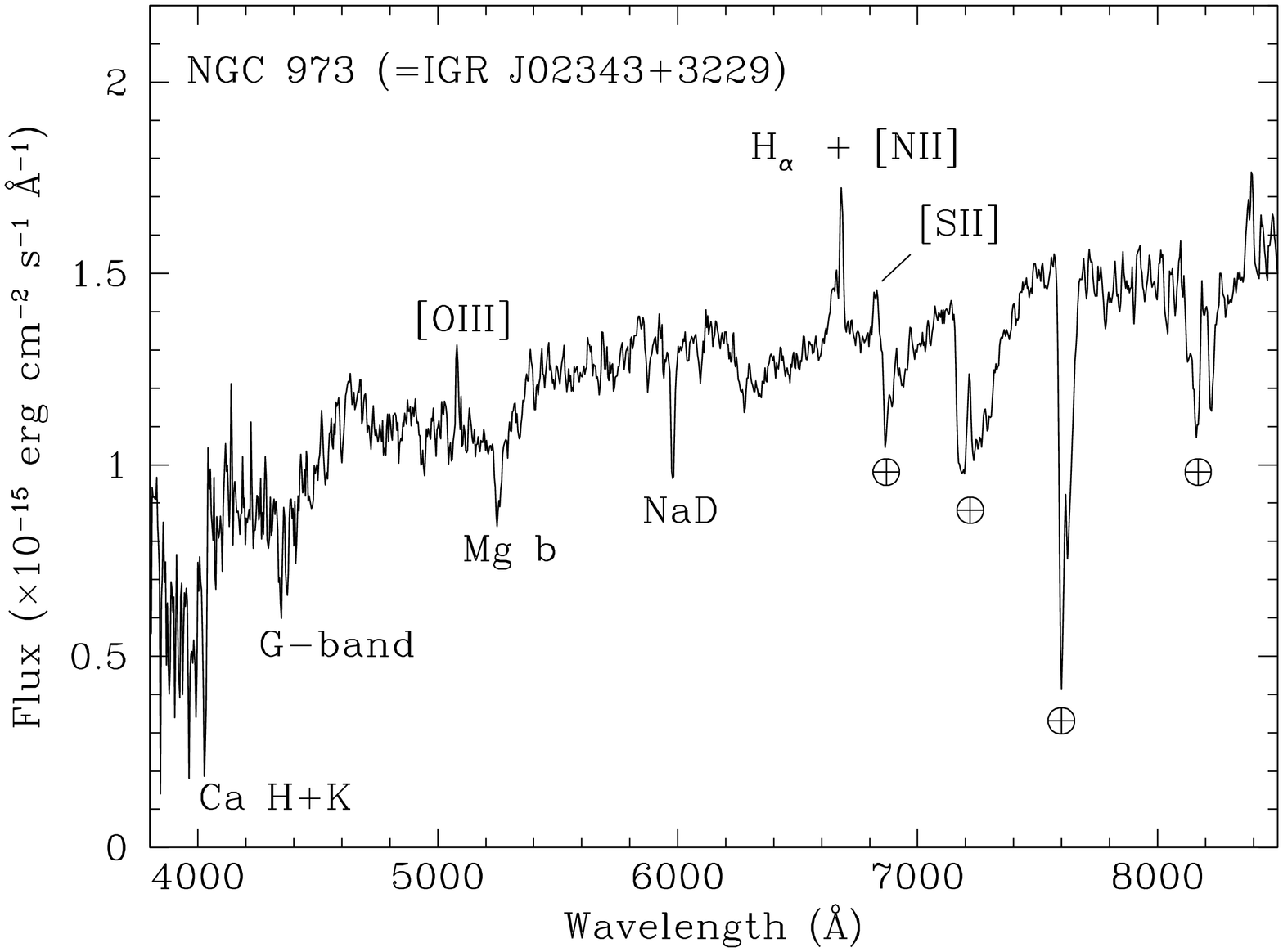,width=9cm,angle=270}}

\vspace{-.9cm}
\mbox{\psfig{file=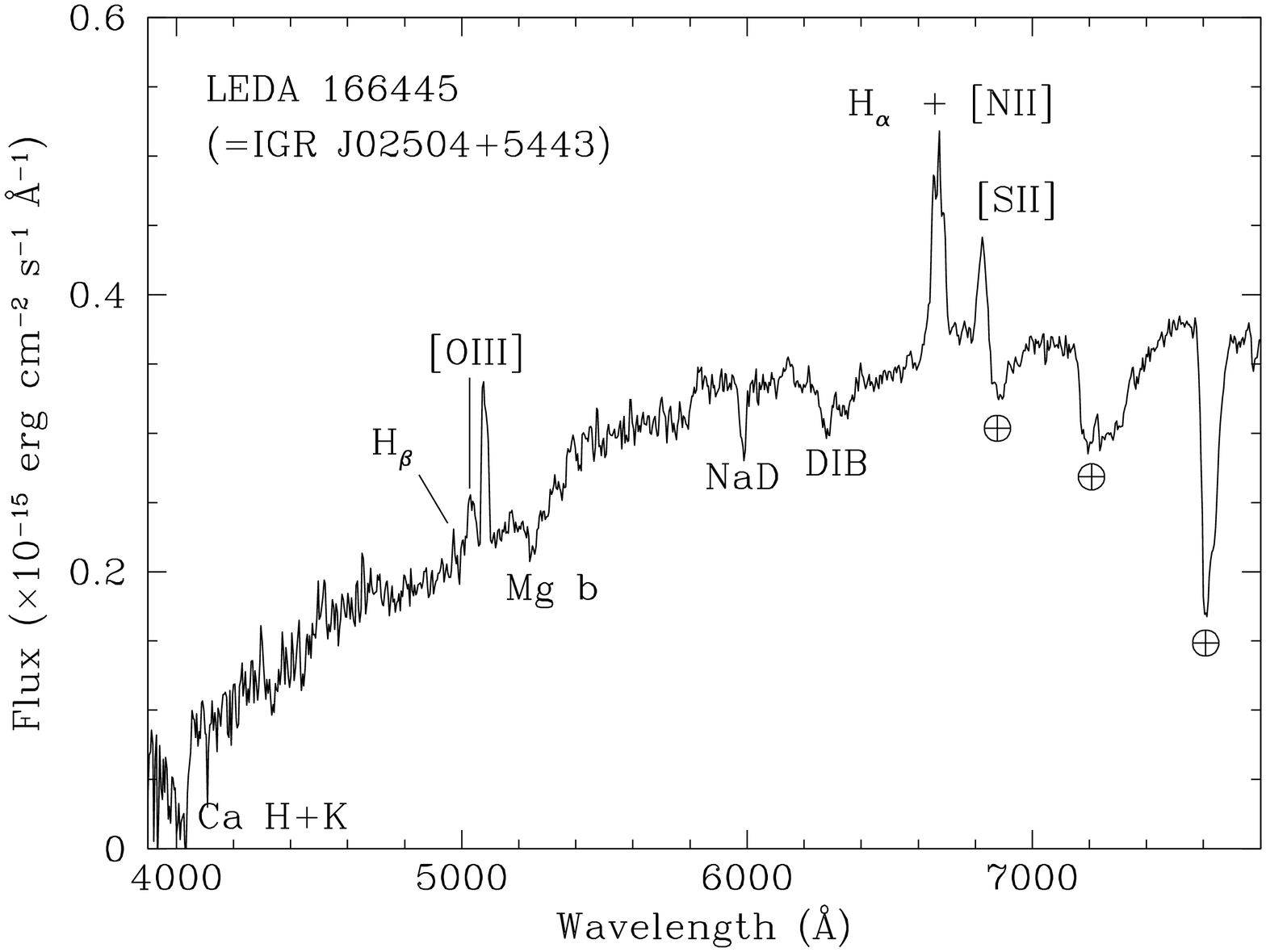,width=9cm,angle=270}}
\mbox{\psfig{file=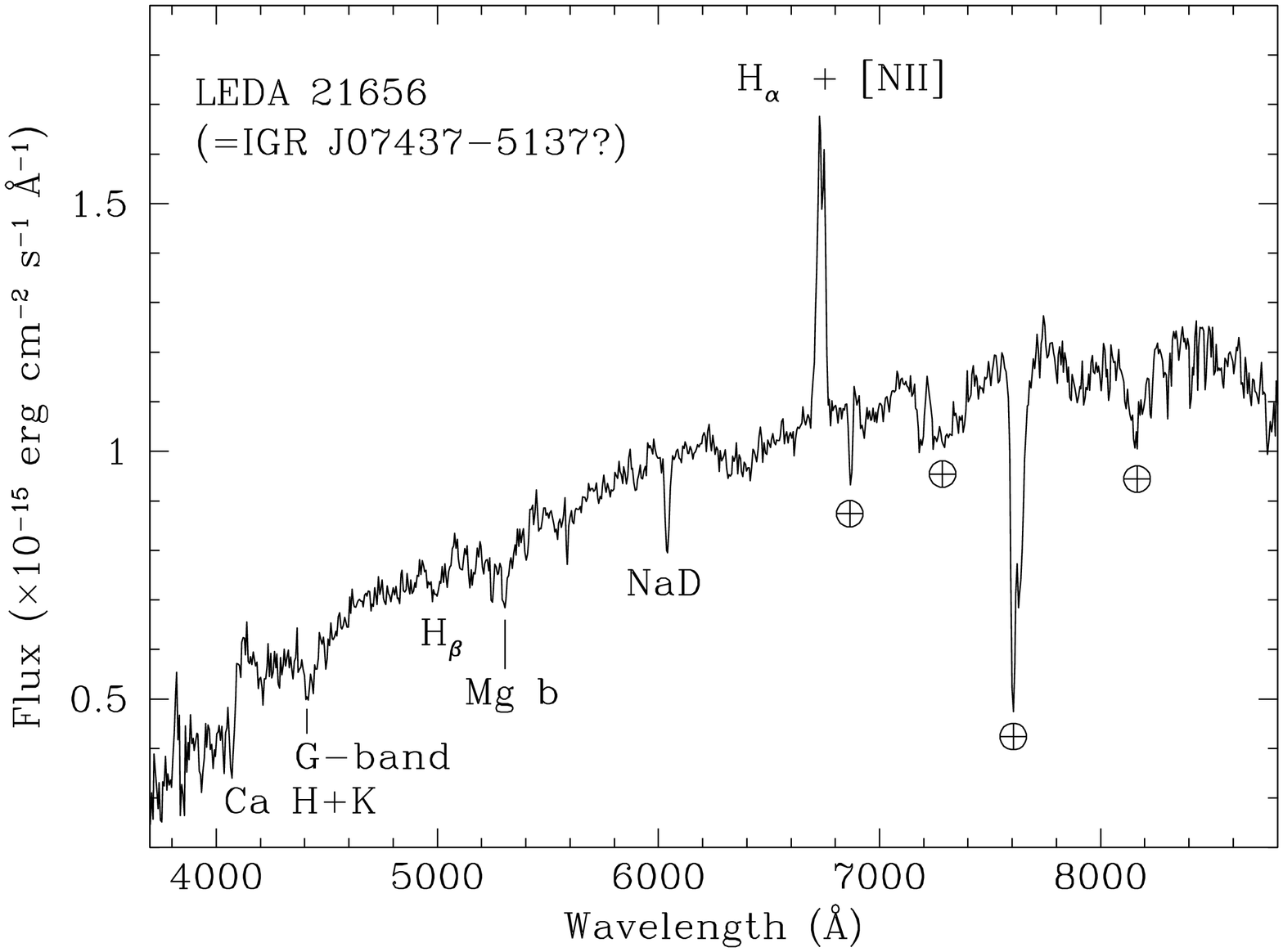,width=9cm,angle=270}}

\vspace{-.9cm}
\mbox{\psfig{file=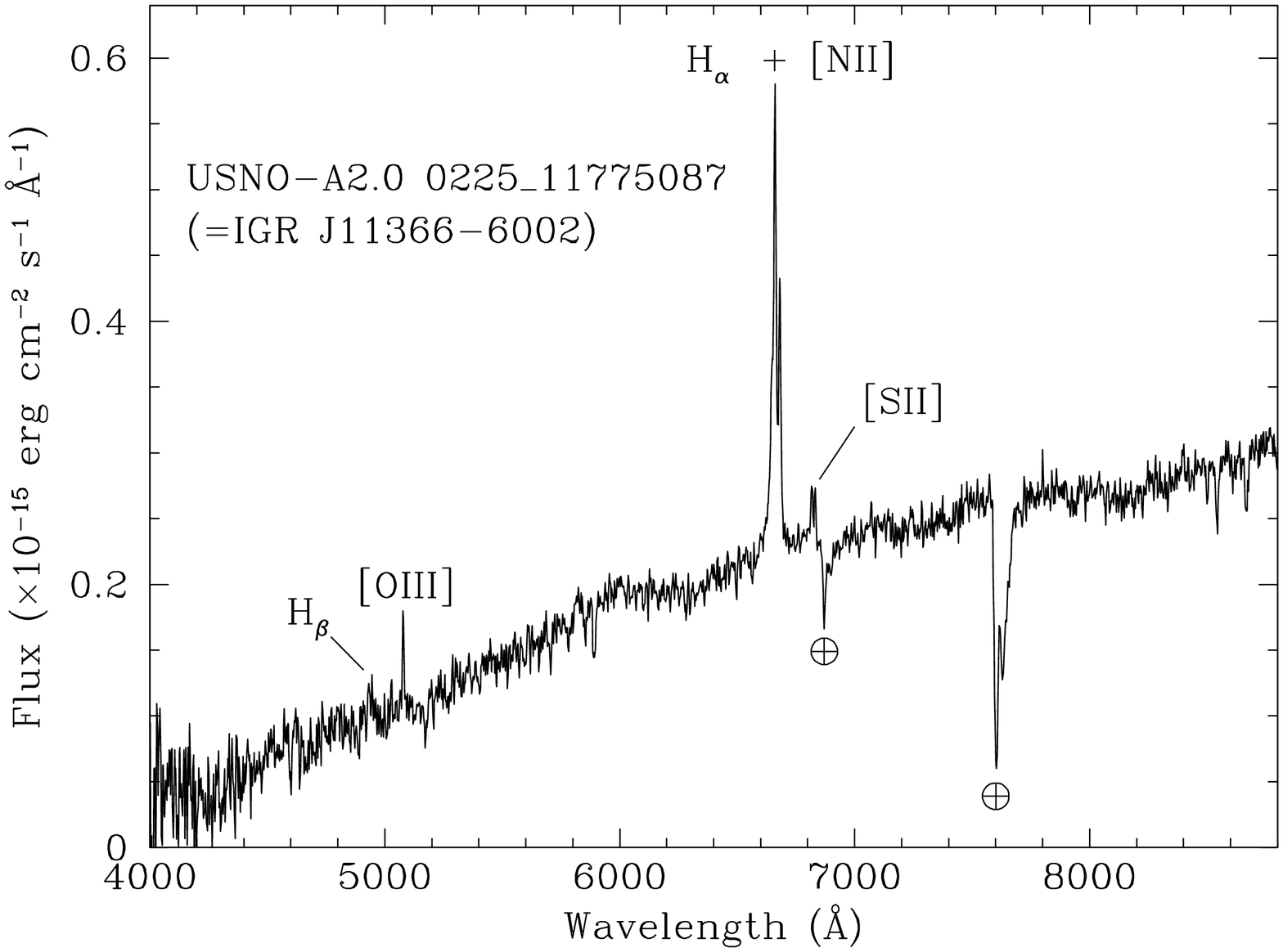,width=9cm,angle=270}}
\mbox{\psfig{file=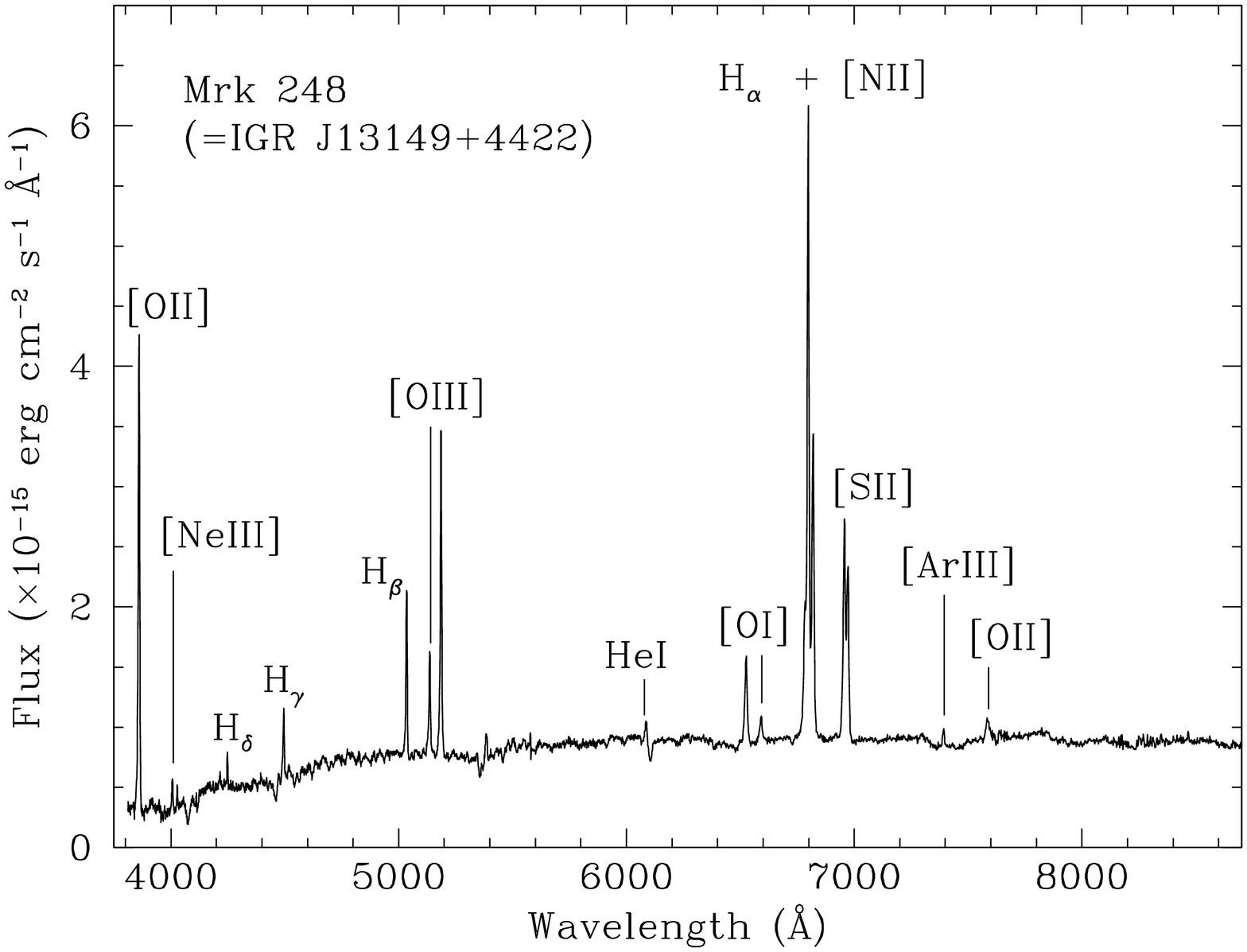,width=9cm,angle=270}}
\vspace{-.5cm}
\caption{Spectra (not corrected for the intervening Galactic absorption)
of the optical counterparts of 8 narrow emission line AGNs belonging to
the {\it INTEGRAL} sources sample presented in this paper.
For each spectrum the main spectral features are labeled. The
symbol $\oplus$ indicates atmospheric telluric absorption bands.
The ESO 3.6m spectra have been smoothed using a Gaussian filter with
$\sigma$ = 5 \AA.
We remark that galaxy LEDA 21656 should be considered as the tentative, 
although likely, counterpart of IGR J07437$-$5137 (see text).}
\end{figure*}

\begin{figure*}[th!]
\mbox{\psfig{file=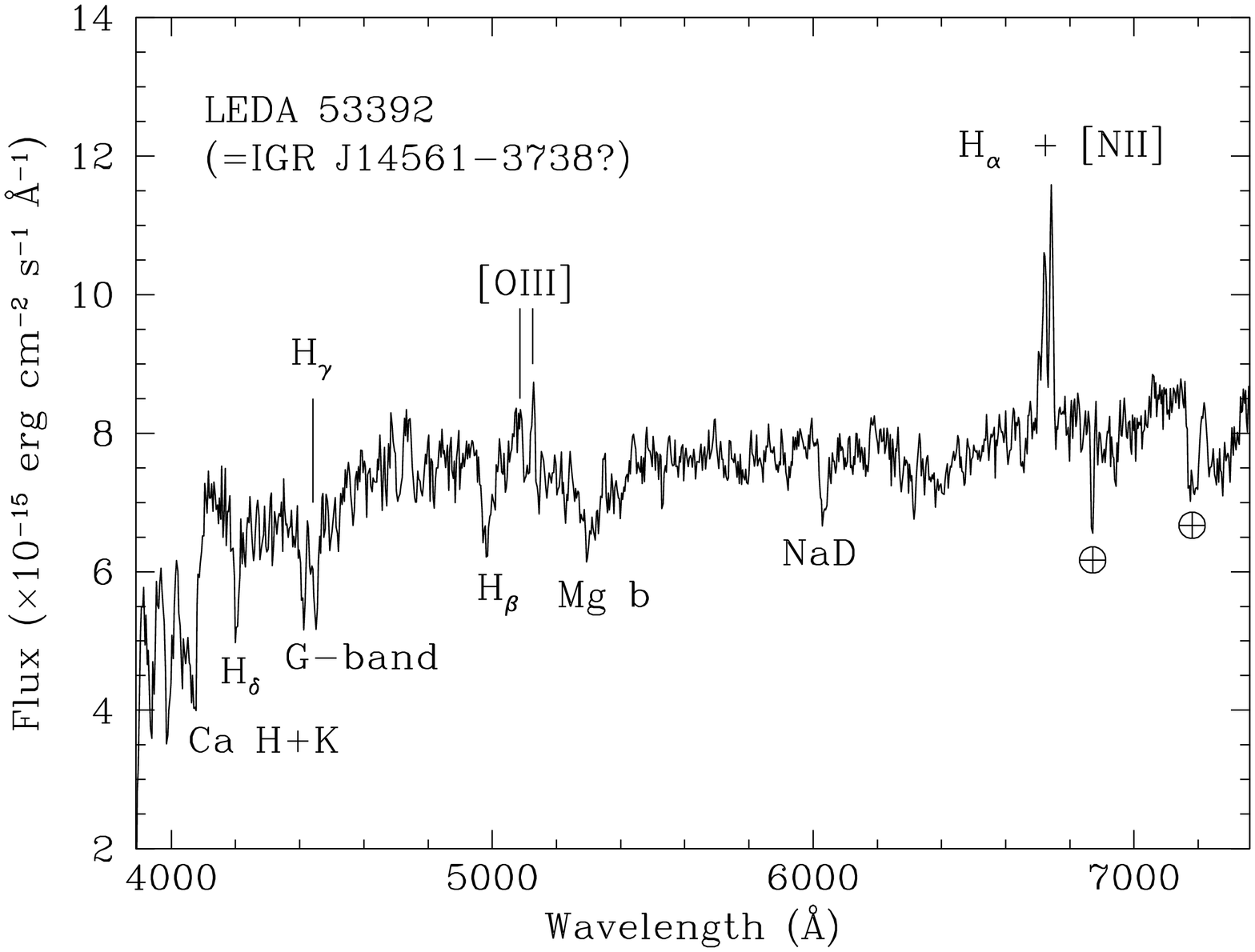,width=9cm,angle=270}}
\mbox{\psfig{file=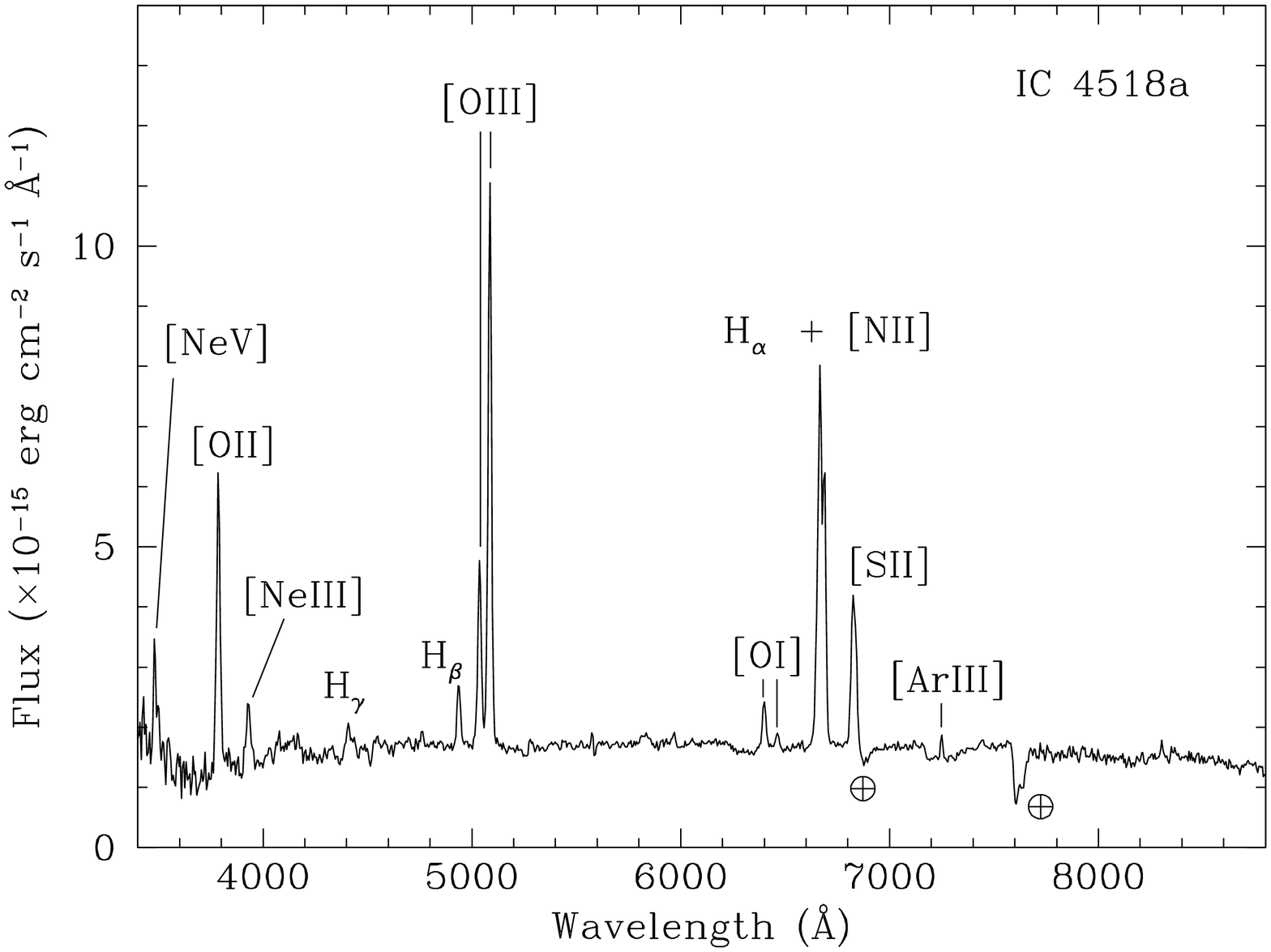,width=9cm,angle=270}}

\vspace{-.9cm}
\mbox{\psfig{file=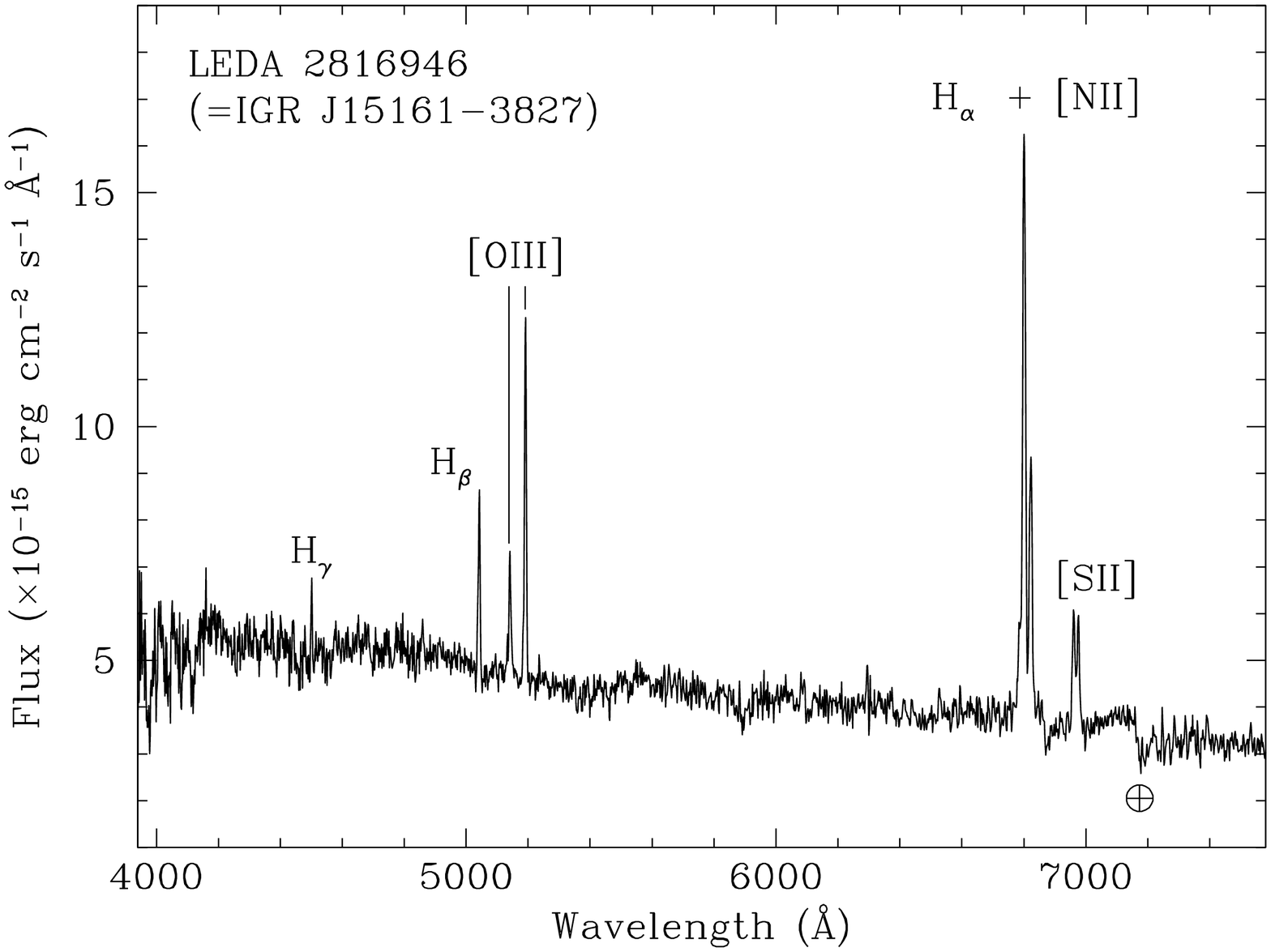,width=9cm,angle=270}}
\mbox{\psfig{file=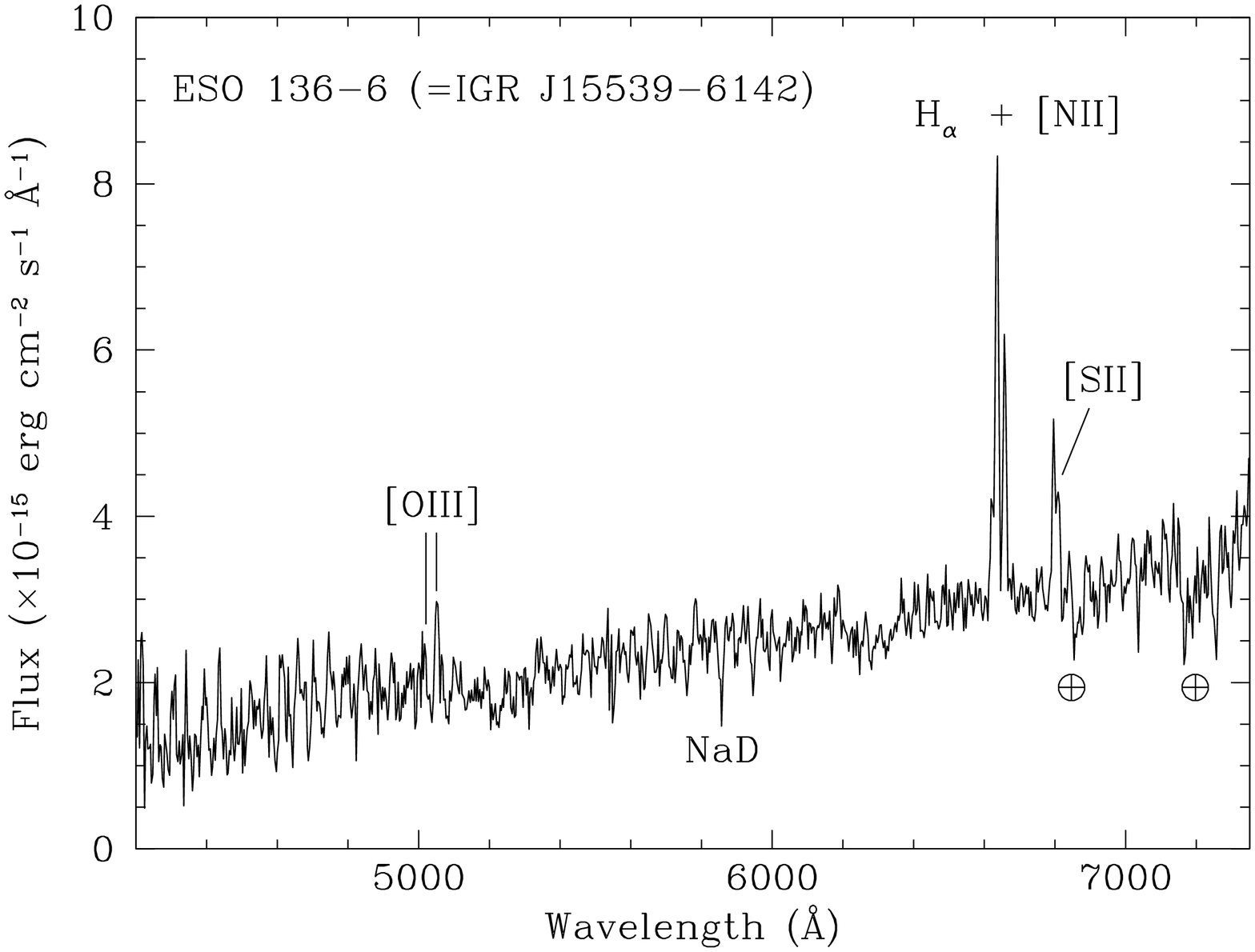,width=9cm,angle=270}}

\vspace{-.9cm}
\mbox{\psfig{file=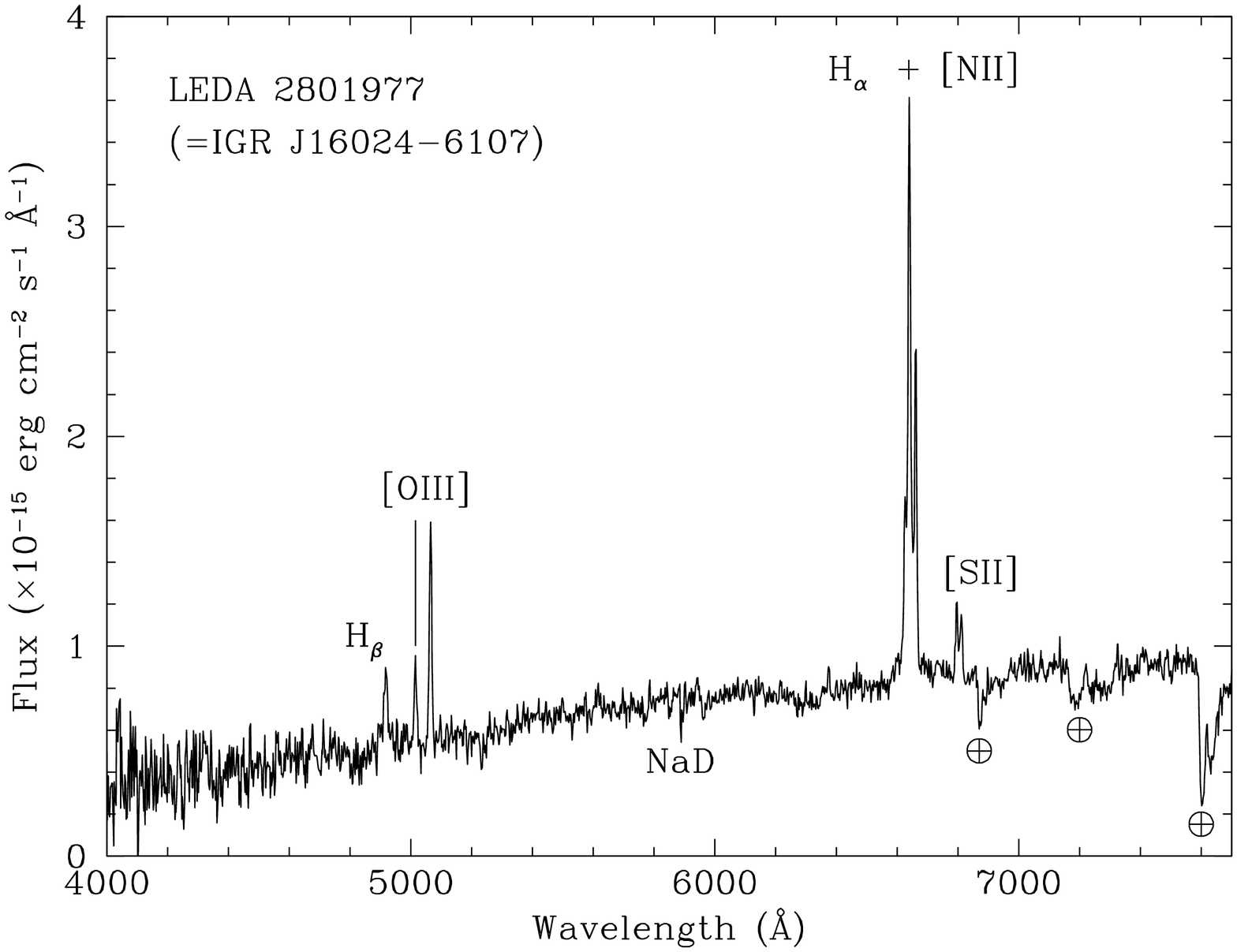,width=9cm,angle=270}}
\mbox{\psfig{file=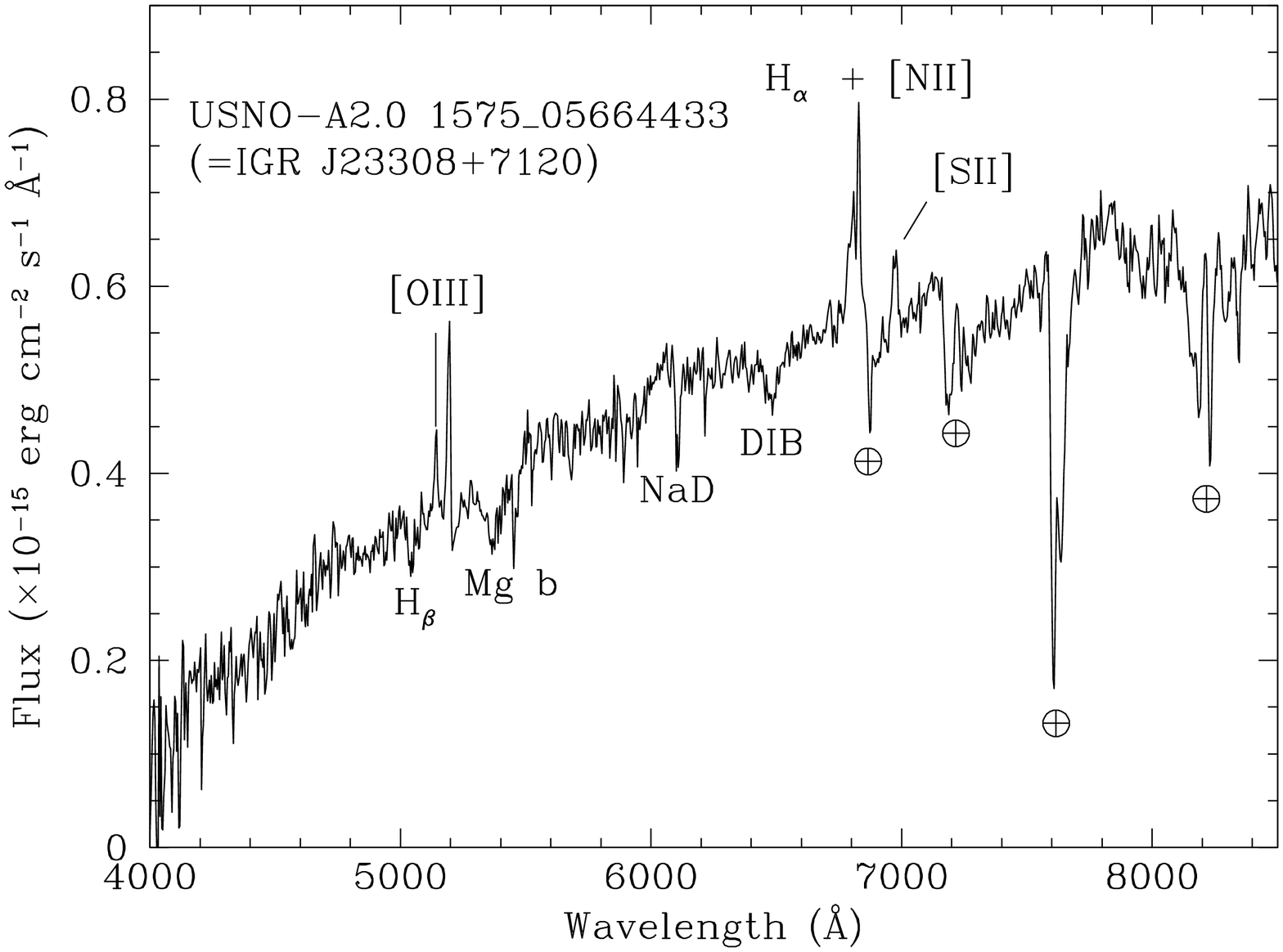,width=9cm,angle=270}}

\vspace{-.9cm}
\parbox{9cm}{
\psfig{file=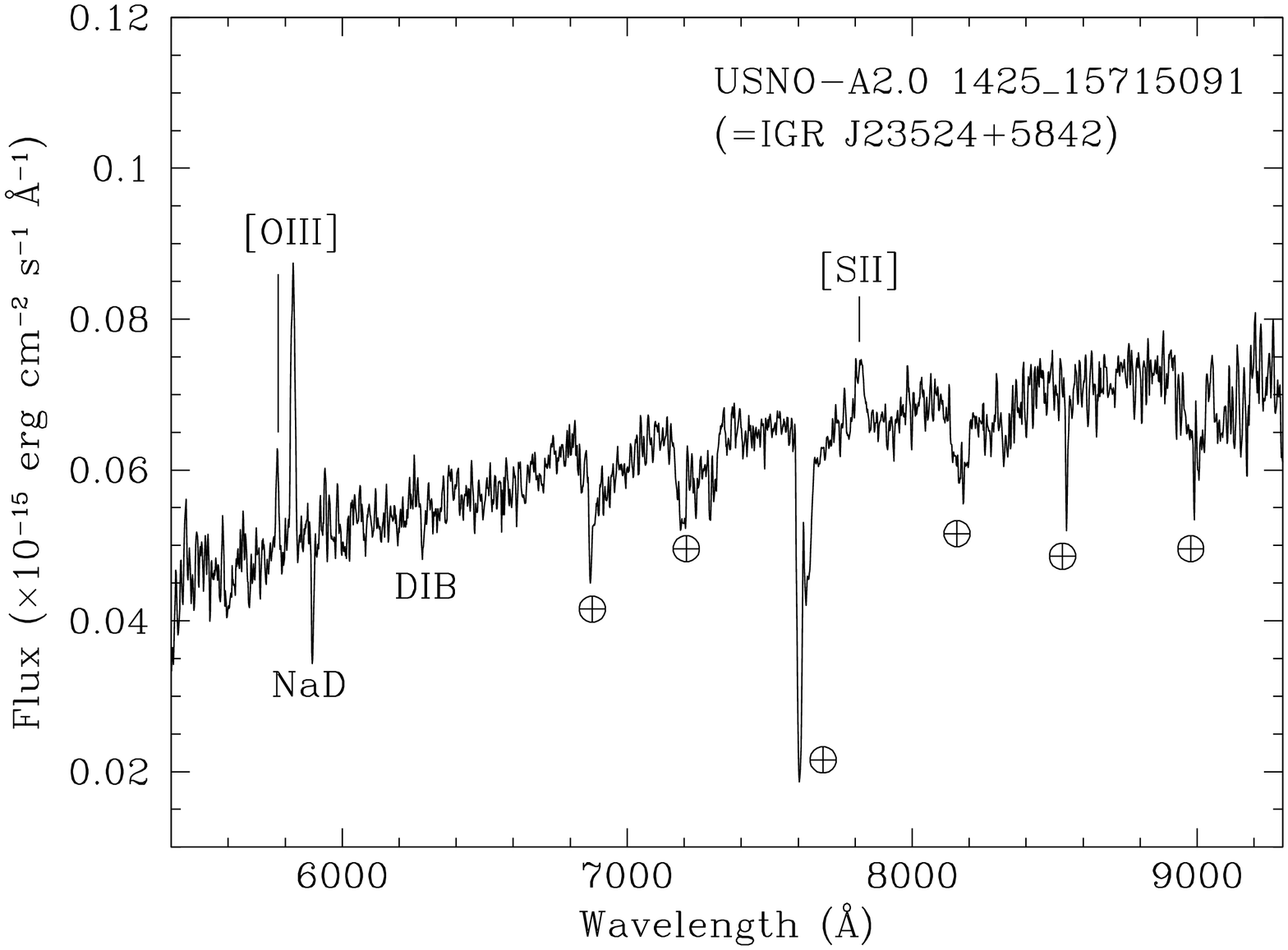,width=9cm,angle=270}
}
\parbox{8cm}{
\vspace{-.6cm}
\caption{Spectra (not corrected for the intervening Galactic absorption)
of the optical counterparts of the remaining 7 narrow emission line AGNs 
belonging to the {\it INTEGRAL} sources sample presented in this paper.
For each spectrum the main spectral features are labeled. The
symbol $\oplus$ indicates atmospheric telluric absorption bands.
The WHT spectra have been smoothed using a Gaussian filter with 
$\sigma$ = 3 \AA, whereas those ESO 3.6m ones have been smoothed 
using a Gaussian filter with $\sigma$ = 5 \AA.
We remark that galaxy LEDA 53392 should be considered as the 
tentative, although likely, counterpart of IGR J14561$-$3738 
(see text).}
}
\end{figure*}

\begin{table*}[th!]
\caption[]{Synoptic table containing the main results concerning the 13
broad emission line AGNs (see Figs. 9 and 10) identified or observed in the 
present sample of {\it INTEGRAL} sources.}
\scriptsize
\begin{center}
\begin{tabular}{lcccccrcr}
\noalign{\smallskip}
\hline
\hline
\noalign{\smallskip}
\multicolumn{1}{c}{Object} & $F_{\rm H_\alpha}$ & $F_{\rm H_\beta}$ &
$F_{\rm [OIII]}$ & Class & $z$ & 
\multicolumn{1}{c}{$D_L$ (Mpc)} & $E(B-V)_{\rm Gal}$ & \multicolumn{1}{c}{$L_{\rm X}$} \\
\noalign{\smallskip}
\hline
\noalign{\smallskip}

IGR J03334+3718 & * & 3.9$\pm$0.2 & 3.3$\pm$0.2 & Sy1.5 & 0.055 & 264.3 & 0.535 & 0.33 (0.1--2.4) \\
 & * & [19$\pm$1] & [17$\pm$1] & & & & & 16.4 (17--60) \\

 & & & & & & & & \\

IGR J06117$-$6625 & * & 3.1$\pm$0.2 & 1.82$\pm$0.09 & Sy1.5 & 0.230 & 1234.8 & 0.063 & 20.0 (0.1--2.4) \\
 & * & [3.7$\pm$0.2] & [2.12$\pm$0.11] & & & & & 529.5 (20--100) \\

 & & & & & & & & \\

IGR J09446$-$2636 & * & 3.5$\pm$0.3 & 2.31$\pm$0.12 & Sy1.5 & 0.1425 & 726.2 & 0.086 & 10.1 (0.1--2.4) \\
 & * & [4.5$\pm$0.5] & [2.82$\pm$0.14] & & & & & 246.1 (17--60) \\

 & & & & & & & & \\

IGR J09523$-$6231 & * & 0.019$\pm$0.004 & 0.19$\pm$0.01 & Sy1.9 & 0.252 & 1369.4 & 0.304 & 42.4 
(0.2--12) \\
 & * & [0.040$\pm$0.008] & [0.40$\pm$0.02] & & & & & 279.4 (20--100) \\

 & & & & & & & & \\

IGR J12131+0700 & * & 0.052$\pm$0.005 & 1.53$\pm$0.05 & Sy1.5/1.8 & 0.2095 & 1111.7& 0.017 & 2.2 (2--10) \\
 & * & [0.064$\pm$0.006] & [1.63$\pm$0.05] & & & & & 290.0 (20--100) \\

 & & & & & & & & \\

IGR J13038+5348 & * & 21$\pm$1 & 4.4$\pm$0.2 & Sy1.2 & 0.030 & 141.5 & 0.020 & 0.079 (0.1--2.4) \\
 & * & [22$\pm$1] & [4.7$\pm$0.2] & & & & & 7.5 (17--60) \\

 & & & & & & & & \\

IGR J13109$-$5552 & * & --- & 0.022$\pm$0.009 & Sy1 & 0.104 & 516.9 & 0.325 & 18.2 (0.2--12) \\
 & * & --- & [0.053$\pm$0.011] & & & & & 77.2 (20--100) \\

 & & & & & & & & \\

IGR J14471$-$6414 & * & 0.12$\pm$0.02 & 0.030$\pm$0.003 & Sy1.2 & 0.053 & 254.3 & 0.814 & 3.7 (2--10) \\
 & * & [1.6$\pm$0.3] & [0.36$\pm$0.04] & & & & & 9.3 (20--100) \\

 & & & & & & & & \\

IGR J16056$-$6110 & * & 8.0$\pm$0.8 & 5.76$\pm$0.03 & Sy1.5 & 0.052 & 249.3 & 0.259 & 0.52 (0.1--2.4) \\
 & * & [16.7$\pm$1.7] & [12.6$\pm$0.6] & & & & & 1.0 (2--10) \\
 & & & & & & & & 11.9 (20--100) \\

 & & & & & & & & \\

IGR J16385$-$2057 & * & 9.0$\pm$0.9 & 1.3$\pm$0.1 & NLSy1 & 0.0269 & 126.6 & 0.500 & 0.53 (0.1--2.4) \\
 & * & [44$\pm$4] & [22$\pm$2] & & & & & 3.1 (20--100) \\

 & & & & & & & & \\

2E 1739.1$-$1210 & * & 2.0$\pm$0.5 & 0.53$\pm$0.05 & Sy1.2 & 0.037 & 175.5 & 0.573 & 0.59 (0.1--2.4) \\
 & * & [16$\pm$3] & [2.2$\pm$0.2] & & & & & 3.7 (0.4--4) \\
 & & & & & & & & 10.1 (20--100) \\

 & & & & & & & & \\

IGR J19405$-$3016 & * & 4.7$\pm$0.3 & 1.47$\pm$0.07 & Sy1.2 & 0.052 & 249.3 & 0.103 & 2.6 (0.1--2.4) \\
 & * & [6.1$\pm$0.4] & [2.0$\pm$0.1] & & & & & 0.82 (2--10) \\
 & & & & & & & & 13.5 (20--100) \\

 & & & & & & & & \\

IGR J21272+4241 & 0.14$\pm$0.01 & 0.068$\pm$0.014 & 0.229$\pm$0.012 & Sy1.8/1.9 & 0.316 & 1788.9 & 
0.412 & 10.1 (2--10)  \\
 & [0.268$\pm$0.013] & [0.18$\pm$0.04] & [0.60$\pm$0.04] & & & & & 260.6 (20--40) \\
 & & & & & & & & $<$180.2 (40--100) \\

\noalign{\smallskip} 
\hline
\noalign{\smallskip} 
\multicolumn{9}{l}{Note: emission line fluxes are reported both as observed and 
(between square brackets) corrected for the intervening Galactic absorption} \\
\multicolumn{9}{l}{$E(B-V)_{\rm Gal}$ along the object line of sight. Line fluxes are in 
units of 10$^{-14}$ erg cm$^{-2}$ s$^{-1}$, whereas X--ray luminosities are in units of} \\
\multicolumn{9}{l}{10$^{43}$ erg s$^{-1}$ and the reference band (between round brackets) 
is expressed in keV. Errors and limits are at 1$\sigma$ and 3$\sigma$ confidence levels,} \\ 
\multicolumn{9}{l}{respectively. The typical error on the redshift measurement is $\pm$0.001 
but for the SDSS and 6dFGS spectra, for which an uncertainty of} \\ 
\multicolumn{9}{l}{$\pm$0.0003 can be assumed.} \\
\multicolumn{9}{l}{$^*$: heavily blended with [N {\sc ii}] lines} \\
\noalign{\smallskip} 
\hline
\hline
\end{tabular} 
\end{center} 
\end{table*}

\begin{table*}[th!]
\caption[]{Synoptic table containing the main results concerning the 15
narrow emission line AGNs (see Figs. 11 and 12) identified or observed in 
the present sample of {\it INTEGRAL} sources.}
\scriptsize
\begin{center}
\begin{tabular}{lcccccrccr}
\noalign{\smallskip}
\hline
\hline
\noalign{\smallskip}
\multicolumn{1}{c}{Object} & $F_{\rm H_\alpha}$ & $F_{\rm H_\beta}$ &
$F_{\rm [OIII]}$ & Class & $z$ &
\multicolumn{1}{c}{$D_L$} & \multicolumn{2}{c}{$E(B-V)$} &
\multicolumn{1}{c}{$L_{\rm X}$} \\
\cline{8-9}
\noalign{\smallskip}
 & & & & & & (Mpc) & Gal. & AGN & \\
\noalign{\smallskip}
\hline
\noalign{\smallskip}

IGR J00040+7020 & 0.58$\pm$0.06 & 0.088$\pm$0.006 & 1.20$\pm$0.06 & Sy2 & 0.096 & 474.6 & 0.843 & 0 & 11.4 (2--10) \\
 & [3.2$\pm$0.3] & [1.12$\pm$0.16] & [13.2$\pm$0.7] & & & & & & 39.1 (20--100) \\

 & & & & & & & & & \\

IGR J00256+6821 & 0.23$\pm$0.02 & $<$0.017 & 0.31$\pm$0.03 & Sy2 & 0.012 & 55.9 & 1.027 & $>$0.63 & 0.019 (2--10) \\
 & [2.5$\pm$0.3] & [$<$0.47] & [7.9$\pm$0.8] & & & & & & 0.53 (20--100) \\

 & & & & & & & & & \\

IGR J01528$-$0326 & 7.8$\pm$0.4 & in abs. & 2.8$\pm$0.8 & likely Sy2 & 0.0167 & 79.3 & 0.029 & --- & 0.30 (2--10) \\
 & [8.2$\pm$0.4] & $''$ & [3.0$\pm$0.8] & & & & & & 2.1 (20--100) \\

 & & & & & & & & & \\

IGR J02343+3229 & 0.10$\pm$0.01 & in abs. & 0.32$\pm$0.05 & Sy2/LINER & 0.015 & 74.7 & 0.099 & --- & 2.6 (17--60) \\
 & [0.125$\pm$0.012] & $''$ & [0.42$\pm$0.06] & & & & & & \\

 & & & & & & & & & \\

IGR J02504+5443 & 0.17$\pm$0.02 & 0.021$\pm$0.006 & 0.29$\pm$0.05 & Sy2& 0.015 & 70.0 & 0.774 & 0.45 & 0.28 (2--10) \\
 & [1.1$\pm$0.1] & [0.25$\pm$0.07] & [3.4$\pm$0.5] & & & & & & 1.9 (20--100) \\

 & & & & & & & & & \\

IGR J07437$-$5137 & 1.12$\pm$0.11 & in abs. & $<$0.1 & Sy2? & 0.025 & 117.5 & 0.299 & --- & 1.5 (20--40) \\
 & [2.2$\pm$0.2] & $''$ & [$<$0.16] & & & & & & $<$1.1 (40--100) \\

 & & & & & & & & & \\

IGR J11366$-$6002 & 0.33$\pm$0.03 & 0.028$\pm$0.008 & 0.055$\pm$0.008 & Sy2/LINER & 0.014 & 65.3 & 0.940 & 0.47 & 0.23 (2--10) \\
 & [2.6$\pm$0.3] & [0.58$\pm$0.17] & [1.05$\pm$0.16] & & & & & & 0.66 (20--100) \\

 & & & & & & & & & \\

IGR J13149+4422 & 4.49$\pm$0.13 & 0.90$\pm$0.04 & 2.04$\pm$0.06 & Sy2/LINER & 0.0353 & 167.2 & 0.019 & 0.52 & 9.7 (0.2--12) \\
 & [4.57$\pm$0.14] & [0.95$\pm$0.05] & [2.17$\pm$0.07] & & & & & & 7.2 (17--60) \\

 & & & & & & & & & \\

IGR J14561$-$3738 & 3.8$\pm$0.4 & in abs. & 1.5$\pm$0.3 & likely Sy2 & 0.024 & 112.7 & 0.084 & --- & 0.62 (3--8) \\
 & [4.5$\pm$0.5] & $''$ & [2.1$\pm$0.4] & & & & & & 0.88 (8--20) \\
 & & & & & & & & & 2.1 (17--60) \\

 & & & & & & & & & \\

IC 4518a & 11$\pm$1 & 2.0$\pm$0.1 & 17.6$\pm$0.5 & Sy2 & 0.016 & 75.9 & 0.157 & 0.44 & 1.5 (20--100) \\
 & [15.8$\pm$1.6] & [3.57$\pm$0.18] & [28.9$\pm$0.9] & & & & & & \\

 & & & & & & & & & \\

IGR J15161$-$3827 & 13.2$\pm$0.7 & 2.69$\pm$0.13 & 6.14$\pm$0.18 & Sy2 & 0.0365 & 173.0 & 0.099 & 0.38 & 0.86 (3--8) \\
 & [16.3$\pm$0.8] & [3.9$\pm$0.2] & [8.5$\pm$0.3] & & & & & & $<$1.3 (8--20) \\
 & & & & & & & & & 5.4 (20--100) \\

 & & & & & & & & & \\

IGR J15539$-$6142 & 5.7$\pm$0.3 & $<$0.34 & 1.4$\pm$0.3 & Sy2 & 0.015 & 69.6 & 0.341 & $>$1.50 & 1.2 (20--100) \\
 & [12.3$\pm$0.6] & $<$0.98 & [4.5$\pm$1.1] & & & & & & \\

 & & & & & & & & & \\

IGR J16024$-$6107 & 3.04$\pm$0.15 & 0.57$\pm$0.06 & 1.01$\pm$0.07 & Sy2 & 0.011 & 52.9 & 0.334 & 0.15 & 0.012 (0.1--2.4) \\
 & [6.6$\pm$0.3] & [2.0$\pm$0.2] & [2.9$\pm$0.2] & & & & & & 0.060 (2--10) \\
 & & & & & & & & & 0.23 (20--40) \\
 & & & & & & & & & $<$0.16 (40--100) \\

 & & & & & & & & & \\

IGR J23308+7120 & 0.17$\pm$0.03 & in abs. & 0.27$\pm$0.03 & likely Sy2 & 0.037 & 175.5 & 0.663 & --- & 0.52 (2--10) \\
 & [0.73$\pm$0.15] & $''$ & [2.0$\pm$0.2] & & & & & & 2.5 (20--40) \\
 & & & & & & & & & $<$2.1 (40--100) \\

 & & & & & & & & & \\

IGR J23524+5842 & --- & $<$0.01 & 0.06$\pm$0.01 & likely Sy2 & 0.164 & 849.9 & 1.290 & --- & 110.8 (20--100) \\
 & --- & $<$0.3 & [1.9$\pm$0.3] & & & & & & \\

\noalign{\smallskip}
\hline
\noalign{\smallskip}

\multicolumn{10}{l}{Note: emission line fluxes are reported both as observed and 
(between square brackets) corrected for the intervening Galactic absorption} \\
\multicolumn{10}{l}{ $E(B-V)_{\rm Gal}$ along the object line of sight. Line fluxes are in 
units of 10$^{-14}$ erg cm$^{-2}$ s$^{-1}$, whereas X--ray luminosities are in units of} \\
\multicolumn{10}{l}{10$^{43}$ erg s$^{-1}$ and the reference band (between round brackets) is 
expressed in keV. Errors and limits are at 1$\sigma$ and 3$\sigma$ confidence levels,} \\
\multicolumn{10}{l}{respectively. The typical error on the redshift measurement is $\pm$0.001 
but for the SDSS and 6dFGS spectra, for which an uncertainty of} \\
\multicolumn{10}{l}{$\pm$0.0003 can be assumed.} \\
\noalign{\smallskip}
\hline
\hline
\end{tabular}
\end{center}
\end{table*}

Globally, 28 objects of our sample show optical spectra which are 
dominated by redshifted broad and/or narrow emission lines typical of 
AGNs. We classify slightly more than half of this subsample (15 objects) 
as genuine or likely Seyfert 2 galaxies; three of them appear to be 
borderline objects between Seyfert 2s and LINERs. The remaining 13 
sources have been identified as Seyfert 1 galaxies: in detail, 4 are 
classified as Seyfert 1.2, 4 as Seyfert 1.5, one as borderline Seyfert 
1.5/1.8, 2 as Seyfert 1.8 or Seyfert 1.9 
and one as narrow-line (NL) Seyfert 1; for the case of IGR J13109$-$5552
only a general Seyfert 1 classification can be given due to its anomalous 
optical spectral appearance (see Fig. 9, lower left panel).

The main observed and inferred parameters for each object are reported 
in Tables 5 and 6 for the broad emission line and narrow emission line AGNs,
respectively. We assumed a null local absorption for Seyfert 1 AGNs.
In these tables, X--ray luminosities were computed from the fluxes reported 
in Bird et al. (2007), in Krivonos et al. (2007), in Landi et al. 
(2007b,c,d,e,f), in the {\it Rossi-XTE} 
All-Sky Slew Survey of Revnivtsev et al. (2004b), in the {\it XMM-Newton}
Slew Survey (Saxton et al. (2008)), or using the {\it ROSAT}/PSPC 
(Voges et al. 1999) and {\it Einstein}/IPC (Harris et al. 1994) 
countrates.

For 15 out of 28 objects reported in this section, the redshift value 
was determined for the first time using the data presented here. 
We remark that the redshift of IGR J16056$-$6110 given in our 
preliminary analysis (Masetti et al. 2007a) was incorrect; here we 
give the correct value.
The redshifts of the remaining 13 sources are consistent with those 
reported in the literature except for galaxy LEDA 21656 (the putative
optical counterpart of IGR J07347$-$5137), for which we measure $z$ = 
0.025 instead of $z$ = 0.0086 as reported in the Hyperleda catalogue
(Prugniel 2005).

For two of the hard X--ray objects in the sample (IGR J06117$-$6625 and 
IGR J12131+0700) we note that no entry with these names is found in any 
catalogue of {\it INTEGRAL} sources. This is because the identifications 
for these two objects, given in Bird et al. (2007) and based on 
positional grounds only, were not correct as these sources were labeled 
there as PKS 0611$-$663 and NGC 4180, respectively.
As a matter of facts, however, the {\it ROSAT} and {\it Swift}/XRT 
positions of these two sources, respectively, indicate that actually
the X--ray emitting objects within the corresponding {\it INTEGRAL} error 
boxes are not those initially reported in the 3$^{\rm rd}$ IBIS Survey,
but rather the ones presented here (see also the preliminary reports of
Masetti et al. 2007a and Landi et al. 2007e). Therefore, here we changed 
their names using the typical notation for the newly-identified {\it 
INTEGRAL} sources.

Moreover, we drop the statement made in Masetti et al. (2007a) concerning 
the possible Seyfert 2 nature of NGC 4180, given that no detectable X--ray 
emission is seen from the nucleus of this galaxy in the corresponding XRT 
pointing. There may be the possibility (as mentioned in Landi et al. 
2007e) that the nucleus of NGC 4180 is extremely absorbed, being thus the 
actual responsible of the high-energy emission detected with {\it 
INTEGRAL} but on the contrary very faint below 10 keV; however, the 
presence of an X--ray emitting broad emission line AGN in the same IBIS 
error box makes this eventuality rather unlikely. Likewise, this latter 
fact suggests that XRT source \#2 in the field of this source (reported by 
Landi et al. 2007e) marginally (if at all) contributes to the hard X--ray 
emission detected by IBIS as IGR J12131+0700, and it hardly is the soft 
X--ray counterpart of this {\it INTEGRAL} source.

It is noted that, for three sources (IGR J03334+3718, IGR J13038+5348 and 
IGR J16024$-$6107), the corresponding putative optical counterpart lies a 
few arcsecs outside the nominal {\it ROSAT} error circle. However, recent 
{\it Swift}/XRT pointings on the two latter sources show that their soft 
X--ray positions are completely consistent with the proposed optical 
counterparts. The same may be true for IGR J03334+3718; besides, the 
optical spectral appearance of its proposed counterpart strongly points to 
the fact that the association proposed by Burenin et al. (2006a), and 
studied in more detail here, is indeed correct. Similarly, the XRT and 
{\it XMM-Newton} positions for source IGR J13109$-$5552 are marginally 
consistent with each other at the 2$\sigma$ level, so we consider them 
to be related to the same source.

As mentioned in Sect. 2, the putative optical counterparts of sources 
IGR J07437$-$5137 and IGR J14561$-$3738 were not chosen via soft 
X--ray catalogues 
cross-correlation. However, all of them are positionally consistent 
with a radio (NVSS or SUMSS) and a far-infrared (IRAS) source. This, 
despite the lack of a catalogued arcsec-sized X--ray counterpart,
makes these optical objects reliable counterpart candidates for these
hard X--ray sources. 
Indeed, the results (Landi et al. 2006, 2007a) for the three objects 
in Paper V without univocal arcsec-sized soft X--ray position and 
selected with similar criteria indicate that this approach is in any 
case quite successful for the selection of candidates lacking archival 
soft X--ray observations of the field of the corresponding 
hard X--ray source.

Nevertheless, we once again stress that the counterparts we propose for 
these sources, although likely, need confirmation through pointed soft 
X--ray observations with satellites capable of arcsec-sized localizations 
(such as {\it Chandra}, {\it XMM-Newton} or {\it Swift}). Keeping these 
caveats in mind, the above associations will be assumed to be correct in 
the following analysis.

Using our [O {\sc iii}] $\lambda$5007 line flux measurements, 
together with the local absorption estimate obtained from the optical 
spectra (see Table 6) and the 2--10 keV fluxes in Landi et al. 
(2007c,d), we can determine the Compton nature of Seyfert 2 AGNs
IGR J00040+7020, IGR J02504+5443, IGR J11366$-$6002 and IGR J16024$-$6107
(see Bassani et al. 1999 for details on the procedure).
It is found that the (2--10 keV)/[O {\sc iii}] flux ratios of these 
objects range between $\sim$30 and $\sim$100: this means that all of them 
are Compton thin Seyfert 2 AGNs.
Likewise, for the case of Seyfert 2 AGN IGR J15161$-$3827, the 
available soft X--ray information in the 3--8 keV range (Revnivtsev et al. 
2004b) allows us to say that this object has a (3--8 keV)/[O {\sc iii}] 
flux ratio of $\sim$9, indicating that it is a Compton thin source as 
well. We remark that Bassani et al. (1999) used X--ray measurements in 
the 2--10 keV band; therefore, the X--ray/[O {\sc iii}] flux ratio measure 
obtained for IGR J15161$-$3827 should be considered a strict lower limit 
of the actual value.

A thorough study of the soft X--ray properties of all sources of 
Table 6 detected below 10 keV with {\it Swift} and other satellites 
is now underway (Landi et al., in preparation) in order to definitely 
determine their Compton nature.

As seen in Fig. 1, the optical counterpart of {\it INTEGRAL} source
IGR J00256+6821 is the western nucleus of the double-nucleus galaxy 
LEDA 136991, as confirmed by the analysis of the optical spectrum 
and by the soft X--ray position afforded with {\it Swift}/XRT.
For completeness of the information, we here report that optical 
spectra of the eastern nucleus, acquired simultaneously with those of
the western one, indicate that both indeed lie at the same redshift;
however, no trace of peculiar (emission) features are found in the
eastern nucleus of LEDA 136991, thus strengthening the conclusion that
only the western nucleus is responsible for the hard X--ray emission.

When one looks at Table 6, one finds that, among the Seyfert 2 AGNs for 
which an estimate of the local absorption can be obtained, IGR J00040+7020 
seems to have no reddening local to the AGN host. This suggests that this 
source may be a ``naked" Seyfert 2 AGN, i.e. a source which lacks the 
broad-line region (BLR; see, e.g., Panessa \& Bassani 2002 and Bianchi et 
al. 2007). A low AGN accretion rate ($\leq$10$^{-3}$: e.g., Nicastro 2000) 
may be the cause of an unformed BLR; in this case, given the relatively 
high luminosity of IGR J00040+7020, a black hole mass of 
$>$1.2$\times$10$^8$ $M_\odot$ is required, assuming an average bolometric 
correction (Risaliti \& Elvis 2004).

Looking at the optical spectrum of the counterpart of IGR J23524+5842
we see that, at the redshift of this source, the H$_\alpha$+[N {\sc ii}]
complex regrettably falls right in the O$_2$ telluric band at 7605 \AA,
so no reliable measurements on these lines are possible.
Nevertheless, the other line ratios which can be computed from
this spectrum allow us to classify this source as a likely Seyfert 2 AGN.

It is also noticeable that the spectrum of IGR J13109$-$5552 shows 
a bumpy continuum which is quite anomalous for AGNs or for their host 
galaxies (see Figs. 9-12).
A possible explanation (as already seen in Papers I and V) may be the 
presence of an interloping optical object in the direction of the
counterpart of this X--ray source. However, the `bumpiness' of this
continuum does not resemble that of any stellar spectral type.
High-resolution imaging and higher S/N spectra are thus desirable to 
shed light on this puzzling source.

Moreover, as already mentioned, we classify IGR J16385$-$2057 as a NL 
Seyfert 1 AGN, as its optical spectrum complies with the criteria of 
Osterbrock \& Pogge (1985) concerning the presence of the Fe {\sc ii} 
bump and the narrowness of the Full Width at Half Maximum (FWHM) of the 
H$_\beta$ emission.

To conclude this Section, following Wu et al. (2004) and Kaspi et al. 
(2000), we can compute an estimate of the mass of the central black hole 
in 11 of the 13 objects classified as Seyfert 1 AGN (this procedure could 
not be applied to IGR J09532$-$6231 as no broad H$_\beta$ emission component 
was detected, and to IGR J13109$-$5552 because no H$_\beta$ emission 
velocities $v_{\rm BLR}$ (measured from the H$_\beta$ emission line FWHM) 
and the corresponding black hole masses for these 11 cases are reported in 
Table 7.

\begin{table}
\caption[]{BLR gas velocities (in km s$^{-1}$) and 
central black hole masses (in units of 10$^7$ $M_\odot$) for 11
Seyfert 1 AGNs belonging to the sample presented in this paper.}
\begin{center}
\begin{tabular}{lcc}
\noalign{\smallskip}
\hline
\hline
\noalign{\smallskip}
\multicolumn{1}{c}{Object} & $v_{\rm BLR}$ & $M_{\rm BH}$ \\
\noalign{\smallskip}
\hline
\noalign{\smallskip}

IGR J03334+3718   & 3600 & 6.2 \\
IGR J06117$-$6625 & 8200 & 88  \\ 
IGR J09446$-$2636 & 3700 & 9.7 \\
IGR J12131+0700   & 3900 & 5.1 \\
IGR J13038+5348   & 4900 & 5.4 \\
IGR J14471$-$6414 & 5300 & 2.4 \\
IGR J16056$-$6110 & 2200 & 1.9 \\
IGR J16385$-$2057 & 1500 & 0.7 \\
2E 1739.1$-$1210  & 8200 & 17  \\
IGR J19405$-$3016 & 5500 & 62  \\
IGR J21272+4241   & 1200 & 0.4 \\

\noalign{\smallskip}
\hline
\hline
\noalign{\smallskip}
\end{tabular}
\end{center}
\end{table}

\subsection{Other sources}

\begin{figure*}[th!]
\mbox{\psfig{file=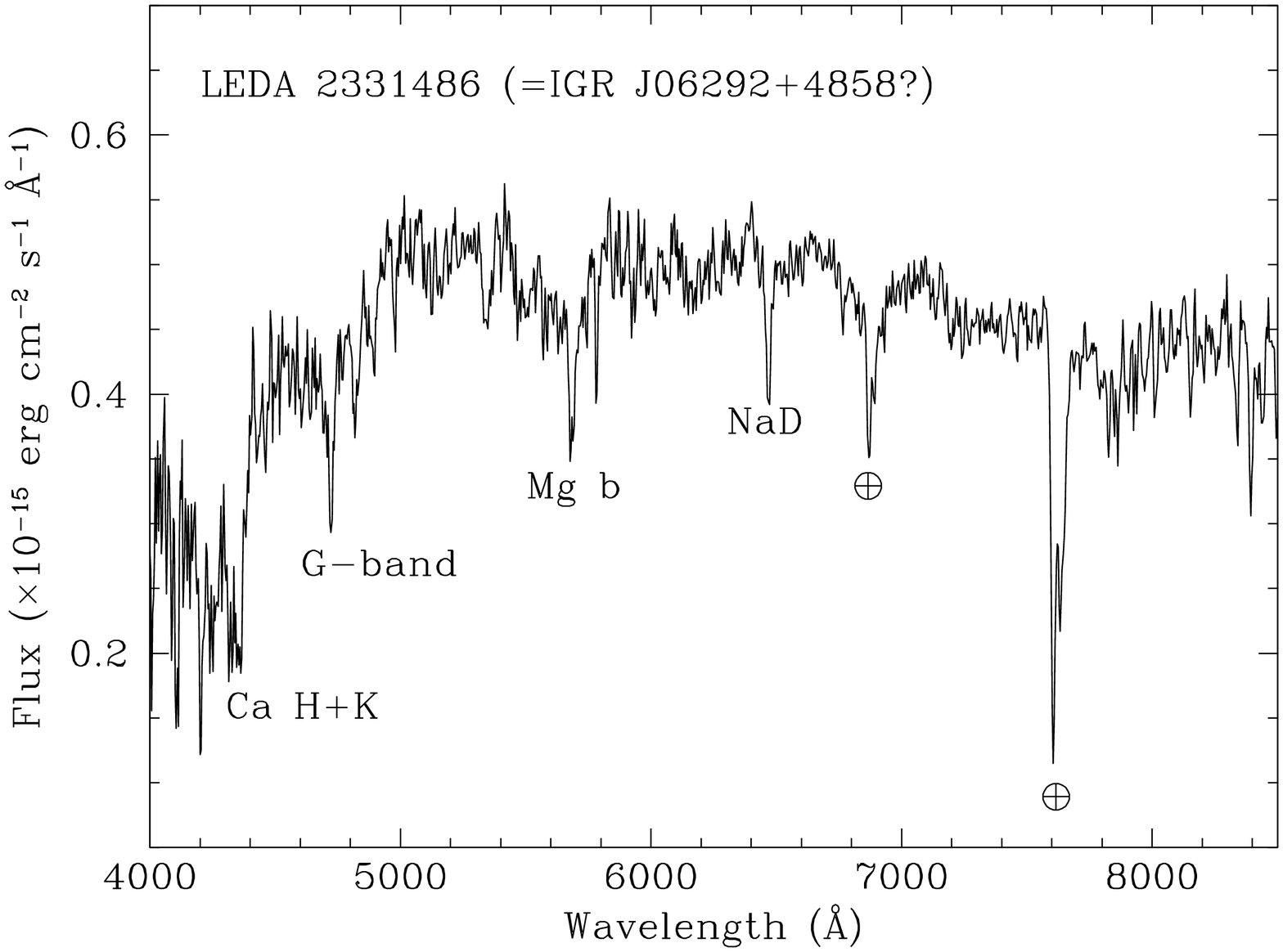,width=9cm,angle=270}}
\mbox{\psfig{file=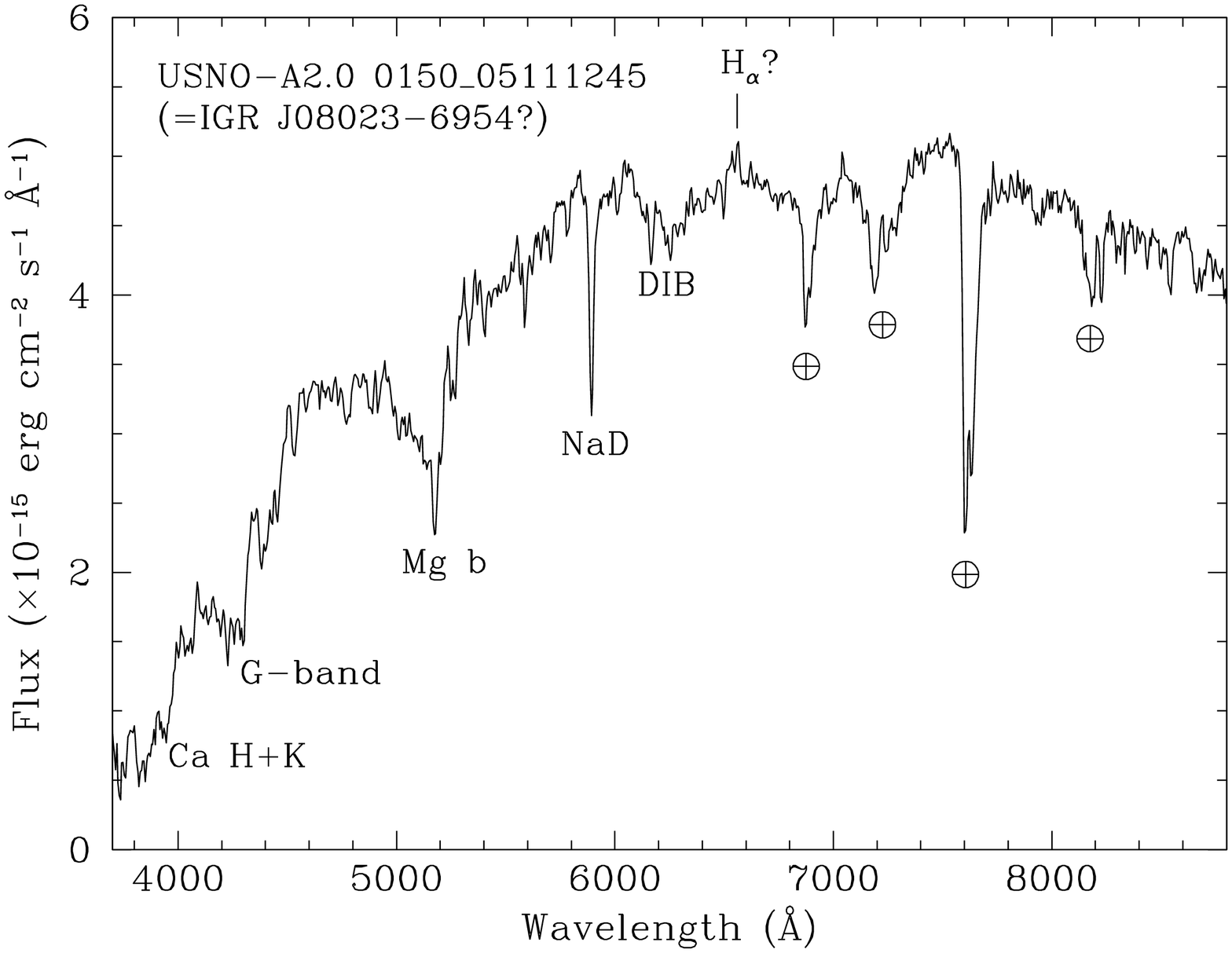,width=9cm,angle=270}}
\vspace{-.5cm}
\caption{Spectra (not corrected for the intervening Galactic absorption) 
of the optical counterparts of the two objects tentatively associated
with {\it INTEGRAL} sources and reported in Sect. 4.4.
For each spectrum the main spectral features are labeled. The 
symbol $\oplus$ indicates atmospheric telluric absorption bands.}
\end{figure*}

We here report on the optical spectra of two objects tentatively
identified as counterparts of two {\it INTEGRAL} sources, IGR 
J06292+4858 and IGR J08023$-$6954. As mentioned in Sect. 2, we
consider these two identifications as tentative due to the fact 
that in one case (IGR J06292+4858) no arcsec-sized soft X--ray 
localization is available, and in the other (IGR J08023$-$6954)
a low S/N ($\la$3$\sigma$) detection with {\it Swift/XRT} was
obtained.

In the case of IGR J06292+4858, we spectroscopically observed the
relatively bright ($R \sim$ 15.3) galaxy LEDA 2331486 located 
within the IBIS error circle of this hard X--ray source.
Our spectroscopy (Fig. 13, left panel) shows that the object
has the characteristics of an early-type galaxy, with absorption
features at redshift $z$ = 0.097$\pm$0.001. 

In order to better classify this source, we followed 
the approach of Laurent-Muehleisen et al. (1998) by using the following 
relevant information: (1) absence of emission lines, with an upper limit 
to their EW of $\sim$5 \AA; (2) absence of strong Balmer absorption 
lines; (3) presence and strength of other absorption features (such as
the G and the Mg {\sc i} bands and the Ca {\sc ii} H+K doublet) superimposed 
on the galaxy continuum; (4) Ca {\sc ii} break contrast at 4000 
\AA~(Br$_{\rm 4000}$), as defined by Dressler \& Shectman (1987), 
with a value $\sim$40\%. All of the above suggests that this
galaxy can be tentatively classified as a possible BL Lac object.
If indeed IGR J06292+4858 and LEDA 2331486 are the same object, we
obtain 20--40 keV and 40--100 keV luminosities of 1.8$\times$10$^{45}$
erg s$^{-1}$ and $<$1.6$\times$10$^{45}$ erg s$^{-1}$,
assuming a luminosity distance $d_L$ = 479.9 Mpc and the fluxes reported
in Bird et al. (2007).

Concerning IGR J08023$-$6954, the optical spectrum of its putative
optical counterpart, star USNO-A2.0 0150\_05111245 (Fig. 13, right panel)
shows the continuum typical of a late-G/early-K star, with the
presence of a weak H$_\alpha$ emission at $z$ = 0, with flux 
(4.2$\pm$0.8)$\times$10$^{-15}$ erg cm$^{-2}$ s$^{-1}$
and EW = 0.86$\pm$0.17 \AA. This suggests a chromospherically active 
star identification (e.g., an RS CVn type system) for this source
if the low S/N XRT detection is indeed real.

Of course, deeper multiwavelegth followup is needed to confirm (or 
disprove) the two tentative identifications above.

\subsection{Statistical considerations}

We can now briefly update the statistical approach made in Paper V
by including the results presented here, along with recent 
identifications of {\it INTEGRAL} sources as HMXBs (Hannikainen et 
al. 2007; Leyder et al. 2007; Masetti et al. 2007h), LMXBs (Masetti 
et al. 2007c) and AGNs (Bassani et al. 2007; Masetti et al. 2008).

It is found that, presently, of the 97 {\it INTEGRAL} sources 
identified through optical or NIR spectroscopy, 55 (57\%) are AGNs
(almost equally divided into Seyfert 1s and Seyfert 2s), 30 (31\%) are 
X--ray binaries (with a large majority, i.e. more than 86\%, of HMXBs), 
and 11 (11\%) are CVs, with 8 of them definitely or likely 
belonging to the IP subclass (see Papers IV-V and the present work)
and 3 of symbiotic star type.

One can compare (see Fig. 14), for instance, these percentages with 
those of the 308 identified objects belonging to the largest 
catalogue of {\it INTEGRAL} sources published up to now, i.e., the 
3$^{\rm rd}$ IBIS Survey (Bird et al. 2007). In this 
survey we have 147 (48\%) X--ray binaries (about half of which are 
HMXBs), 118 (38\%) AGNs and 23 (7\%) CVs, with most (perhaps all) 
of them of magnetic nature (IPs or Polars; see Barlow et al. 2006).

These numbers confirm once more the effectiveness of this method of 
catalogue cross-correlation plus optical spectroscopy followup in 
revealing the AGN and CV nature of a large number of unidentified {\it 
INTEGRAL} sources, while it may introduces a bias against detecting X--ray 
binaries: this is however quite understandable, given that, as {\it 
INTEGRAL} is sensitive to the hard X--ray radiation, it can easily detect 
sources deeply embedded in the Galactic Plane dust and therefore heavily 
obscured (as already stressed by, e.g., Walter 2007). Thus, we foresee 
that the use of NIR spectroscopy followup will reveal a larger number of 
absorbed X--ray binaries.

We conclude by remarking that, of the 97 optical and NIR spectroscopic 
identifications mentioned here, 78 were obtained within the framework of 
our spectroscopic follow-up program (Papers I-V, the present work, and 
references therein).

\begin{figure}[h!]
\hspace{-0.5cm}
\psfig{file=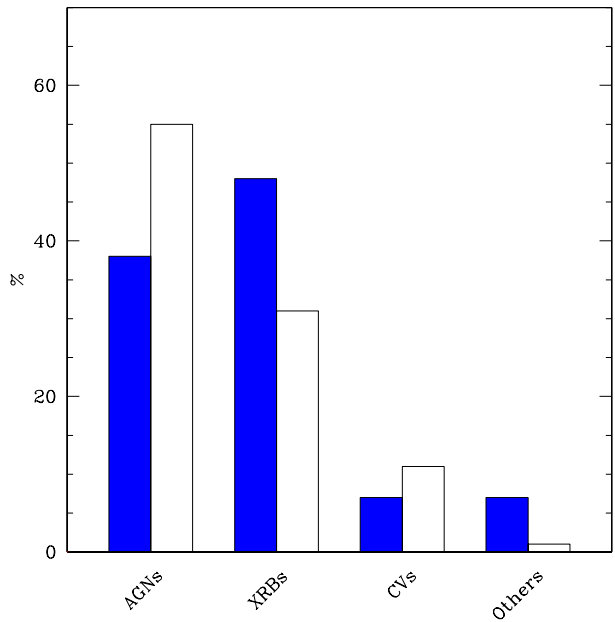,width=9.5cm}
\vspace{-0.7cm}
\caption{Histogram, subdivided into source types, showing the percentage 
of {\it INTEGRAL} objects of known nature and belonging to the 
3$^{\rm rd}$ IBIS Survey (Bird et al. 2007; left-side, 
darker columns), and {\it INTEGRAL} sources from various surveys 
and identified through optical or NIR spectroscopy (right-side, lighter 
columns).}
\end{figure}

\section{Conclusions}

In our continuing effort of identification of {\it INTEGRAL} sources by 
means of optical spectroscopy (Papers I-V), we have identified and 
studied 39 hard X--ray objects of unknown or poorly explored nature 
by means of a multisite campaign at 8 different telescopes and
using the archival data of 2 spectroscopic surveys.

We found that the selected sample is made of 29 AGNs (13 of which are of 
Seyfert 1 type, 15 are Seyfert 2 AGNs and 1 is possibly a BL Lac), 5 HMXBs, 
2 LMXBs, 1 magnetic CV, 1 symbiotic star and 1 active star. In terms 
of relative sizes of these groups, we notice the overwhelming majority 
(74\%) of AGNs in the present sample.

We recall that, in four cases (IGR J06292+4858, IGR J07437$-$5137, 
IGR J08023$-$6954 and IGR J14561$-$3738), only a tentative albeit 
likely optical counterpart was given because of the lack of a definite  
arcsec-sized soft X--ray position. Thus, for them an observation with 
sufficient S/N to be achieved using soft X--ray satellites affording 
arcsec-sized localizations (such as {\it Chandra}, {\it XMM-Newton} 
or {\it Swift}) is needed to confirm the proposed association.

The results presented in this work further indicate the high effectiveness 
of this method of catalogue cross-correlation plus optical spectroscopy to 
pinpoint the nature of the still unidentified {\it INTEGRAL} sources.
We now plan to extend this approach to NIR spectroscopy.

\begin{acknowledgements}

We thank Silvia Galleti for Service Mode observations at the Loiano 
telescope; Hripsime Navasardyan for Service Mode observations at the
Asiago Telescope; Don Pollacco Neil Mahoney for Service Mode observations 
at the WHT; Stefano Bernabei, Ivan Bruni, Antonio De Blasi and Roberto 
Gualandi for night assistance at the Loiano telescope;
Claudio Aguilera, Arturo G\'omez, Edgardo Cosgrove and Alberto Miranda
for day and night assistance at the CTIO telescope; 
Antonio De Franceschi for night assistance at CASLEO; 
Gaspare Lo Curto and Alessandro Ederoclite for the support at the ESO 
telescopes; Ariel S\'anchez and Andr\'es Gonz\'alez for night 
assistance at the ESO 3.6m telescope. We also thank the anonymous referee 
for useful remarks which helped us to improve the paper. 
This research has made use of the ASI Science Data Center Multimission 
Archive; it also used the NASA Astrophysics Data System Abstract Service, 
the NASA/IPAC Extragalactic Database (NED), and the NASA/IPAC Infrared 
Science Archive, which are operated by the Jet Propulsion Laboratory, 
California Institute of Technology, under contract with the National 
Aeronautics and Space Administration. 
This publication made use of data products from the Two Micron All 
Sky Survey (2MASS), which is a joint project of the University of 
Massachusetts and the Infrared Processing and Analysis Center/California 
Institute of Technology, funded by the National Aeronautics and Space 
Administration and the National Science Foundation.
This research has also made use of data extracted from the 6dF 
Galaxy Survey and the Sloan Digitized Sky Survey archives;
it has also made use of the ESO Science Archive operated at Garching bei 
M\"unchen, Germany, of the SIMBAD database operated at CDS, Strasbourg, 
France, and of the HyperLeda catalogue operated at the Observatoire de 
Lyon, France.
The authors acknowledge the ASI and INAF financial support via grant 
No. I/023/05/0.
NM and SAC were Visiting Astronomers at CASLEO, operated under agreement 
between the Consejo Nacional de Investigaciones Cient\'{\i}ficas
y T\'ecnicas de la Rep\'ublica Argentina and the National Universities 
of La Plata, C\'ordoba and San Juan. The CCD and data acquisition system
at CASLEO have been partly financed by R.M. Rich through U.S. NSF grant
AST-90-15827.
GER and SAC were supported by grants PICT 03-13291 BID 1728/OC-AR
(ANPCyT), PIP 5375 (CONICET), and AYA2007-68034-C03-01 (FEDER funds).
LM is supported by the University of Padua through grant No. 
CPDR061795/06. 
NM thanks ESO for the pleasant hospitality in Santiago de Chile
during the preparation of this paper.
\end{acknowledgements}

\end{document}